\documentclass{aa}  

\usepackage{graphicx}
\usepackage{txfonts}
\usepackage{graphicx}	% Including figure files
\usepackage{amsmath}	% Advanced maths commands
\usepackage{amssymb}	% Extra maths symbols
\usepackage{xcolor}
\usepackage{booktabs}
\usepackage[switch]{lineno}

\usepackage[hidelinks]{hyperref}
\hypersetup{
colorlinks=true,
linkcolor=blue,
filecolor=blue,
citecolor = black,
urlcolor=cyan,
}
\newcommand\gio[1]{{#1}}

\begin{document}

   \title{Cygnus OB2 as a test case for particle acceleration in young massive star clusters}
   \authorrunning{Menchiari et al.}
   \titlerunning{Cygnus OB2 as a test case for particle acceleration in YMSCs}

   %\subtitle{I. Overviewing the $\kappa$-mechanism}

   \author{S. Menchiari \inst{1}\fnmsep\inst{2}\fnmsep\thanks{\email{stefano.menchiari@inaf.it}}, G. Morlino\inst{1}, E. Amato\inst{1}, N. Bucciantini\inst{1} \and M.\ T.\ Beltr\'an\inst{1}}

   \institute{INAF - Osservatorio Astrofisico di Arcetri, Largo Enrico Fermi 5, Firenze, Italy
             \and
             Università degli studi di Siena, Dipartimento di scienze fisiche della terra e dell'ambiente, Via Roma 56 - 53100 Siena             
             }

   \date{Received December 1, 2023}

% \abstract{}{}{}{}{} 
% 5 {} token are mandatory
 
\abstract
{In this paper, we focus on the scientific case of Cygnus OB2, a northern sky YMSC located towards the Cygnus X star-forming complex. 
We consider a model that assumes cosmic ray acceleration occurring only at the termination shock of the collective wind of the YMSC and address the question of whether, and under what hypotheses, hadronic emission by the accelerated particles can account for the observations of Cygnus OB2 obtained by Fermi-LAT, HAWC and LHAASO. 
In order to do so, we carefully review the available information on this source, also confronting different estimates of the relevant parameters with ad hoc developed simulations.
Once other model parameters are fixed, the spectral and spatial properties of the emission are found to be very sensitive to the unknown properties of the turbulent magnetic field. Comparison with the data shows that our suggested scenario is incompatible with Kolmogorov turbulence. Assuming Kraichnan or Bohm type turbulence spectra, the model accounts well for the Very High Energy (VHE) data, but fails to reproduce the centrally peaked morphology observed by Fermi-LAT, suggesting that additional effects might be important for lower energy $\gamma$-ray emission. We discuss how additional progress can be made with a more detailed and extended knowledge of the spectral and morphological properties of the emission.}
   {}
   {}
   {}
   {}


   \maketitle
%
%-------------------------------------------------------------------
%\linenumbers
\section{Introduction}

The origin of cosmic rays (CRs) is a century-long enigma that does not have a final solution yet. There is a general consensus that supernova remnants (SNRs) are the main factories of Galactic CRs, based on  both theoretical arguments and observational pieces of evidence. From a theoretical point of view, SNRs are capable of accelerating particles through the diffusive shock acceleration process \citep{Drury:1983} that occurs at the shock front driven by the supersonic motion of the supernova ejecta as they expand in the interstellar medium. In addition, assuming that 10\% of the kinetic energy released in a supernova explosion goes into accelerating particles, a supernova rate of $\sim2$ explosions per century is sufficient to sustain the measured CR luminosity \citep{BaadeCRsSNRsOrigin1934}. From the observational side, X-ray observations of several SNRs have confirmed the presence of relativistic electrons with inferred energies up to tens of TeV, the maximum allowed by efficient radiation loss limited acceleration at the SNR shock \citep{KoyamaXraySNR1995, ReynoldsSNR2008, HelderSNR2012}, while the detection of the pion-bump at $\sim 1$ GeV has proven the hadronic nature of the $\gamma$-ray emission form several SNRs \cite[see][for a review]{Funk:2017}.

However, the SNR paradigm still fails to properly account for some fundamental properties of CRs.
The first issue concerns their composition: CR measurements at Earth have revealed some anomalous isotopic ratios, the most noticeable being the $^{22}$Ne/$^{20}$Ne, which is found $\simeq 5.3$ times higher than the Solar value \citep{WiedenbeckNe1981, BinnsNe2008}.
An additional problem is the maximum energy attainable at SNR shocks. To explain the {\it knee} in the CR spectrum, protons needs to be accelerated at least up to $\sim 1$\,PeV, while observations suggest that SNRs only achieve much lower energies, by at least one order of magnitude. Such observations are indeed in agreement with theoretical models, which predict maximum energies up to $\sim 100$\,TeV for typical SNRs \citep{Bell:2013, Cardillo:2015}, unless extreme conditions are assumed \citep{Cristofari+2022}. 
%A second issue concerns the CR composition at Earth: measurements have revealed anomalous composition of few isotopic ratios, the most noticeable being the $^{22}$Ne/$^{20}$Ne, which is found $\simeq 5.3$ times higher than the \nb{S}olar value \citep{WiedenbeckNe1981, BinnsNe2008} \nb{[how robust is the Solar value as representative of current ISM?]}. 

One possible solution to the above-mentioned problems is that in addition to SNRs, other classes of sources may be contributing to the production of Galactic CRs. In particular, the chemical composition requires that at least a fraction of CRs should come from the acceleration of massive star wind material \citep{Prantzos:2012}. Since the 80s, stellar winds have been proposed as possible CR factories \citep{Casse-Paul:1980, Cesarsky-Montmerle:1983}, especially in the context of Young Massive Star Clusters (YMSCs). The consensus around this hypothesis has rapidly grown in the last decade, as several experiments have detected both high-energy and very-high-energy extended $\gamma$-ray emission in coincidence with various YMSCs, such as Cygnus OB2 \citep{ackermannCocoonFreshlyAccelerated2011a, BartoliIdentificationTeVGammaray2014, HAWCcoll:2021, Astiasarain+2023}, Westerlund 1 \citep{abramowskiDiscoveryExtendedVHE2012, HESSWesterlund12022}, Westerlund 2 \citep{yangDiffuseRayEmissionin2018} NGC 3603 \citep{sahaMorphologicalSpectralStudy2020}. The most common idea is that the energy powering these objects results from the combination of the powerful winds blown by the hundreds of massive stars hosted in the clusters. CR acceleration can be achieved in different ways. One possibility is that CRs are accelerated inside the cluster core, where particle energies are increased by multiple crossings of the shocks induced by wind-wind collisions, \citep{reimerNonthermalHighEnergy2006} or by efficient scattering on the magnetic turbulence caused by such collisions \citep{Bykov_review:2020}. On the other hand, in the case of compact YMSCs, the winds from massive stars can combine together to create a collective cluster wind. In this case, particle acceleration can result from diffusive shock acceleration at the wind termination shock (TS)  \citep{Morlino+2021}. In clusters older than $\sim$ 5 Myr,  multiple supernova explosions are expected to be the dominant energy source for CR production \citep{VieuCRinSuperbubble2022}.

%From the perspective of the last mentioned model, it is interesting to note that the recent HESS observations of the YMSC Westerlund 1 \citep{HESSWesterlund12022} have detected a shell-like morphology $\gamma$-ray emission. This peculiar shape could be caused by high-energy particles begin accelerated at the termination shock, which would favour the acceleration model suggested by \cite{Morlino+2021}.

The potential contribution of YMSCs as CR sources could indeed be the missing piece of the CR origin puzzle, since these sources can potentially account for the CR properties that the SNR paradigm fails to explain: the composition of the YMSC wind material can in principle account for the anomalous $^{22}$Ne to $^{20}$Ne ratio \citep{GuptaNeMSC2020, Tatischeff+2021}; in addition, the maximum achievable energy in these systems can be as high as a few PeV. In the case of acceleration taking place at the wind TS, as shown by \cite{Morlino+2021}, depending on the plasma turbulence spectrum, CRs can smoothly reach PeV energies if $\sim$10 \% of the YMSC wind luminosity ends in the turbulent magnetic field and particle diffusion is close to Bohm-like. An even higher energy could be reached by SNR shocks propagating inside the wind-blown bubble, where  the turbulence level has been preliminarily enhanced by the same stellar winds \citep{Vieu-Reville-Aharonian:2022}. 
It is important to stress that the conditions needed to reach such high energies in YMSCs have not been verified yet: in particular the particle acceleration efficiency and the properties of magnetic turbulence in these systems are currently unknown. In fact, on the theory side, a wind-blown bubble is a very complex system whose proper modelling requires numerical simulations able to take into account several physical processes \cite[see, e.g.][]{LancasterFragmentation2021}. On the observational side, on the other hand, the most direct diagnostics of the relevant physical conditions is offered by the spectro-morphological analysis of the emission in the $\gamma$-ray range, where radiation of hadronic origin may be detected.

%$\gamma$-rays observations are of the highest importance in that the simultaneous measurement of flux, spectrum and the spatial $\gamma$-ray profile can shed light on above mentioned physical conditions \nb{[what does it means?]}. 

In this paper, we consider the YMSC Cygnus OB2, and we try to model, under the assumption of pure hadronic processes, the $\gamma$-ray emission detected by several experiments. We assume that particles are accelerated at the collective wind TS using the model developed by \cite{Morlino+2021}. A similar analysis has been already presented by \cite{Blasi-Morlino:2023}. This paper complements the former work in several aspects. Firstly, we carefully model the local gas distribution as inferred by the measured column density. Secondly, we illustrate the analysis of the stellar population behind the estimate of the total wind power that these two work share, an aspect that is usually treated in a very cursory way. Finally, rather than fixing {\it a priori} a model for particle transport, we consider the magnetic turbulence as an unknown and attempt at fitting the $\gamma$-ray emission from the Cygnus OB2 region (often referred to as the Cygnus {\it cocoon}) with different underlying turbulence models. More specifically, we consider all the most common models of magnetic turbulence in Astrophysics, Kolmogorov's, Kraichnan's and Bohm's, and use $\gamma$-ray observations to discriminate which is viable.

The paper is structured as follows: in \S~\ref{sec:CygOB2}, we describe the general properties of the Cygnus OB2 star cluster. Then, in \S~\ref{sec:model} we present the properties of the expected CR distribution close to a YMSC. In \S~\ref{sec:GammaEmission}, we calculate its associated $\gamma$-ray emission comparing it with available data. A detailed discussion of the analysis is reported in \S~\ref{sec:disc} and conclusion are finally drawn in \S~\ref{sec:conc}.

\section{The Cygnus OB2 association and its surrounding}
\label{sec:CygOB2}
Cygnus OB2 (Cyg~OB2) is one of the most massive and compact OB associations in the Milky Way ($l \simeq 80.22^\circ$, $b \simeq0.79^\circ$) harboring hundreds, possibly thousands, of massive stars. The first study of its population was carried out by \cite{ReddishOB21966}, who inferred a total of 400--3000 OB stars, by counting them on the Palomar Sky Survey plates. \cite{KnodlsederOB22000} found a compatible result using near-infrared images, with an estimated total population of 2600$\pm$400 OB stars, among which 120$\pm$20 O-type stars. Such numbers are probably overestimated due to the problematic background subtraction and the highly patchy extinction pattern towards the association: in a more recent work, that better accounted for these issues, \cite{WrightOB2SFH2010} estimated a total content of $\sim$1200 OB stars, with $\sim$75 O-type stars. 

In terms of spatial distribution, the cluster is compact, and the stars show a centrally peaked radial profile, with a high stellar density at the center of the association, similar to the YMSCs observed in the Large Magellanic Cloud \citep{KnodlsederOB22000}.
%The radial distribution of stars from the observations seems to follow a compact and peaked profile, with a high stellar density at the \nb{center} of the association, similar to the YMSCs observed in the Large Magellanic Cloud \citep{KnodlsederOB22000}.
%Due to the peaked morphology of Cyg~OB2, 
A large fraction of the stars is enclosed in a region of radius $\sim$14 pc. A recent census of this central core has revealed the presence of 3 Wolf-Rayet stars and 166 OB stars, among which 52 are classified as O-type \citep{WrightMassiveStarPopOB22015}. 
 
Several different estimates have been made of the age of Cyg~OB2.
The presence of O-type dwarf stars and high-luminosity blue supergiants in the sample of 85 OB stars selected by \cite{HansonStudyCygOB22003} suggests that Cyg~OB2 should not be older than a few Myr, with a most likely age of 2$\pm1$ Myr. On the other hand, a study of the population of A-type stars indicated the presence of a group of 5--7 Myr old stars, located mainly in the southern part of the association \citep{DrewEarlyAStars2008}. In parallel, X-ray analysis of low-mass stars seems to point to an age of 3--5 Myr \citep{WrightOB2SFH2010}. \cite{WrightMassiveStarPopOB22015} found a typical age of 2--3 Myr and 4--5 Myr by applying, respectively, non-rotating and rotating evolutionary stellar models to a selected sample of 169 OB stars. All these results might be reconciled in a scenario of continuous star formation activity starting $\sim$7 Myr ago and extending to $\sim$1 Myr ago, with a possible peak of star formation around 4--5 Myr ago. 
This is also compatible with the results found by \cite{ComeronNewOB2Members2012}, who suggested a continuous star-forming activity in the region for the last 10 Myr. 

The distance of Cyg~OB2 is a subject of debate. Right after the discovery of the association, \cite{JohnsonObscuredO-Association1954} inferred a distance of $\sim$1500 pc using spectroscopic observations of 11 stars. In the first comprehensive investigation of the Cyg~OB2 population, \cite{ReddishOB21966} estimated a distance of 2100 pc. In the early '90s, independent studies,  based on the method of spectroscopic parallax, evaluated a distance of $\sim$1700 pc \citep{TorresDodgenPhotometryOB1991, MasseyMassiveStarsCYGOB21991}. Currently, the most commonly adopted value is the one determined by \cite{HansonStudyCygOB22003}, who calculated a distance of 1400$\pm$80 pc after a detailed analysis of the absolute magnitude and extinction of 14 OB stars. This estimate is compatible with the position of some molecular clouds in the Cygnus-X region, whose distance has been calculated using maser parallaxes \citep{ryglParallaxesProperMotions2012}. Moreover, the value is also in agreement with the recent result of 1330 $\pm$ 60 pc based on the parallaxes of some eclipsing binaries \citep{KiminkiPredictingGAIA'sParallax2015}. Finally, Gaia's parallaxes from the second data release \citep{BerlanasDisentanglingSpatialSubstructure2019} point out that the association might be composed of two main subgroups, the first located at a distance of $\sim$1350 pc and the second at $\sim 1755\pm 320$~pc.
In this work we assume a distance of 1400 pc in order to retain consistency with \cite{WrightMassiveStarPopOB22015} which used this distance to estimate the spectral types of stars. %We will comment on the possible implications of a larger distance in \S~\ref{sec:disc}. 

%Finally, it is worth stressing that close to Cyg~OB2 there is a second prominent stellar cluster, NGC~6910, located at a distance of $\sim$ 1730~pc and with an estimated age of 5-10 Myr  and mass between 200 and 700 $M_{\odot}$ \citep{Cantat-Gaudin+2020}. The cluster is also characterized by a shallow initial mass function \citep[$\propto m^{-1.7}$, see][]{Kaur+2020} \nb{[is this relevant??]}. \cite{Astiasarain+2023} has also considered NGC~6910 as a possible contributor to the $\gamma$-ray emission in the Cygnus cocoon region, however, as we show in Appendix~\ref{App:Alternative_LwMdot} we estimate a stellar wind luminosity two orders of magnitude lower than for Cyg~OB2. Hence we decided to exclude NGC~6910 from our analysis. 

%\nb{[I believe that we should place here a section describing the Gas distribution. In the intro it was made clear that we improved on previous results on 3 ways: better stellar population, better turbulence model, better gas distribution. In sec 4 there is a specific discussion on the way we modelled the local target, but we lack an intro on the line of the one on the stellar population.]}

\begin{figure*}
\includegraphics[width=\textwidth]{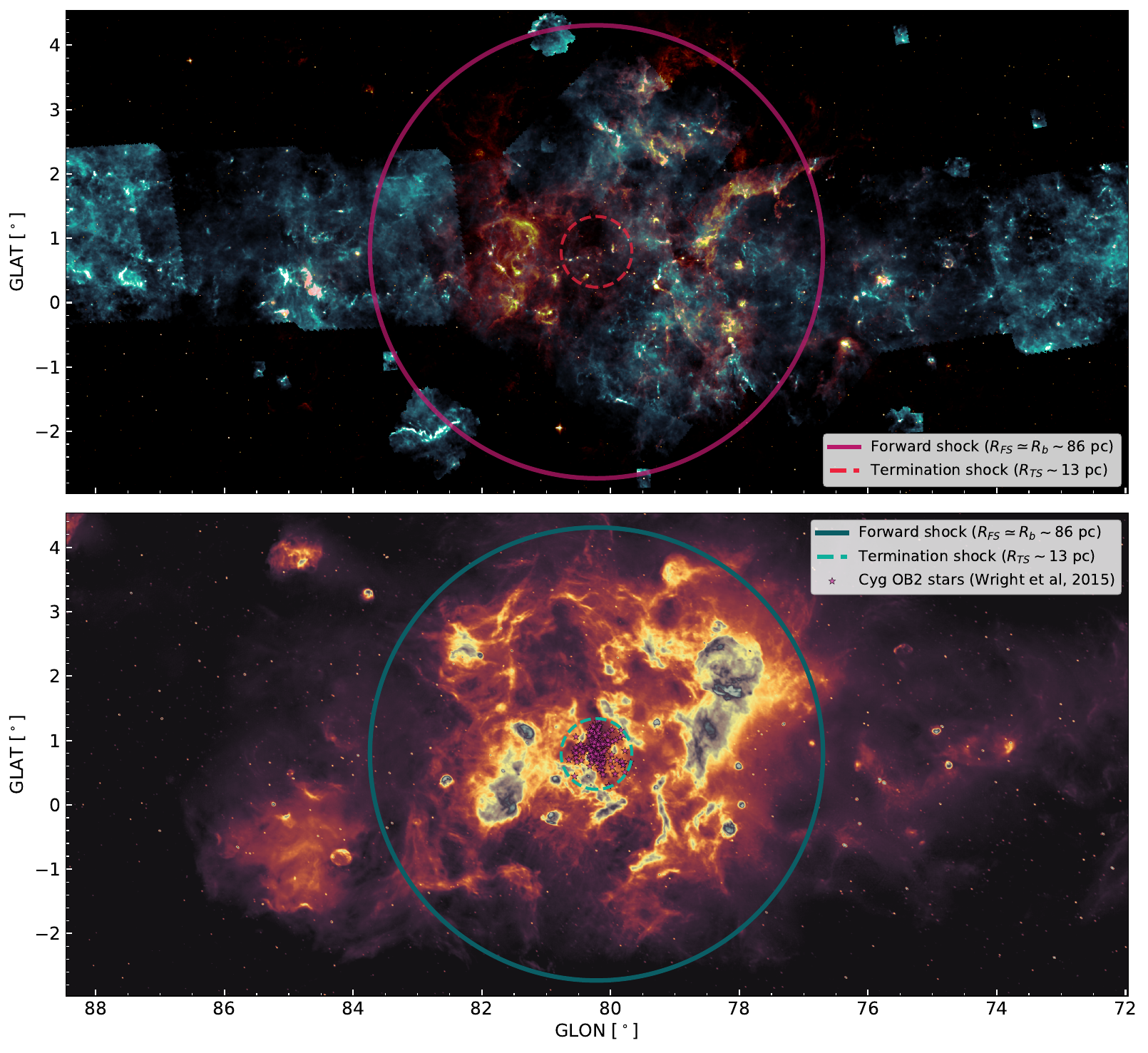}
\caption{Upper panel: Combination of observations at 500 $\mu$m and 8 $\mu$m from the Herschel SPIRE and MSX telescopes of the Cygnus X star-forming complex. Radiation at 500 $\mu$m (light blue scale) indicates emission from cold dust, while the 8 $\mu$m (orange scale) traces photon-dominated regions. Lower panel: Continuum observations at 1.42 GHz from the Canadian Galactic Plane Survey (CGPS) \protect\citep{taylorCANADIANGALACTICPLANE2003}. A large fraction of this emission is likely thermal free-free radiation by material possibly ionized by Cyg~OB2 \protect\citep{EmigFilamentaryStructuresIonizedGasCygX2022}. Color scales are given in arbitrary units. See the online document for a colored version of the image.}
\label{fig:CygusX}
\end{figure*}

 A very important ingredient for our analysis is the gas density in the Cygnus region, since this serves as a target for relativistic protons producing hadronic emission and plays a role in several of our estimates.
 Cyg~OB2 is located at the center of the Cygnus-X star-forming complex \citep{ReipurthStarFormationAndYCinCyg2008}, an extended ($\sim10^\circ$) radio structure hosting numerous molecular clouds \citep{schneiderNewViewCygnus2006}, HII regions \citep{DickelHII1969}  and several other OB associations \citep{UyanikerCygSB2001} (see Figure~\ref{fig:CygusX}). Cygnus-X was initially detected in radio between the 40s and the 50s \citep{HeyFluctuationRF1946, BoltonDiscreteSourcesCyg1948, PiddingtonRFRadCyg1952}, and the first census of radio sources was carried out by \cite{DownesMWObsCygX1966}, who identified 27 strong continuum radio emitters at 5 GHz, later renamed DR sources (DR1 to DR27). From a morphological point of view, the Cygnus-X complex is divided into two major regions: the CygX-North, which includes radio sources from DR17 to DR23, and the CygX-South, which is composed of the radio sources from DR4 to DR15. A large fraction of all these DR sources are part of the same complex, as confirmed both by \cite{schneiderNewViewCygnus2006}, through an analysis of the molecular clouds associated with them, and by \cite{ryglParallaxesProperMotions2012}, who estimated with high accuracy the distance of some of these molecular clouds through precise maser parallax measurements. Interestingly, some of the DR sources are likely located very close to the Cyg~OB2 star cluster. According to \cite{schneiderNewViewCygnus2006}, the molecular clouds DR18 and DR20W exhibit a cometary morphology pointing towards Cyg~OB2, indicating a direct interaction with the radiation and stellar winds from the cluster. Similarly, a recent study of the neutral hydrogen towards the DR21 molecular filament has revealed that the DR21 cloud has formed due to the compression between two HII regions, G081.920+00.138 and the one created by Cyg~OB2 \citep{LiCygXNFilament2023}.
The $\gamma$-ray emission from the most massive of those clouds can be used as an additional diagnostic of the particle acceleration and transport in the bubble, as we discuss in \S~\ref{sec:future}.

Finally, it is worth stressing that close to Cyg~OB2 there is a second prominent stellar cluster, NGC~6910, located at a distance of $\sim$ 1730~pc and with an estimated age of 5--10 Myr and mass $\sim$775~$M_{\odot}$ \citep{Kaur+2020}. \cite{Astiasarain+2023} have also considered NGC~6910 as a possible contributor to the $\gamma$-ray emission in the Cygnus cocoon region, however, as we show in Appendix~\ref{App:Alternative_LwMdot} we estimate a stellar wind luminosity two orders of magnitude lower than for Cyg~OB2. Hence, we decided to exclude NGC~6910 from our analysis. 

%%% SECTION %%%
\section{Particle acceleration model}
\label{sec:model}
\subsection{The wind blown bubble}
\label{subsec:BubbleStructure}
It is well established that the powerful winds from a YMSC can excavate a cavity inside the interstellar medium (ISM). If the cluster is compact enough, namely if the size of the wind termination shock of single stars is larger than the average distance between stars, one can describe the structure and evolution of this cavity as that of a bubble generated by the wind of a single massive star, whose dynamics was investigated by \cite{Weaver-McCray:1977}.

Stellar wind bubbles can be divided in three distinct regions: a central zone, filled with the cold supersonic wind coming from the central source (in our case the cluster core); a cavity composed of hot shocked wind material; an outer shell made of swept-up ISM, piled up in front of the forward shock (FS, propagating in the ISM). The wind region and the cavity are separated by a termination shock (TS) located at a radius $R_{\rm ts}$, while a contact discontinuity (CD), located at $R_{\rm cd}$, separates the cavity from the swept up shell. The  forward shock is at $R_{\rm fs}$. 

The bubble evolution undergoes three different stages that depend on the onset of radiative losses. The typical cooling timescale as a function of the density ($n$) and temperature ($T$) of a hot plasma can be written as \citep{DraineISMPhysics2011}:
\begin{equation}
t_{\rm cool} \simeq 60 \left(\frac{n}{0.001 \ \rm cm^{-3}} \right)^{-1} \left(\frac{T}{10^6 \rm \ K} \right)^{1.7} \rm Myr \,.
\end{equation}
where $n$ and $T$ are normalized to values typical of the shocked wind and the numerical coefficient refers to Solar metallicity. 
During the first 1--10 kyr radiative losses are negligible everywhere, hence, the expansion is fully adiabatic \citep{MenchiariPhDThesis2023}. After this initial stage, radiative losses lead to the collapse of the dense outer shell, such that one can assume $R_{\rm cd} \simeq R_{\rm fs}\equiv R_{\rm b}$, while the hot diluted cavity would continue to expand adiabatically for tens of Myr if conduction were negligible (see below). 
Eventually, in the late phase, losses begin to impact the evolution of the cavity as well.

Given the age of Cyg~OB2 (a few Myrs), the bubble produced by the cluster should be in the intermediate stage of evolution. In such a phase, the bubble size can be obtained by solving the momentum equation of the swept-up shell together with the equation describing the energy variation of the hot gas in the cavity. This leads to the following expression for the bubble radius \cite[see also][]{Morlino+2021}:
\begin{equation} \label{eq:Rb}
  R_{\rm b} =\left( \frac{125}{154 \pi} \right)^{1/5} 
  L_{\rm w}^{1/5} \rho_0^{-1/5} t_{\rm age}^{3/5} 
\end{equation}
where $\rho_0$ is the density of the local ambient medium in which the bubble expands, $t_{\rm age}$ is the age of the YMSC and $L_{\rm w}$ is the wind kinetic luminosity.

As for the position of the TS, this can be calculated from balance between the ram pressure of the cluster wind and the pressure of the hot gas in the cavity, which gives:
\begin{equation}
\label{eq:Rts}
R_{\rm ts} = \sqrt{\frac{(3850 \pi)^{2/5}}{28 \pi}} \, \dot{M}^{1/2} v_{\rm w}^{1/2} L_{\rm w}^{-1/5} \rho_0^{-3/10} t_{\rm age}^{2/5} 
\end{equation}
where $\dot{M}$ and $v_{\rm w}$ are the mass loss rate and the velocity of the cluster wind, respectively.

Even though radiative losses in the hot cavity are negligible for the first tens of Myr, thermal conduction and turbulent mixing between the hot cavity and the dense swept-up shell can enhance the importance of cooling, resulting in a modification of the system evolution.  
An approximate way to account for the impact of these processes on the bubble structure is to parameterise the rate of energy losses ($L_{\rm th}$) as a fraction $\zeta_{\rm th}$ of the wind luminosity, such that $L_{\rm th} = \zeta_{\rm th} L_{\rm w}$ \citep{El-Badry+2019, Blasi-Morlino:2023}. In this way, Equations~\eqref{eq:Rb} and \eqref{eq:Rts} retain their validity after substituting $L_{\rm w} \rightarrow (1-\zeta_{\rm th}) L_{\rm w}$. This implies that the bubble radius will decrease by a factor  $(1-\zeta_{\rm th})^{1/5}$ while the TS radius will increase by $(1-\zeta_{\rm th})^{-2/5}$. For example, for $\zeta_{\rm th} = 0.5$, $R_{\rm b}$ ($R_{\rm ts}$) will decrease (increase) by 13\% (30\%)\footnote{Notice that this approximate description is only valid for moderate cooling losses. For $\zeta_{\rm th} \sim 1$, instead, one needs to solve the full set of equations.}. 
A further enhancement of the cooling effect may be produced by turbulent mixing. \cite{LancasterFragmentation2021} use 3D HD-simulations to show that turbulent mixing can be very effective, producing a bubble evolution closer to the momentum conserving solution than to the energy conserving solution adopted here. Their results suggest a smaller bubble size by a factor $\sim 2$ at few Myr \cite[see also][]{Dwarkadas:2023}. However, in that work, cooling is probably overestimated, due to the neglect of magnetic fields, which tend to reduce turbulent mixing and suppress HD instabilities in general \cite[see, e.g.][]{Orlando+2008}. 
Hence, for the purpose of the present work, we consider the adiabatic solution an adequate approximation.

The properties of the collective wind can be estimated using mass and momentum conservation such that the total mass loss rate and wind speed are:
\begin{align}
  \dot{M} &= \sum_i \dot{M}_i  \label{eq:Mdot} \\
  v_{\rm w} &= \dot{M}^{-1} \, \sum_i \dot{M}_i \, v_{\rm w, \it i} \label{eq:vw}    
\end{align}
where $\dot{M}_i$ and $v_{\rm w, \it i}$ are the mass loss rate and wind speed of single stars, respectively. The total luminosity will then be $L_{\rm w} = \dot{M} v_{\rm w}^2 / 2$. For the specific case of Cyg~OB2, we can use the sample of star selected by \cite{WrightMassiveStarPopOB22015}.
To evaluate $\dot{M}_i$, we use two different approaches: a theoretical one, as described by \cite{YungelsonEvolutionFateMassiveStars2008}, and a data-driven one, using empirical relations summarized by \cite{RenzoSystematicSurveyMdot2017}. The detailed calculation is described in Appendix~\ref{Apndx:MassLossRatio} where we show that the mass loss rate and luminosity range in an interval $\dot{M} \in [0.7\times 10^{-4}, \, 1.5\times 10^{-4}]$ M$_\odot$\,yr$^{-1}$ and $L_{\rm w} \in [1.5\times 10^{38}, \, 2.9 \times 10^{38}]$\,erg\,s$^{-1}$. It is worth noticing that this range of values of $L_{\rm w}$ is compatible with other estimates in the literature based on different approaches. No estimates of $\dot{M}$ are available. 
Unless differently specified, in the rest of the paper we will use the following reference values for the mass loss rate and wind luminosity
\begin{align}
  &\dot{M}_{\rm ref} = 10^{-4}\rm  M_\odot \,yr^{-1} \\  
  &L_{\rm w, ref}=2\times 10^{38} \,.
\end{align}

Assuming that the age of Cyg~OB2 is $t_{\rm age}=3$ Myr, and that the bubble has expanded in a medium with an average density\footnote{The assumed density is typical of giant molecular clouds as shown in the analysis by \cite{MenchiariPhDThesis2023}.} of $\rho_0 = 20\,m_p$ cm$^{-3}$ (with $m_p$ the proton mass), the sizes of the cavity and of the TS are $R_{\rm b} \simeq 86$\,pc and $R_{\rm ts} \simeq 13$\,pc, respectively. Figure~\ref{fig:CygusX} shows the expected size of the bubble compared to the size of the Cygnus-X star-forming complex and the distribution of OB stars observed by \cite{WrightMassiveStarPopOB22015}. Noticeably, the projected size of $R_{\rm b}$ is compatible with the extension of the bubble-like structure observed in 21 cm continuum. Notice also that the estimated ranges of $\dot{M}$ and $L_{\rm w}$ lead to an uncertainty on $R_{\rm b}$ and $R_{\rm ts}$ of $\sim$ 20\%. 

Finally we notice that the bubble evolution may be affected by possible SN explosions that could have occurred in the recent past. However, no clear signatures of SNRs have been detected in the Cygnus OB2 association, and, if the actual age of Cyg~OB2 is $\sim 3$ Myr, as assumed here, the estimated energy injected by SNe should be anyway negligible with respect to the wind input (see Appendix \ref{App:Alternative_LwMdot}).

%%% SUB-SECTION %%%
\subsection{Cosmic ray distribution}
\label{subsec:CRRadialDistrib}
Let us now consider the scenario in which CRs are accelerated at the TS of the collective wind. An analytical solution to this problem has been presented by \cite{Morlino+2021} while a fully numerical approach has been developed by \cite{Blasi-Morlino:2023}\footnote{\gio{In the numerical approach by \cite{Blasi-Morlino:2023} the 1D transport equation~\eqref{eq:TransportEq} is discretized and solved numerically on a grid. The main advantage of that approach is that it allows one to deal with any radial dependence of the fluid velocity downstream of the TS, while the analytical solution presented by \cite{Morlino+2021} is only valid for $u\propto r^{-2}$.}}. Hence, the reader is referred to those papers for further details. 
\gio{Here, we will use the former approach. However, in that work, the distribution function at the TS was found recursively. This would be computationally too expensive to implement in the current work, which aims at an exploration of the parameter space. Therefore, in order to speed up the calculation when performing a fit to the $\gamma$-ray data, we adopt an analytical fit to the full solution.} 

The system can be briefly described as follows: particles are accelerated at the TS via the diffusive shock acceleration process, and subsequently escape from the acceleration site experiencing advection and diffusion in the hot bubble until they reach the forward shock. From there, CRs are free to leave the system by diffusing in the unperturbed ISM. Given the slow expansion rate of the cavity, the system can be assumed to be stationary, and if one assumes a radial symmetry, the distribution $f(r, E)$ of CRs can be found by solving the following steady-state transport equation in 1D spherical symmetry:
\begin{equation}
\label{eq:TransportEq}
 \frac{\partial }{\partial r} \left [ r^2 D(r,p) \frac{\partial f}{\partial r} \right ] - r^2 u(r) \frac{\partial f}{\partial r} + \frac{d [r^2 u(r)]}{dr} \frac{p}{3} \frac{\partial f}{\partial p} +r^2 Q(r,p) = 0
\end{equation}
where $p$ is the particle momentum, $u(r)$ is the plasma speed, and $D(r,p)$ is the diffusion coefficient. The source term $Q(r,p)$ describes particle injection taking place at the TS:
\begin{equation}
 Q(r,p)= \frac{\eta_{\rm inj} n_{\rm w} v_{\rm w}}{4 \pi p^2_{\rm inj}} \, \delta(p-p_{\rm inj}) \, \delta(r-R_{\rm ts})\ ,
\end{equation}
where $n_{\rm w}$ is the density immediately upstream of the TS, $v_{\rm w}$ is the speed of the cold wind, and $\eta_{\rm inj}$ is the fraction of particle flux that undergoes the acceleration process. 

The solution to Equation~\eqref{eq:TransportEq} can be found as follows. First, the equation is solved separately in the wind region (upstream the TS), in the cavity (downstream the TS) and in the unperturbed ISM. Then, the three solutions are joined using particle flux conservation at $r=R_{\rm ts}$ and $r=R_{\rm b}$. Finally, one needs to specify two boundary conditions: one at the center of the bubble, where the flux is assumed to be null, and the other at infinity where $f$ matches the distribution of the galactic CR sea, $f_{\rm gal}$. The expression for $f_{\rm gal}$ is taken from the Galactic CR proton spectrum as measured by AMS-02 data \citep{AMS02-protons:2015}. 
The CR radial distribution in the wind region ($f_{\rm w}$), in the cavity ($f_{\rm b}$) and outside the bubble ($f_{\rm out}$) is found to be respectively:
\begin{subequations}
\label{eq:CRdistribFunc}
\begin{equation}
\label{eq:FUpstream}
  f_{\rm w}(r,p)\simeq f_{\rm ts}(p) \cdot \exp \left [ -\int_r^{R_{\rm ts}} \frac{v_{\rm w}}{D_{\rm w}(r',p)}  dr' \right ]  
\end{equation}
\begin{equation}
 f_{\rm b}(r,p)=  \frac{f_{\rm ts}(p) \left[e^{\alpha} + \beta(e^{\alpha_{\rm b}} - e^{\alpha}) \right] + f_{\rm gal}(p) \,\beta (e^\alpha-1)}{1+\beta(e^{\alpha_{\rm b}}-1)} 
\end{equation}
\begin{equation}
 f_{\rm out}(r,p)=f_{\rm b}(R_{\rm b}, p)\frac{R_{\rm b}}{r}+f_{\rm gal}(p) \left ( 1- \frac{R_{\rm ts}}{r} \right )
\end{equation}
\end{subequations}
where
\begin{subequations}
\begin{equation}
\alpha = \alpha(r, p)=\frac{v_{\rm b} R_{\rm ts}}{D_{\rm b}(p)} \left ( 1 -  \frac{R_{\rm ts}}{r} \right )
\end{equation}
\begin{equation}
\alpha_{\rm b}=\alpha(r=R_{\rm b},p)
\end{equation}
\begin{equation}
\beta = \beta(p) = \frac{D_{\rm out}(p) R_{\rm b}}{v_{\rm b} R_{\rm ts}^2} \,.
\end{equation}
\end{subequations}
The terms $D_{\rm w}$, $D_{\rm b}$ and $D_{\rm out}$ are the diffusion coefficients in the wind region, in the cavity and outside the bubble, respectively, while $v_{\rm b}=v_w/4$ is the plasma speed immediately downstream of the TS (assumed to be strong). 
Notice that Equation~\eqref{eq:FUpstream} is a first-order approximation to the full solution presented by \cite{Morlino+2021}, but it is fully adequate for the aim of this work. $f_{\rm ts}$ is the distribution at the TS, which can be expressed as:
\begin{equation}
    f_{\rm ts}(p) = \frac{3 n_{\rm w} v_{\rm w}^2 \epsilon_{\rm cr}}{4 \pi \Lambda_p (m_pc)^3 c^2 } \left( \frac{p}{m_p c} \right)^{-s} \, e^{-\Gamma(p)} \,.
    \label{eq:fts}
\end{equation}
where $c$ is the speed of light, $\epsilon_{\rm cr}$ is defined as the fraction of wind luminosity converted into CR luminosity, namely $L_{\rm cr} = \epsilon_{\rm cr} L_{\rm w}$, while the function $\Lambda_    p$ is the usual normalization factor:
\begin{equation} \label{eq:Lambda}
 \Lambda_p =\int_{x_{\rm inj}}^{\infty} x^{2-s} e^{-\Gamma(x)} \left( \sqrt{1+x^2}-1 \right ) dx \,,
\end{equation}
with $x=p/(m_p c)$.
The function $\Gamma(p)$ contains information about the spherical geometry and the diffusion coefficients in both the wind region and the hot cavity. $\Gamma(p)$ has a complicated expression defined recursively \citep{Morlino+2021}, but a good approximation is: 
\begin{equation} \label{eq:Gamma_approx}
 e^{\Gamma(p)} \simeq \left [ 1 + a_1 \left (\frac{p}{p_{\max}} \right )^{a_1} \right ] e^{- a_3 (p/p_{\max})^{a_4}} \ ,
\end{equation}
where $p_{\max}$ is the {\it upstream} maximum momentum defined by the condition that the upstream diffusion length is equal to the shock radius, i.e. $D_{\rm w}(p_{\max})= v_{\rm w} R_{\rm ts}$\footnote{We recall that $p_{\max}$, as defined here, is not the actual maximum momentum of the system, which is determined by the combination of both the diffusion length upstream and downstream \cite[see][]{Morlino+2021}.}.
$a_1...a_4$ are parameters that mainly depend on the adopted diffusion model and are obtained by fitting Equation~\eqref{eq:Gamma_approx} to the full solution. The values of these coefficients are reported in Table~\ref{tab:fTSparamenters} for the three different diffusion models used in this work.
\begin{table}
\begin{center}
\begin{tabular}[c]{l c c c c}  
\toprule \toprule
 Models  & a$_1$ & a$_2$ & a$_3$ & a$_4$ \\
 \midrule
 Kolmogorov & 10 & 0.308653 & 22.0241 & 0.43112\\
 Kraichnan & 5 & 0.448549 & 12.52 & 0.642666\\
 Bohm & 8.94 & 1.29597 & 5.31019 & 1.13245\\
\bottomrule %\bottomrule
\end{tabular}
\caption{Parameters values used to calculate the exponential function in Equation~\eqref{eq:Lambda} for different assumption of the diffusion coefficient.}
\label{tab:fTSparamenters}
\end{center}
\end{table}

A final remark before the end of this section concerns proton energy losses. Equation~\eqref{eq:TransportEq} does not account for energy losses due to inelastic interactions which, in general, are important if the average density in the bubble is $\gtrsim 10$ cm$^{-3}$ \citep{BlasiCygOB2Gamma2023}. In particular, the shape of proton spectrum can be significantly affected only for $n_b\gtrsim 20$ cm$^{-3}$ while heavier nuclei can suffer substantial losses even at lower density, given that the inelastic cross section increases with the atomic number $A$ as $\propto A^{0.7}$. Considering only the mass injected by the stellar winds, the bubble density would be extremely low (e.g. $n_b\sim 5 \times 10^{-3}$\,$cm^{-3}$ for $\dot{M} =10^{-4}$ M$_\odot$\,yr$^{-1}$ and $R_{\rm b}=86$ pc), but it can substantially increase due to the evaporation of the shell \citep{CastorInterstellarBubbles1975}. The mass evaporation rate from the cold shell, given in terms of the cluster properties \citep{MenchiariPhDThesis2023} is:
\begin{equation}
\label{eq:ShellMdot}
\begin{split}
 \dot{M}_{s} \simeq 2\times 10^{-4} \left(\frac{L_{\rm w}}{10^{37} \rm \ erg \ s^{-1}} \right)^{27/35} \left(\frac{\rho_0}{10 \, m_p \rm \ cm^{-3}} \right)^{-2/35} \\ \left(\frac{t_{\rm age}}{1 \rm \ Myr} \right)^{6/35} \rm M_\odot yr^{-1}  \ .
\end{split}
\end{equation}
Using again the parameters of Cyg~OB2 the bubble density can rise up to  $\sim 0.1$ cm$^{-3}$, still too small to be relevant for inelastic collisions.
A larger density can be obtained if the shell fragments in compact dense clumps (dense enough that the wind is not able to blow them away) or if there were pre-existing molecular clouds in the bubble region. Whether or not clumps affect the final CR spectrum depends on a combination of their filling factor and the value of the diffusion coefficient inside the clumps.
In \S~\ref{sec:GammaEmission} we will show that the column density measured in the Cygnus region thorough $^{12}$CO and HI lines suggests an average density $\lesssim 8.5$\,cm$^{-3}$, supporting the presence of clumpy structure in the region. 
With such a value, energy losses of protons remain small enough that the resulting $\gamma$-ray spectrum should not be significantly affected \citep{Blasi-Morlino:2023}.

%%% SUB-SECTION %%%
\subsection{Diffusion coefficient in the wind bubble}
\label{sec:diff}
The particle distribution function strongly depends on the diffusion coefficient determined by the magnetic turbulence inside the bubble, which remains the most uncertain parameter of the problem. As discussed in \cite{BlasiCygOB2Gamma2023} the CR-self generated turbulence is not strong enough to allow efficient particle confinement. Most probably the turbulence is generated by the wind itself through MHD instabilities due to inhomogeneity and non-stationarity of the wind flow. Hence, we assume that in the upstream some fraction $\eta_{\rm B}$ of the wind luminosity is converted into magnetic flux
\begin{equation}
  4\pi r^2 v_{\rm w} \frac{\delta B_{\rm w}^2}{4\pi} = \eta_{\rm B} \frac{1}{2} \dot{M} v_{\rm w}^2 \,.
\end{equation}
At the TS we assume that the magnetic field is compressed by the shock, so that, for a strong shock and an isotropic turbulent field, we have $\delta B_{\rm b} = \sqrt{11} \, \delta B_{\rm w}(R_{\rm ts})$. 
The type of diffusion associated to the turbulence depends on the details of the cascade. We consider here both Kolmogorov (Kol) and Kraichnan-type (Kra) of cascade such that the diffusion coefficients are 
\begin{align}
  D_{\rm Kol}(E)=\frac{1}{3} v_p r_{\rm L}^{1/3} L_{\rm inj}^{2/3} \label{eq:DK41}\\
  D_{\rm Kra}(E)=\frac{1}{3} v_p r_{\rm L}^{1/2} L_{\rm inj}^{1/2} \label{eq:DKra}
\end{align}
where $v_p$ is the particle velocity, $r_{\rm L}=c p/(e \delta B_{\rm inj})$ is the particle Larmor radius , $L_{\rm inj}$ the turbulence injection scale, and $\delta B_{\rm inj}=\delta B(L_{\rm inj})$. For particles having $r_{\rm L} > L_{\rm inj}$ we assume $D\propto p^2$ as predicted by quasi-linear theory and confirmed by numerical simulations \cite[see, e.g.][]{Subedi+2017}. We do expect $L_{\rm inj}$ to be comparable to the size of the cluster core ($L_{\rm inj} \in [1, \, 2]$ pc). However, more spatial scales may be connected to the generation of turbulence, from the TS size, down to the typical distance between stars. If the power injected at all those scales is of the same order, one would rather expect a flatter turbulence, similar to Bohm's. For this reason we explore here also Bohm-like diffusion, described as
\begin{equation} \label{eq:DBohm}
  D_{\rm Bohm}(E)=\frac{1}{3} v_{\rm p} r_{\rm L} . %(\delta B) .
\end{equation}

\begin{figure}
\centering
\includegraphics[width=\columnwidth]{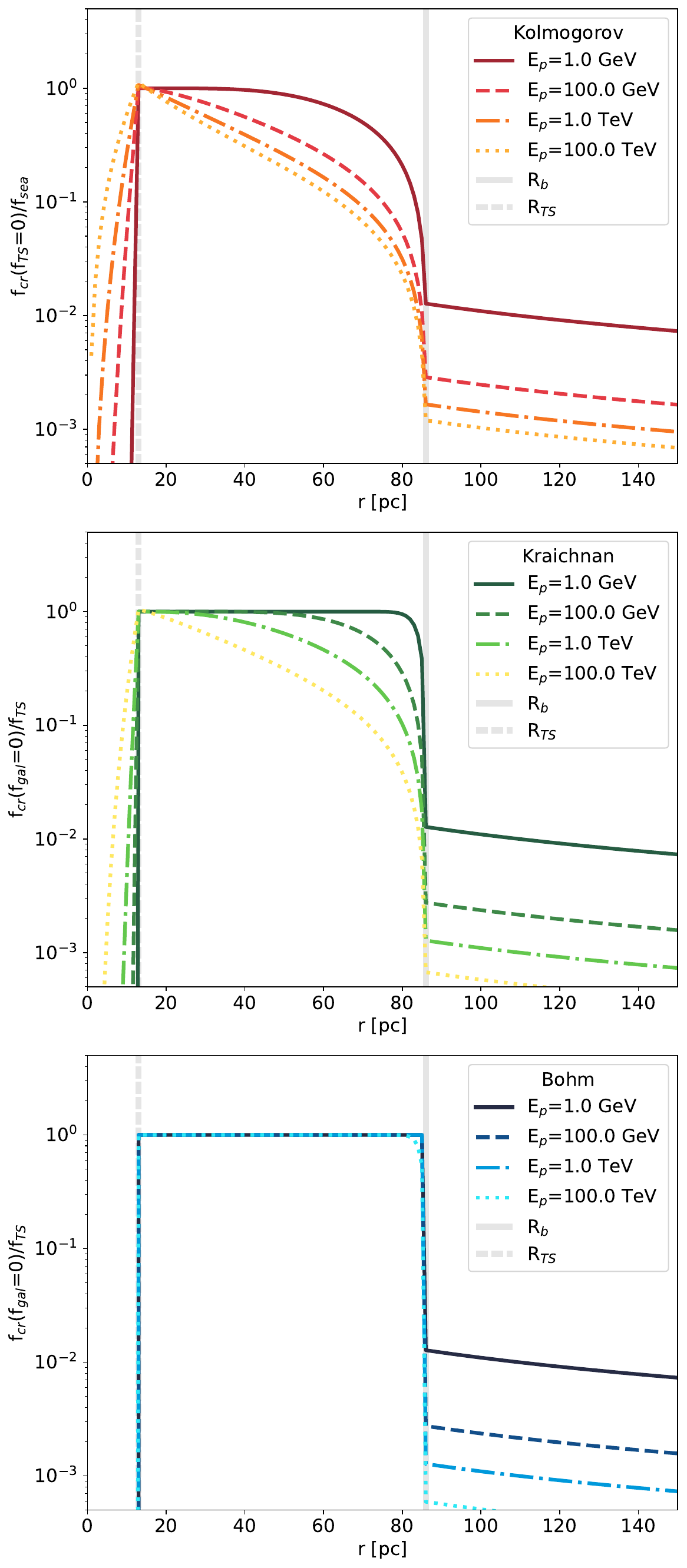}
\caption{Radial CR distribution normalized to its value at the termination shock in the case of Kolmogorov (top panel), Kraichnan (central panel), and Bohm (bottom panel) turbulence. In these plots, the contribution of the galactic CR sea to $f_{cr}$ has not been considered.}
\label{fig:f_cr}
\end{figure}

The impact of these three different types of diffusion on the spatial profile of CRs is shown in Figure~\ref{fig:f_cr}, where we assume $f_{\rm gal}=0$.
The profiles are calculated using a wind luminosity $L_{\rm w, ref}$, and a particle spectral index at injection $s=4$. One can readily see that Kolmogorov turbulence results in a decreasing profile for $r>R_{\rm ts}$, while for Bohm diffusion the profile is flat at all relevant energies for $R_{\rm ts}< r < R_{\rm b}$. The Kraichnan case is intermediate, resulting in a distribution that is flat below $\sim 10$\,GeV and peaked above such energy. The departure from a flat distribution occurs when diffusion dominates over advection. The transition energy at which this happens can be estimated comparing the time scales of the two processes: the advection time scale
\begin{equation}   \label{eq:t_adv}
  t_{\rm adv}=\int_{R_{\rm ts}}^{R_{\rm b}} \frac{dr^\prime}{u(r^\prime)} = \frac{R_{\rm ts}}{3 v_{\rm b}} \left( \frac{R_{\rm b}^3}{R_{\rm ts}^3} - 1 \right)\ ,
\end{equation}
and the the diffusion time scale,
\begin{equation} \label{eq:t_diff}
  \tau_{\rm d}(r,E)=\frac{(R_{\rm b}-R_{\rm ts})^2}{4 D_{\rm b}(E)} \, .
\end{equation}
Using the reference values for Cyg~OB2 discussed in \S~\ref{subsec:BubbleStructure}, we find that $t_{\rm adv} \simeq \tau_{\rm d}$ at energies of $\sim 20$ GeV, $\sim 2$ TeV, and $\sim 300$ TeV for Kolmogorov, Kraichnan, and Bohm cases, respectively. 

In addition to the spatial profile, the diffusion coefficient also determines the maximum achievable energy. As discussed by \cite{Morlino+2021} the  maximum energy results from the combination of three elements: the confinement time upstream and downstream of the shock, plus effective jump conditions at the shock due to the spherical geometry. The latter condition determines the sharpness of the cutoff. 
It is useful to report here the maximum energy given by the upstream condition, i.e. $D_{\rm w}(E_{\max})/v_{\rm w} = R_{\rm ts}$, because it is the one that appears in Equation~\eqref{eq:Gamma_approx}.
Considering Equations~\eqref{eq:Rts}, \eqref{eq:vw}, and the expression for the diffusion coefficients, the maximum energies for the three cases are:
\begin{align}
 E_{\max}^{\rm Kol} &\simeq 1.2 \left(\frac{\eta_{\rm B}}{0.1}\right)^{1/2} \left( \frac{\dot{M}}{10^{-4} \rm M_\odot yr^{-1}}\right )^{-3/4} \left( \frac{L_{\rm w}}{10^{39} \rm \, erg\,s^{-1}} \right )^{37/20} \nonumber \\ 
   & \left( \frac{\rho_0}{20\ m_p \, \rm \ cm^{-3}} \right )^{-3/5} \left(\frac{t_{\rm age}}{3 \rm \ Myr}\right )^{4/5} \left( \frac{L_{c}}{2 \rm \ pc} \right )^{-2} \rm PeV \label{eq:EmaxK41} \,;    
\end{align}
\begin{align}
 E_{\max}^{\rm Kra} &\simeq 2.84 \left(\frac{\eta_{\rm B}}{0.1}\right)^{1/2} \left( \frac{\dot{M}}{10^{-4} \rm M_\odot yr^{-1}}\right )^{-5/10} \left( \frac{L_{\rm w}}{10^{39} \rm \ erg \, s^{-1}} \right )^{13/10} \nonumber\\   
   & \left( \frac{\rho_0}{ 20\, m_p \rm \ cm^{-3}} \right )^{-3/10} \left(\frac{t_{\rm age}}{3 \rm \ Myr}\right )^{2/5} \left( \frac{L_{c}}{2 \rm \ pc} \right )^{-1} \rm PeV \label{eq:EmaxKra}   \,;
\end{align} 
\begin{align}
 E_{\max}^{\rm Bohm} &\simeq 10 \left(\frac{\eta_{\rm B}}{0.1}\right)^{1/2} \left( \frac{\dot{M}}{10^{-4} \rm M_\odot yr^{-1}}\right )^{-1/4} \left( \frac{L_{\rm w}}{10^{39} \rm \ erg\, s^{-1}} \right )^{3/4} \rm PeV \label{eq:EmaxB}   \,.
\end{align}
In \S~\ref{sec:AnalysisProcedure}, we will separately consider the three different types of diffusion, comparing the resulting $\gamma$-ray emission with observations of the Cygnus Cocoon. 

% ======= Stesse equazioni per Emax ma rifatte meglio (da aggiustare, troppo lunghe e in funzione di v_{\rm w}) ==========
%\begin{equation*}
%\label{eq:EmaxK41}
%E_{max}^{K41} =10^{14} \eta_{\rm b}^{1/2} \left( \frac{\dot{M}}{10^{-4} \rm M_\odot yr^{-1}}\right )^{11/10} \left( \frac{v_{\rm w}}{10^3 \rm \ km s^{-1}} \right )^{37/10}  \left( \frac{\rho_0}{m_p \rm \ cm^{-3}} \right )^{-3/5} \left(\frac{t_{age}}{10 \rm \ Myr}\right )^{4/5} \left( \frac{L_c}{2 \rm \ pc} \right )^{-2} \rm eV
%\end{equation*}
%\begin{equation*}
%\label{eq:EmaxKra}
%E_{max}^{Kra} = 4\times 10^{14} \eta_{\rm b}^{1/2} \left( \frac{\dot{M}}{10^{-4} \rm M_\odot yr^{-1}}\right )^{4/5} \left( \frac{v_{\rm w}}{10^3 \rm \ km s^{-1}} \right )^{13/5}  \left( \frac{\rho_0}{m_p \rm \ cm^{-3}} \right )^{-3/10} \left(\frac{t_{age}}{10 \rm \ Myr}\right )^{2/5} \left( \frac{L_c}{2 \rm \ pc} \right )^{-1} \rm eV
%\end{equation*}
%\begin{equation*}
%\label{eq:EmaxB}
%E_{max}^{Bohm} = 7.53 \times 10^{15} \eta_{\rm b}^{1/2} \left( \frac{\dot{M}}{10^{-4} \rm M_\odot yr^{-1}}\right )^{1/2} \left( \frac{v_{\rm w}}{10^3 \rm \ km s^{-1}} \right )^{3/2} \rm eV
%\end{equation*}
% =================================================================================================================

%%% SECTION %%%
\section{Hadronic $\gamma$-ray emission}
\label{sec:GammaEmission}
Accelerated protons give rise to $\gamma$-ray emission through the decay of neutral pions produced in inelastic collisions. The resulting spectrum and morphology will depend on the distributions of both CRs and target medium. The differential $\gamma$-ray flux coming from a given volume $V$ of the sky is given by
\begin{equation}
\label{eq:GammaFluxGeneral}
 \phi_\gamma(E_\gamma)=\frac{1}{4 \pi d^2} \iint c f_{\rm cr}(r, E_p) n(\vec r) \frac{d \sigma (E_p, E_\gamma) }{dE_p} dE_p dV, 
\end{equation}
where $d$ is the distance of the considered volume, $E_p$ and $E_\gamma$ are the CR kinetic energy and the $\gamma$-ray energy respectively, while $\sigma$ is the cross-section for $\gamma$-ray production \cite[which we take from][]{Kafexhiu:2014}. The CR distribution, $f_{\rm cr}(E_p,r)$, is the one described in \S~\ref{subsec:CRRadialDistrib}, while $n(\vec r)$ represents the gas density in and around the star cluster. As discussed above, the presence of clumps of neutral material is likely to dominate the mass of the target, especially if located within the bubble, hence, we will determine the amount of gas from observations, rather than relying on the theoretical profile of the bubble density distribution.

\begin{figure}
\centering
\includegraphics[width=\columnwidth]{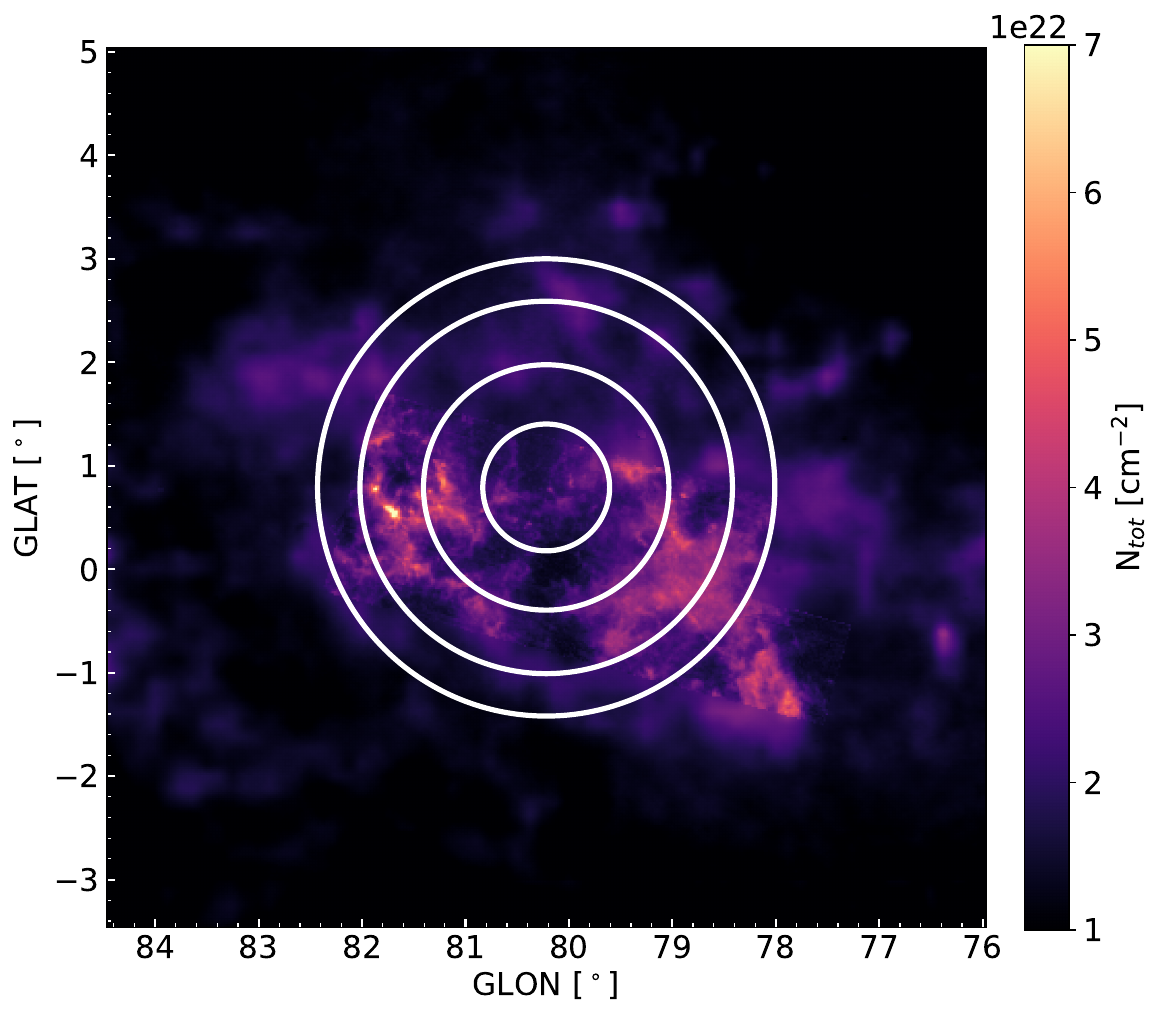}
\includegraphics[width=\columnwidth]{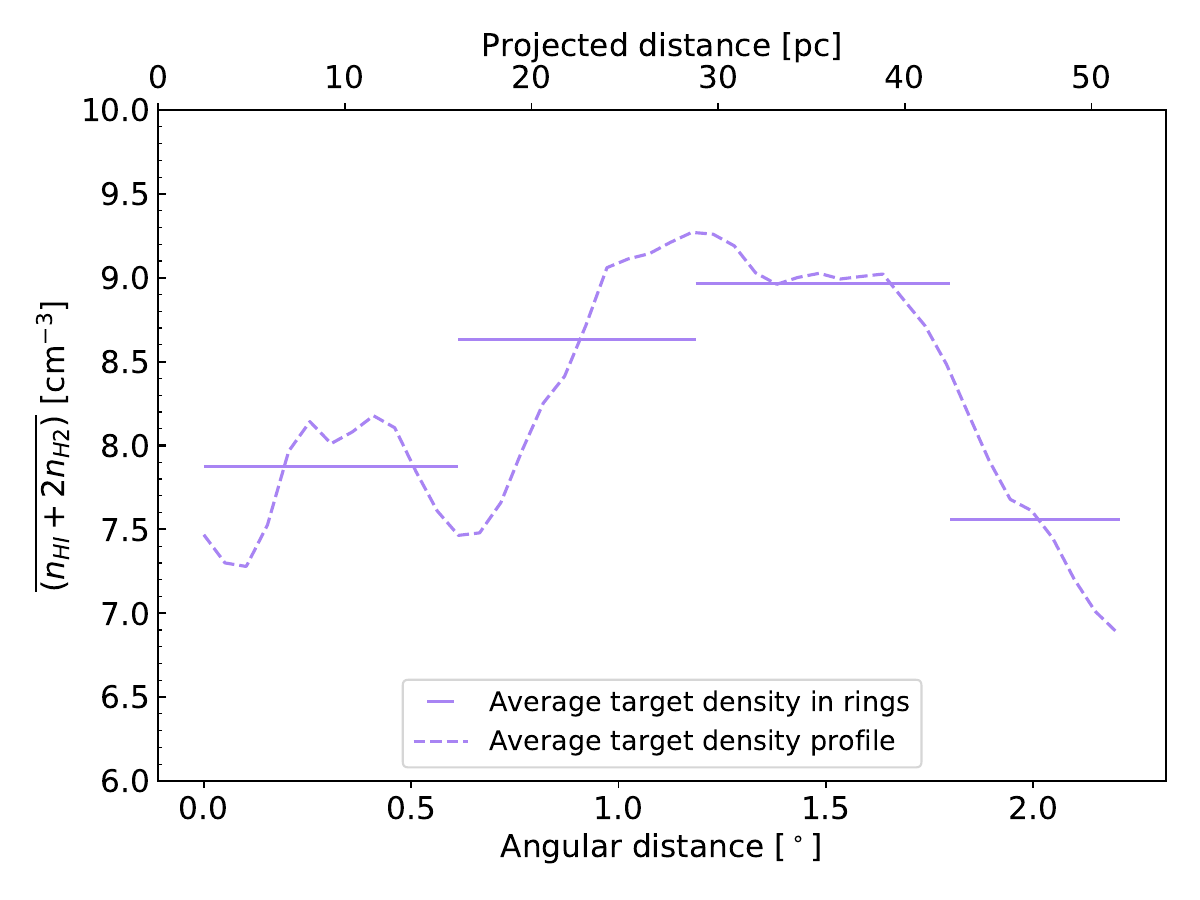}
\caption{\textit{Top panel}: Total target column density considering the interstellar medium in the neutral and molecular phases. The column density of HI has been evaluated using the 21 cm line observation from the CGPS, while for the molecular gas, we used a combination of $^{12}$CO observations from the Nobeyama radio telescope and the data from the $^{12}$CO galactic plane survey by \protect\cite{dameMilkyWayMolecular2001}. All the observations are kinematically cut selecting only the gas between -20 and 20 km\,s$^{-1}$. White rings represent the regions used by \protect\cite{aharonianMassiveStarsMajor2019a} and \protect\cite{HAWCcoll:2021} to probe the $\gamma$-ray radial profile. The radii of the different circles correspond to angular sizes of 0.61$^\circ$, 1.19$^\circ$, 1.8$^\circ$ and 2.21$^\circ$. \textit{Bottom panel}: Average gas density profile in the Cygnus region for each annulus.}
\label{fig:ISM_Map}
\end{figure}

Estimating the amount of gas in the Cygnus region based on observations is not straightforward because this region is located at a galactic longitude where the differential galactic rotation results in low radial velocities, with values compatible with the typical gas dispersion motion, thus preventing a reliable modelling of the ISM profile along the line of sight \citep{schneiderNewViewCygnus2006}. For the sake of simplicity, we will here assume that the gas is uniformly distributed along the line of sight in a range $\Delta z=\pm400$ pc around Cyg~OB2 position, hence, without considering the density profile of the bubble (in \S~\ref{sec:disc} we will further discuss the uncertainty due to such an assumption). Our choice of $\Delta z$ follows from the total extent of 800 pc for the distribution of dust towards the Cygnus-X star forming complex \citep{GreenDust3dMaps2019}. We  further assume that the target medium only contains molecular and neutral gas, therefore neglecting the contribution of ionized gas. To quantify the amount of neutral hydrogen, we use 21~cm line data from the Canadian Galactic Plane Survey \citep{taylorCANADIANGALACTICPLANE2003}. For the molecular component, we use high-resolution observations of $^{12}$CO J(1--0) spectral line from the Nobeyama radio telescope \citep{takekoshiNobeyama45mCygnus2019} in combination with the data from the composite galactic survey of \cite{dameMilkyWayMolecular2001}. Although the kinetic ambiguity prevents an accurate location of the gas along the line of sight, we apply to all observations a velocity cut between -20 km\,s$^{-1}$ and 20 km\,s$^{-1}$, so as to remove the contribution from gas located in the Perseus and Outer arms. We calculate the neutral hydrogen column density using the standard approach described by \cite{ToolsOfRadioAstronomy}:
\begin{equation}
\left [ \frac{N_{\rm HI}}{\rm cm^2} \right ] = - 1.8224 \times 10^{18} \left[ \frac{T_{\rm s}}{K} \right] \int_{-20}^{20} \log \left( 1 - \frac{T^{\rm HI}_{\rm br} (v)}{T_s-T_{\rm bg}} \right) \left[ \frac{dv}{\rm km\,s^{-1}} \right],
\end{equation}
where $T^{\rm HI}_{\rm br}$ is the observed line brightness temperature, $T_{\rm s}$ is the spin temperature assumed to be 150~K, and $T_{\rm bg}=2.66$~K is the brightness temperature of the cosmic microwave background at 21~cm. For the molecular hydrogen, we estimate the column density using the standard $X_{\rm CO}$ conversion factor:
\begin{equation}
    N_{\rm H_2} = X_{\rm CO} \int_{-20}^{20} T_{\rm br}^{\rm CO}(v) \left[ \frac{dv}{\rm km\,s^{-1}} \right], 
\end{equation}
with $X_{\rm CO}=1.68 \times 10^{20}$~mol.~cm$^{-2}$~km$^{-1}$~s~K$^{-1}$ as found by \cite{ackermannCocoonFreshlyAccelerated2011a}. The total gas column density is $N_{\rm tot}=(N_{\rm HI}+2N_{\rm H_2})$ with an average density of $n(r)=N_{\rm tot}/\Delta z$. The resulting column density is shown in Figure~\ref{fig:ISM_Map}, together with the mean gas density averaged over the angle. These are compatible with other estimates \citep{ackermannCocoonFreshlyAccelerated2011a,aharonianMassiveStarsMajor2019a}.
The average gas density in the entire region is $n_{\rm gas}= 8.5$ if one considers the gas uniformly distributed in the whole cylinder of 800\,pc size. However, the density in the bubble could be smaller if the majority of the gas is located outside of it.

Because $\gamma$-ray data available in the literature are provided for annuli around the center of Cyg~OB2, it is useful to express the $\gamma$-ray flux as a function of the radial coordinate projected along on the sky, $l$. Considering that
$r=\sqrt{l^2+z^2}$, where $z$ is the distance from the center of the cluster projected along the line of sight, the $\gamma$-ray flux from an annulus of thickness $\Delta l$ is 
\begin{equation}
\label{eq:GammaFluxRadial}
\phi_\gamma(l, \Delta l, E_\gamma)= \int_l^{l+\Delta l} \overline{n}(l') \xi(l', E_\gamma)  l' dl' , 
\end{equation}
where $\overline{n}(l')$ is the average ISM density profile along the line of sight calculated from the column density maps and $\xi(l', E_\gamma)$ is defined as
\begin{equation}
\xi(l', E_\gamma) =\frac{1}{d^2} \iint c f_{\rm cr}(l',z, E_p) \frac{d \sigma (E_p, E_\gamma) }{dE_p} dE_p dz.
\end{equation}
By varying the integration boundaries in $z$, Equation~\eqref{eq:GammaFluxRadial} can also be used to estimate the total $\gamma$-ray flux from the Cocoon.

%%% SUB-SECTION %%%
\subsection{Comparison with observations}
\label{sec:AnalysisProcedure}
By comparing  the spectrum and morphology of the $\gamma$-ray emission with the predictions of our model, we can find best fit values of $L_{\rm w}$, $s$, and $\epsilon_{\rm cr}$\footnote{The parameters are allowed to vary in the following ranges: $L_{\rm w} \in [10^{37}, 5\times 10^{39}]$ erg\,s$^{-1}$, $s \in [1.8, 2.6]$ and $\epsilon_{\rm cr} \in [10^{-3}, 10^{-1}]$}. Then, the overall validity of our proposed model is evaluated by looking at the consistency between the best fit value of $L_{\rm w}$ and the one inferred from the stellar population as calculated in \S~\ref{sec:CygOB2}.
%To test the validity of our model,
%we compare the spectral and morphological properties of the expected $\gamma$-ray emission  with current available observations provided by several experiments. The general idea is to find what are the best parameters in terms of $L_{\rm w}$, $s$, and $\epsilon_{\rm cr}$ that can reproduce the observed $\gamma$-ray spectrum. Then, {\it a posteriori}, we check if the obtained $L_{\rm w}$ is consistent with the one inferred from the stellar population as calculated in \S~\ref{sec:CygOB2}. For the sake of simplicity, if not differently specified, we
In order to simplify the analysis, we fix all the other source parameters to the values discuss in \S~\ref{sec:CygOB2}  $t_{\rm age} = 3$\,Myr, $\dot{M}=\dot{M}_{\rm ref} yr^{-1}$, $\rho_0 = 20 m_p \rm \, cm^{-3}$, $\eta_{\rm B} = 0.1$, $L_{\rm inj}= 2$\,pc and distance $d= 1.4$~kpc. GeV data of the Cygnus Cocoon are from Fermi-LAT 4FGL (4FGL J2028.6+4110e) \citep{AbdollahiFermi4FGL2020}, TeV data are from Argo (ARGO J2031+4157) \citep{BartoliIdentificationTeVGammaray2014} and HAWC (HAWC J2030+409) \citep{HAWCcoll:2021}. The Cygnus region was also detected by LHAASO (J1908+0621) in 2021 \citep{CaoLHAASOPlaneSurvey2021} but the  detailed spectrum has not been published yet\footnote{The collaboration recently posted a new preprint  \citep{LHAASOcoll:2023} but spectral data points and corresponding errors are not provided, hence, we cannot properly include them in our analysis.}. 

It is to be noticed that we do not include the $\gamma$-ray emission produced by the foreground CR sea (i.e. we assume $f_{\rm gal}=0$ in Equations \eqref{eq:CRdistribFunc}), as the diffuse component has been already subtracted from the data that we use.

The best fit parameters are found by $\chi^2$ minimization of the spectral properties over a region of 2.2$^\circ$ radius centered on the stellar cluster (corresponding to a projected radius of $\sim 54$ pc).
Given that the spectrum of extended sources is usually fitted with a 2D symmetric Gaussian, with wings extending beyond the region of interest, the flux must be properly rescaled \footnote{Gaussian widths are: 2.0$^\circ$ for 4FGL J2028.6+4110e, 1.8$^\circ$ for ARGO J2031+4157 and 2.13$^\circ$ for HAWC J2030+409, which lead to rescaling factors of 0.45, 0.53 and 0.41 respectively.}.

%In the search for the best parameters, we use the following approach. We fit through $\chi^2$ minimization the observed spectral energy distribution extracted from a region of 2.2$^\circ$ centered on the stellar cluster (corresponding to a projected radius of $\sim 54$ pc). The parameters are left to vary in the following ranges: $L_{\rm w} \in [10^{37}, 5\times 10^{39}]$ erg\,s$^{-1}$, $s \in [1.8, 2.6]$ and $\epsilon_{\rm cr} \in [10^{-3}, 10^{-1}]$. 

%An important aspect concerning observations is that the spectrum of extended sources is usually fitted with a 2D symmetric Gaussian profile and the provided flux includes also the fraction in the Gaussian tales which is not measured. Hence, to be consistent with the estimated flux, we decided to rescale the data points to account only for the flux coming from a region of 2.2$^\circ$. Different experiment used different Gaussian width: 2.0$^\circ$ for 4FGL J2028.6+4110e, 1.8$^\circ$ for ARGO J2031+4157 and 2.13$^\circ$ for HAWC J2030+409, which lead to the rescaling factors of 0.45, 0.53 and 0.41, respectively. In addition, for the calculation of the $\chi^2$,

Once we obtain the CR distribution that best describes the observed spectrum, we compare the corresponding $\gamma$-ray radial profile with the ones obtained from Fermi-LAT  \citep{aharonianMassiveStarsMajor2019a} and HAWC data \citep{HAWCcoll:2021}. To this purpose, we calculate the total $\gamma$-ray luminosity in four rings centered on Cyg~OB2, with projected sizes of 0$\--$15 pc, 15$\--$29 pc, 29$\--$44 pc, and 44$\--$54 pc (see Figure~\ref{fig:ISM_Map}). 
The corresponding surface brightness in the energy range $[E_{-}, E_{+}]$ is:
\begin{equation}
\label{eq:GammaLum}
 {\rm SB}_{\gamma} = \frac{L_{\gamma}}{\Omega_{\rm ring}} = \frac{4 \pi d_{\rm OB2}^2}{\Omega_{\rm ring}} \int_{E_{-}}^{E_{+}} E_\gamma \phi_\gamma(E_\gamma) dE_\gamma  \,,
\end{equation}
where $\Omega_{\rm ring}$ is the annulus area and the energy ranges are $[10, 300]$\,GeV for the Fermi-LAT and $[1, 250]$\,TeV for HAWC.

\begin{figure}
\includegraphics[width=\columnwidth]{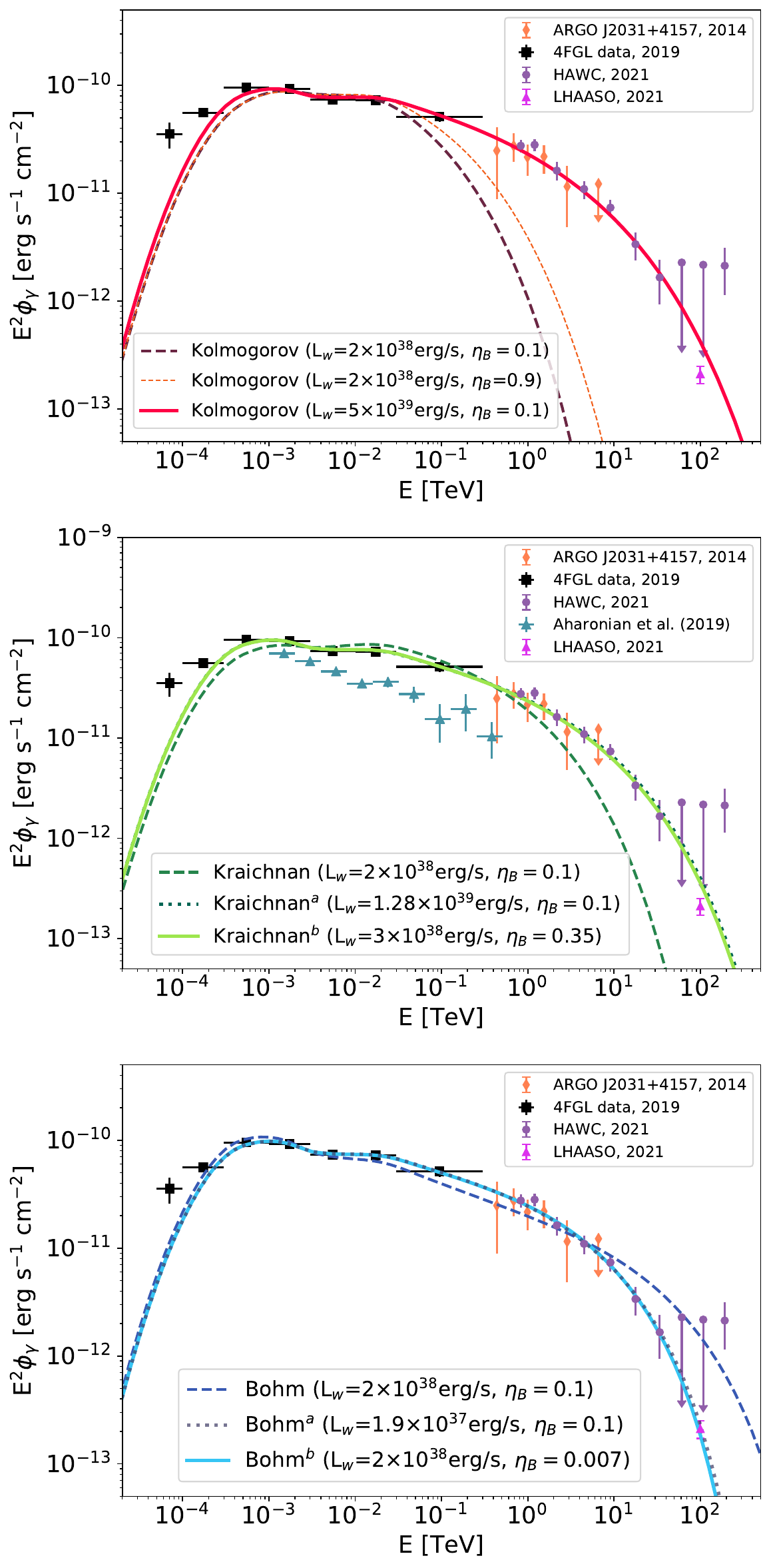}
\caption{Spectral energy distribution of the expected hadronic $\gamma$-ray emission extracted from a region of 2.2$^\circ$ centered on Cyg~OB2. Lines in different panels refer to Kolmogorov (top panel), Kraichnan (middle panel) and Bohm cases (bottom panel). Solid lines always represent the best fit curves, whose underlying parameters are summarized in Table~\ref{tab:BestFit}. Other line-types correspond to the values of the parameters listed in the legenda inside each panel.}
\label{fig:E2GammaFlux}
\end{figure}

\begin{table*}
\begin{center}
\begin{tabular}[c]{l c c c c c c c c c}  
\toprule \toprule
Models  &  $L_{\rm w}$ & $s$ & $\epsilon_{\rm cr}$ & $\eta_{\rm B}$ & $\delta B_{\rm b}$ & $E_{\rm max}$ & $R_{\rm ts}$ & $R_{\rm b}$ & $\bar{\chi}^2_{\min}$ \\
  & [erg\,s$^{-1}$] &  & [\%] &  & [$\mu$G] & [PeV] & [pc] & [pc] & \\
\midrule
Kolmogorov & $5 \cdot 10^{39}$ & 4.17 & $0.4$ & 0.1 (fixed) & 60 & 23 & 16 & 163 & 0.66\\

Kraichnan$^a$ & $1.28 \cdot 10^{39}$ & 4.23 & $0.7 $ & 0.1 (fixed) & 45 & 3.97 & 14 & 124 & 0.39\\

Kraichnan$^b$ & $3 \cdot 10^{38}$ (fixed) & 4.23 & $1.8 $ & 0.35 (fixed) & 63 & 3.8 & 12.5 & 93 & 0.48\\

Bohm$^a$  & $1.9 \cdot 10^{37}$ & 4.27 & $13 $ & $0.1$ (fixed) & 3 & 0.51 & 12 & 53 & 0.27 \\
Bohm$^b$ & $2 \cdot 10^{38}$ (fixed) & 4.27 & $2.2$ & $2.4 \cdot 10^{-3}$ & 5 & 0.47 & 13 & 86 & 0.25\\
\bottomrule %\bottomrule
\end{tabular}
\caption{Best fit parameters values and main system properties for three different models. For the Kolmogorov, Kraichnan$^{a}$ and Bohm$^{a}$ cases, the fitted parameters are $L_{\rm w}$, $s$, and $\epsilon_{\rm cr}$. For the Kraichnan$^{b}$ case also $L_w$ is fixed, while for Bohm$^b$  we fix the wind luminosity and vary $s$, $\epsilon_{\rm cr}$ and $\eta_{\rm B}$. The remaining parameters are, instead, derived from the fit.}
\label{tab:BestFit}
\end{center}
\end{table*}

\subsubsection{Kolmogorov case}
\label{subsec:K41Case}
Assuming Kolmogorov-like diffusion, we find that the minimum $\chi^2$ corresponds to $L_{\rm w} = 5 \times 10^{39}$ erg\,s$^{-1}$, which is the upper boundary of our interval. The $\chi^2$ behaviour makes it clear that even larger values would be preferred.
%We start with the case of CR distribution calculated using the Kolmogorov-like diffusion. After applying our analysis method, we found out that the $\chi^2$ as a function of $L_{\rm w}$ does not have an absolute minimum in the considered interval but has a monotonically decreasing trend. This implies that the best fit requires $L_{\rm w} > 5 \times 10^{39}$ erg\,s$^{-1}$. 
%Even if we could not find a global minimum, the highest luminosity that we allow already provides a reasonable fit, as can be seen in the top panel of Figure~\ref{fig:E2GammaFlux}
%, where the best fit values of the other parameters are $s = 4.17$ and $\epsilon_{\rm cr} = 4 \times 10^{-3}$. Given that this cluster wind luminosity is more than one order of magnitude higher than the value  estimated in \S~\ref{sec:CygOB2}, Kolmogorov-like diffusion is strongly disfavoured.
%It follows that to adequately reproduce the observed spectrum, the required wind luminosity has to be more than one order of magnitude higher than the value  estimated in \S~\ref{sec:CygOB2}, implying that the Kolmogorov case is strongly disfavoured.  
While not corresponding to a global minimum, $L_{\rm w} = 5 \times 10^{39}$ erg\,s$^{-1}$ provides a good fit to the data (see top panel of Figure~\ref{fig:E2GammaFlux}), with reasonable values of the other parameters (see Table\ref{tab:BestFit}). However, such a luminosity is more than a factor 10 higher than the value estimated in \S~\ref{sec:CygOB2}, thus making Kolmogorov-like diffusion incompatible with our model: Kolmogorov turbulence does not provide efficient confinement of the particles resulting into a low maximum energy and a very broad cutoff.
As a result, a cutoff that in $\gamma$-rays shows at $\sim 100$\,TeV corresponds to a proton maximum energy of $\simeq 20$\, PeV (see Table \ref{tab:BestFit}).

Indeed by fixing $L_{\rm w}$ to the best value inferred from the stellar population, i.e. $L_{\rm w, ref}$, the corresponding $\gamma$-ray spectrum dramatically fails in reproducing the data (Figure~\ref{fig:E2GammaFlux}). We also verified that no good agreement can be achieved by increasing the level of magnetic turbulence while keeping $L_{\rm w}=L_{\rm w, ref}$: assuming the unrealistically high value of $\eta_{\rm B} = 0.9$, the calculated flux remains well below the observations for $E_{\gamma} \gtrsim 1$\,TeV (Figure~\ref{fig:E2GammaFlux}).

\subsubsection{Kraichnan case}
\label{subsec:KraichnanCase}
%For Kraichnan-like diffusion the global $\chi^2$ minimum corresponds to: $L_{\rm w}=1.28 \times 10^{39}$\,erg\,s$^{-1}$, $\epsilon_{C\rm R} = 7 \times 10^{-3}$ and  $s = 4.23$ (case labelled as Kraichnan$^a$ in Table~\ref{tab:BestFit} and shown with solid line in middle panel of Figure~\ref{fig:E2GammaFlux}) \nb{[Again what is what must be stated in the caption of figures and tables not somewhere god knows where in the text!!!!!!]}. 
For Kraichnan-like diffusion the global $\chi^2$ minimum corresponds to $L_{\rm w}=1.28 \times 10^{39}$\,erg\,s$^{-1}$ (case labelled as Kraichnan$^a$ in Table~\ref{tab:BestFit}, shown by the solid line in the middle panel of Figure~\ref{fig:E2GammaFlux}), which is  still a factor of $\sim 6$ higher than $L_{\rm w, ref}$ and a factor of 4 larger than the maximum derived in Appendix~\ref{Apndx:MassLossRatio}. Again, such a high wind luminosity is required in order to fit the very high energy part of the spectrum. A wind luminosity of $1.28 \times 10^{39}$\, erg\,s$^{-1}$ corresponds to $E_{\max}=3.97$ PeV in Equation~\eqref{eq:EmaxKra}. 
In principle, the spectrum can still be reproduced with a lower wind power by modifying the other cluster parameters appropriately: using Equation~\eqref{eq:EmaxKra}, and setting $L_{\rm w}=3 \times 10^{38}$ erg\,s$^{-1}$, $\dot{M}=0.7 \times 10^{-4}$ M$_\odot$ yr$^{-1}$, $\eta_{\rm B}=0.35$, $L_{\rm inj}=0.7$ pc, we find a similar value of $E_{\max}$. With this choice of parameters, the best fit values for the spectral index and the efficiency of CR production are $s \simeq 4.23$ and $\epsilon_{cr} \simeq 0.018$, respectively (case labelled as Kraichnan$^b$ in Table~\ref{tab:BestFit}, shown by the dotted line in the middle panel of Figure~\ref{fig:E2GammaFlux}). Notice that the parameter values are rather extreme but still within the limits inferred from the stellar population study. 

Concerning the morphological study of the emission, since the CR distribution is very similar in both cases, we illustrate the spatial properties of the $\gamma$-ray emission only for case Kraichnan$^b$.

\begin{figure}
\includegraphics[width=\columnwidth]{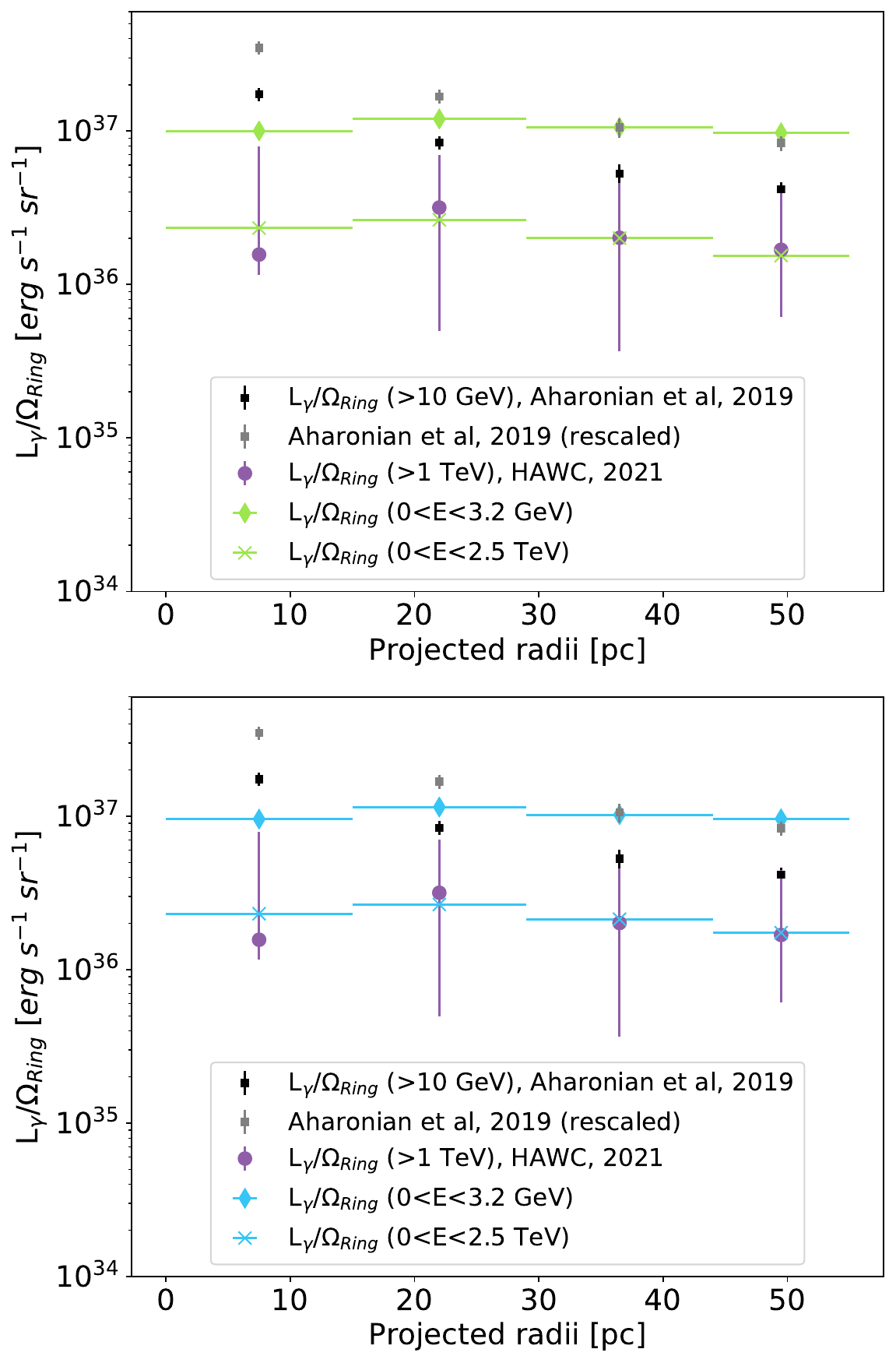}
\caption{Comparison with observations of the expected $\gamma$-ray surface brightness radial profile for the Kraichnan (upper panel) and Bohm (bottom panel) cases. Black squares correspond to the measurements by Fermi-LAT \citep{aharonianMassiveStarsMajor2019a}, while purple circles indicate the surface brightness observed by HAWC \citep{HAWCcoll:2021}. Grey squares represent Fermi-LAT measurements rescaled by a factor of 2. Diamonds and crosses indicate the surface brightness values based on our model, at energies of $10<E_\gamma<316$ GeV and $1<E_\gamma<251$ TeV respectively.}
\label{fig:GammaRadProfile}
\end{figure}

The top panel of Figure~\ref{fig:GammaRadProfile} shows the comparison of our results with the radial profile derived from Fermi-LAT and HAWC data. Concerning the former, one can clearly see that our predicted flux is not in agreement with that  estimated by \cite{aharonianMassiveStarsMajor2019a}. 
First of all, we should note that the spectrum by \cite{aharonianMassiveStarsMajor2019a} is a factor $\sim 2$ lower than the one published in the 4FGL (used in this analysis; see middle panel of Figure~\ref{fig:E2GammaFlux}). In addition, even when this factor of 2 is taken into account, the resulting  radial profile (gray points in top panel of Figure~\ref{fig:GammaRadProfile}) decreases too fast from the inner to the outer region.
This trend is especially surprising in light of HAWC data, showing a flat radial profile. As we will discuss in more detail in Sec.\ref{sec:disc}, any advection-diffusion model would predict a flatter profile at lower energies.

\subsubsection{Bohm case}
\label{subsec:BohmCase}
Finally, for the Bohm case we found the global $\chi^2$ minimum for $L_{\rm w} = 1.9 \times 10^{37}$ erg\,s$^{-1}$, with corresponding best fit values of $s \simeq 4.27$ and $\epsilon_{cr} \simeq 0.13$ (case Bohm$^a$ in Table~\ref{tab:BestFit} and solid line in bottom panel of Figure~\ref{fig:E2GammaFlux}). In contrast with the previous cases, the required wind luminosity is lower by almost one order of magnitude with respect to the lower bound of the range of values estimated in \S~\ref{sec:CygOB2}. In fact, Bohm-like turbulence is the most effective at confining particles and has the strongest energy dependence, resulting in the highest maximum energy (for fixed $L_{\rm w}$) and sharpest cut-off. Furthermore, the low diffusion coefficient within the bubble prevents the escape of high-energy particles from the system, thus avoiding the suppression of emission at very high energies. This can be seen by noticing that the spectrum obtained for $L_{\rm w}=L_{\rm w, ref}$ overshoots the observations in the cutoff region (dashed line in the Figure~\ref{fig:E2GammaFlux}).

We then tried to fit the spectrum fixing $L_{\rm w}=L_{\rm w, ref}$, while allowing $\eta_{\rm B}$ to vary.
%\nb{Like we did for} Kolmogorov and Kraichnan cases, we tryed to fit the spectrum fixing the  wind luminosity to $L_{\rm w} = 2 \times 10^{38}$ erg\,s$^{-1}$, \nb{based on stella population,} while allowing the value of $\eta_{\rm B}$ to vary.  
We found the global $\chi^2$ minimum for $\eta_{\rm B}=2.4 \times 10^{-3}$, with an associated spectral index and CR efficiency of $s \simeq 4.27$ and $\epsilon_{cr} \simeq 0.022$ (case Bohm$^b$ in Table~\ref{tab:BestFit} and dotted line in Figure~\ref{fig:E2GammaFlux}).
The corresponding maximum energy is $E_{\max} \simeq 470$ TeV, only slightly lower than the value found for the case Bohm$^a$. 

\gio{
\subsubsection{Parameter degeneracy}
The solutions presented in the previous sections are not unique because there is a degeneracy in the parameter space: for each assumed diffusion coefficient, a class of solutions with a similar $\chi^2$ value can be identified. \gio{These solutions are characterized by relations between the parameters that can be approximated by the  analytical expressions we provide below.}
%approximate analytical expression between parameters, as we show below.

Once the diffusion coefficient has been chosen, the shape of the $\gamma$-ray emission is mainly determined by three quantities: \gio{the spectral index of the injected particle distribution}, $s$, the flux at low energies, $\phi_0$, and the value of the maximum energy, $E_{\max}$. 
We need to look for the dependencies of these observables on the unknown parameters, $L_w$, $\epsilon_{\rm cr}$ and $\eta_{\rm B}$. 

The spectral slope has a very mild dependence on those parameters: it is mainly determined by the diffusion coefficient. Hence, if we fix the latter, $s$ is uniquely determined. The flux normalization, instead, is roughly proportional to $\sim f_{\rm ts} n_{\rm gas} R_b$. The linear dependence on $R_b$ results from assuming that the bubble radius is larger than the size of the observed region (as is the case for all solutions discussed above) hence the emission is integrated over a cylinder. Now, considering the dependencies of $f_{\rm ts}$ (Eq.\ref{eq:fts}) and $R_b$ (Eq.\ref{eq:Rb}), we obtain $\phi_0 \sim \epsilon_{\rm cr}\, L_w^{3/5}$, where we have assumed that the wind speed remains roughly constant by changing the wind luminosity, which is a good approximation for YMSCs. The maximum energy scales instead as $E_{\max} \sim \eta_{\rm B}^{1/2}\, L_w^b$ where $b= \frac{11}{10}$, $\frac{4}{5}$ or $\frac{1}{2}$ for Kolmogorov, Kraichnan or Bohm diffusion, respectively. Therefore, once an optimal solution has been identified, having parameters $\{L_{w,0}, \epsilon_{\rm cr,0}, \eta_{\rm B,0} \}$, then, by varying the wind luminosity in the range of allowed values, all solutions having 
\begin{equation} \label{eq:degeneracy}
  \begin{cases} 
  \epsilon_{\rm cr} = \epsilon_{\rm cr,0} \, (L_w/L_{w,0})^{-3/5} \\
  \eta_{\rm B} = \eta_{\rm B,0} \,  (L_w/L_{w,0})^{-2 b}
  \end{cases}
\end{equation}
will provide an acceptable fit to the $\gamma$-ray spectrum. On the other hand, if the wind luminosity is known, then, for each diffusion coefficient, there is only one single solution.
}

\section{Discussion}
\label{sec:disc}

\subsection{Spectral energy distribution}
\label{sec:disc_spectrum}
Table \ref{tab:BestFit} summarizes our results. 
The first thing to notice is that all cases require a particle injection slope $s>4$. This is softer than the standard spectral index expected from diffusive shock acceleration at strong shocks, such as the ones found in compact and powerful star clusters. This result is not too surprising: also young SNRs, hosting very strong shocks, typically show spectra steeper than $p^{-4}$, a feature that can result from several different mechanisms. In the case of SNRs a possible explanation is connected with the magnetic waves that scatter off particles: if the wave speed is mainly directed away from the shock (either upstream or downstream): the effective compression factor felt by particles is reduced, resulting into a steeper spectrum. Such an effect has been found for shocks where the magnetic turbulence is self-generated by CRs \citep{Haggerty-Caprioli:2020, Cristofari+2022} however it is not clear whether the same result applies for magnetic turbulence generated by MHD instabilities, as seems the case for YMSCs.
Another possibility is that the shock is affected by a non perfect central symmetry of the SC. In the specific case of the Cyg~OB2 association, 50\% of the brightest stars are enclosed in a volume of radius $\sim$5 pc. Hence, the presence of stars outside this volume and closer to the position of the TS can affect the structure of the shock itself. Finally, strong cooling of the wind bubble can also reduce the velocity jump if the bubble temperature drops below $\sim 10^6$\,K.

Note also that, the best fit spectral index is substantially steeper than the value of 4.08 inferred in \cite{Blasi-Morlino:2023}. The reason is due to the different cross section used for pp scattering. While \cite{Blasi-Morlino:2023} used the parametrization by \cite{kelner2006}, here we adopted the one derived by \cite{Kafexhiu:2014} which is more accurate at low energies (the most relevant in determining the spectral slope). An additional difference is that here we have neglected the effect of energy losses due to inelastic proton-proton collisions, which, on the contrary, have been extensively studied by \cite{Blasi-Morlino:2023}. However, as shown in \S~\ref{sec:GammaEmission}, the upper limit to the total average gas density in the bubble is $\sim 8.5$ cm$^{-3}$, a small enough value so that the energy losses only marginally affect the spectrum of protons.

Another important result is the low value inferred for the CR acceleration efficiency, which is of order of $1\%$ (with the exception of the case Bohm$^a$ for which we found $\epsilon_{\rm cr}=13\%$), a factor 3--10 smaller than what usually estimated for SNR shocks. However, remembering that the $\gamma$-ray flux is proportional to $\epsilon_{\rm cr}\times n_{\rm gas}$, the acceleration efficiency is likely underestimated by our calculation if, as discussed in \S~\ref{sec:GammaEmission}, the target gas is mainly distributed outside the wind bubble.
On the contrary, the additional uncertainties on $\epsilon_{\rm cr}$ due to Cyg~OB2 distance \cite[e.g. if the cluster is indeed at 1760 pc, as proposed by][]{BerlanasDisentanglingSpatialSubstructure2019}, is negligible in terms of physical interpretation.

A significant difference between the Kraichnan and Bohm models is the nominal value of the maximum energy reached by the particles ($\simeq 4$ PeV for Kraichnan and $\simeq 0.5$ PeV for Bohm). Despite such a difference, the $\gamma$-ray spectra of the two cases are remarkably similar, with an exponential cutoff feature in the $\gamma$-ray spectrum that in both cases is located at a few tens of TeV. This, once more, is due to a shallower shape of the cutoff in the Kraichnan case compared to the Bohm one \citep{Morlino+2021}.
This fact highlights the very important role played by the modeling of particle transport when deriving the maximum energy achieved by the accelerator from the spectrum of its $\gamma$-ray emission. 

\gio{Finally, we notice that all cases presented here underestimate the $\gamma$-ray flux at energies $\lesssim 100$\,MeV with respect to the values reported in the Fermi-LAT 4FGL catalog. The most plausible explanation for such a discrepancy is that the background adopted in the Fermi analysis is underestimated. Indeed, the analysis used for the 4FGL is generic and not specialized to the case of the Cygnus region. In fact, other analyses, like the ones presented by \cite{aharonianMassiveStarsMajor2019a} and \cite{Astiasarain+2023}, have not included data below a few hundreds MeV, exactly because the background subtraction is highly uncertain.
Another possible explanation is that an additional contribution, possibly coming from electrons' Inverse Compton Scattering or bremsstrahlung, may be present. As discussed in \S~\ref{sec:IC}, while dominant leptonic emission at the highest energies is highly disfavoured, currently a substantial leptonic contribution at the lowest $\gamma$-ray energies cannot be ruled out.}

\subsection{Spatial profile}
\label{sec:disc_profile}
\begin{figure}
\centering
\includegraphics[width=\columnwidth]{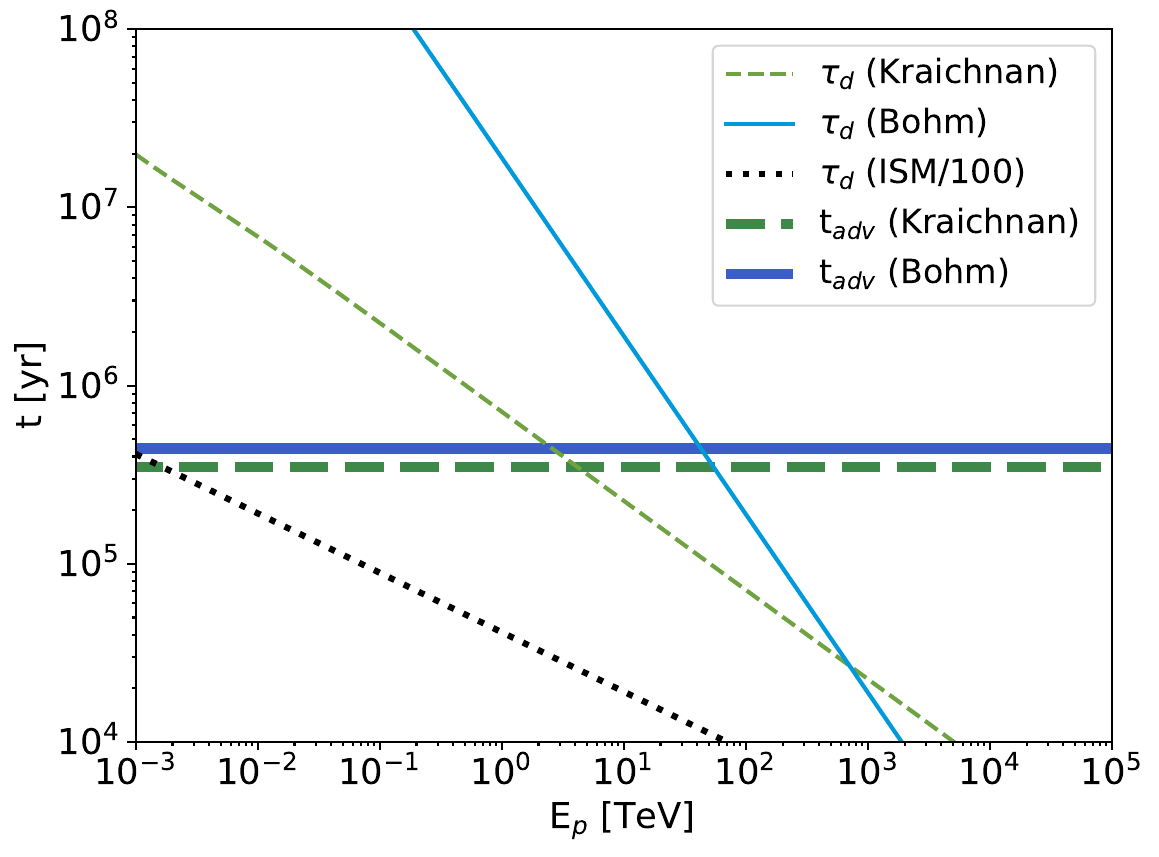}
\caption{Comparison between diffusion (thin lines) and advection (thick lines) timescales for the Kraichnan (green dashed lines) and Bohm (blue solid lines) cases. The black dotted line shows the diffusion time in the case of a diffusion coefficient with the same slope as that in the standard interstellar medium normalized a factor of 100. All timescales are calculated considering only the region used for the morphological analysis, i.e. substituting $R_{\rm b}$ with 54 pc in Equations~\eqref{eq:t_adv} and \eqref{eq:t_diff}.}
\label{fig:DiffTime}
\end{figure}

As shown in \S~\ref{sec:AnalysisProcedure}, Bohm and Kraichnan cases predict a flat morphology in the $\gamma$-ray emission profile, in both the Fermi-LAT and HAWC energy bands. The uniformity of the profile is a result of the advection dominated transport of CRs, which makes the distribution constant in space. While this result is expected in the case of Bohm-like turbulence (see Figure~\ref{fig:f_cr}), for the Kraichnan case in the HAWC energy band it might seem at odds with naive expectations, because for $E\gtrsim 10$\,TeV the diffusion time becomes shorter than the advection one, producing a decrease of the CR density. The reason why we do not see such a decrease in the $\gamma$-ray profile is that we are probing the morphology in a projected sky area of about 54\,pc, smaller than the size of the bubble, which for our best-fit Kraichnan case is $R_{\rm b} \simeq 93$ pc. 
As described in \S~\ref{subsec:CRRadialDistrib}, since the advection velocity scales as $\propto r^{-2}$, areas close to the TS will be characterised by high advection velocities, resulting in advection dominated transport. This can be easily understood by looking at Figure~\ref{fig:DiffTime}, which shows the time-scales of advection and diffusion in a region of size 54 pc. For the Kraichnan (Bohm) case, advection dominates up to energies of $\sim 10$\,TeV (100\,TeV). Note that if the distance of Cyg~OB2 were 1760 pc, the region tested would be of 67 pc, still smaller than the size of the bubble, hence, we would not expect a significant change in the radial morphology.

Indeed, our constant CR distribution, after convolution with the gas distribution, is consistent with HAWC observations, while it disagrees with the radial morphology inferred from Fermi-LAT data \citep{aharonianMassiveStarsMajor2019a}. The centrally peaked profile shown by the latter cannot be reconciled with the HAWC morphology within our model: if advecion is dominant in the HAWC band, it must dominate also in the Fermi-LAT band, independently of the model parameters.

In fact, based on Fermi-LAT data, \cite{aharonianMassiveStarsMajor2019a} deduced a CR distribution decreasing with radius as $1/r$, as one would expect in the case of continuous injection of CRs from a central source and diffusion dominated transport. These authors used the $\gamma$-ray luminosity at $E_{\gamma} \geq 1$\,TeV to infer the diffusion coefficient at $\sim 10$ TeV, obtaining a value $\sim 100$ times smaller than the average value in the ISM. If we extrapolate this finding down to 100 GeV (the proton energy corresponding to the $\gamma$-ray emission at  $E_\gamma \approx$ 10 GeV), assuming the Kolmogorov scaling $\propto E^{1/3}$, we obtain $D(100\,{\rm GeV})\sim 10^{27}\, \rm cm^{2}s^{-1}$. Recalling that diffusion dominates if $\tau_{\rm diff} \ll \tau_{\rm adv}$, we can put an upper limit on the average advection speed that reads $ v_{\rm adv} \ll D(p) / L$. Quantitatively:
\begin{equation}
    v_{\rm adv} \ll 60 \left[\frac{D(100\,\rm GeV)}{10^{27}\, \rm cm^2 s^{-1}}\right] \left( \frac{\theta}{2.2\, \rm ^\circ}\right)^{-1} \left( \frac{d}{1.4\, \rm kpc}\right)^{-1}\, \rm km\,s^{-1}\,.
\end{equation}
where $\theta$ is the angular distance from Cyg~OB2. Such a value is much smaller than the average flow speed in the bubble in standard models of bubbles around YMSCs, implying that a pure diffusion regime is in tension with the existence of a bubble. 

On the other hand, the model we propose naturally explains the HAWC morphology, which is not easy to reconcile with a pure diffusive profile. If a wind bubble exists, indeed, additional processes not included in our model must be responsible for the Fermi-LAT morphology. A possibility is that, in the Fermi-LAT band, the contribution of leptonic Inverse Compton emission might be relevant: leptons accelerated at the TS would suffer energy losses while advected outward and a centrally peaked profile is naturally expected. Another possible scenario that can accommodate both HAWC and Fermi-LAT data is one in which, upstream of the TS, particle acceleration up to tens of GeV is contributed by single star winds. If there is sufficient target for $\gamma$-ray emission before the TS, these CRs would provide an additional, centrally peaked low energy contribution. A quantitative assessment of these scenarios is left for future work.

%%% SUBSECTION %%%
\gio{
\subsection{A leptonic contribution?} 
\label{sec:IC}
We have neglected so far a possible leptonic contribution to the $\gamma$-ray emission. The main reason for this is that recent studies, based on different pieces of observational evidence, tend to disfavour the dominance of leptonic emission beyond 1 TeV. First of all, \cite{Guevel+2023} considered the lack of non-thermal X-ray emission: upper limits on the X-ray flux suggest that the IC scattering does not contribute more than $\sim 1/4$ of the observed $\gamma$-ray emission at 1 TeV, assuming an equipartition magnetic field of $\sim 20\, \mu$G. In addition, analyzing the last 10 yr of IceCube data, \cite{Neronov-neutrini:2023} find a 3$\sigma$ excess of neutrino events from an extended Cygnus Cocoon, at a flux level compatible with the assumption that the observed $\gamma$-ray emission above $\sim 1\,$TeV is due to hadronic interactions. 

Moreover, there is also a theoretical motivation that strongly disfavours the possibility that leptonic emission dominates the $\gamma$-ray spectrum at the highest energies if acceleration is limited to the TS, as we assume in this work: the maximum energy that electrons can achieve at the wind TS is too low due to energy losses. This can be easily shown equating the acceleration time, $t_{\rm acc}= 8 D/v_w^2$ with the energy loss time, $t_{\rm loss}^{-1} = t_{\rm syn}(B)^{-1} + t_{\rm IC}(U_{\rm CMB} + U_{\star})^{-1}$ where the IC losses account for the CMB ($U_{\rm CMB}$) and star light ($U_*$) photon fields at the location of the termination shock, $R_{\rm TS}$. 
For starlight, we use $U_{\star} = \mathcal{L}/(4\pi R_{\rm TS}^2\,c)$, where $\mathcal{L}$ is the total  luminosity of the star cluster estimated by integrating the stellar luminosity from Eq.~\eqref{eq:MergedMLR} over the stellar IMF: this leads to $U_{\star} \simeq 30\, \rm eV\,cm^{-3}$. We estimate $t_{\rm IC}$ using the full Klein-Nishina cross section in the approximate expression obtained by \cite{Khangulyan+:2014}.
In order to obtain a conservative upper limit to the electron maximum energy, we minimize the ratio between the time-scales for acceleration and losse. By using the Bohm diffusion coefficient and the upper limit for the wind speed, which is $\sim 2800\, \rm km\,s^{-1}$, we find that $E_{e, \max} \simeq 180$\,TeV is found for a magnetic field of $\sim 1.6 \,\mu$G. If, instead, we chose the Kraichnan diffusion with $L_c = 2$\,pc as adopted in \S~\ref{sec:diff}, we get $E_{e, \max} \simeq 40$\,TeV for a magnetic field of $1.9\, \mu$G. We notice that these values of the magnetic field are one order of magnitude below equipartition. Values closer to equipartition, $B\sim 20
\, \mu$G, would limit the electron energy to $<70$ TeV even in the most favourable scenario of Bohm diffusion. Hence, in the most optimistic scenario of Bohm diffusion, IC emission can reach $\sim 100$\,TeV, but with a sharp cutoff due to the Klein-Nishina suppression. Hence, it would be very difficult to account for the LHAASO data up to $\sim 1$\,PeV.

Moving to lower energies, we first notice that the spectral shape below $\sim 1$\,GeV in the Fermi-LAT data suggests the presence of the pion bump, a spectral signature difficult to explain with leptonic emission only, without invoking some fine-tuning either in the particle spectrum (suitably broken power-law) or in the combination of the contributing radiation backgrounds. In addition, if the IC scattering is not negligible at the highest energies, the typical shape of IC emission would imply that the same process only marginally contributes at energies below $\sim 10$\,GeV \citep{HAWCcoll:2021}.
An additional contribution at lower energies could come from the non-thermal bremsstrahlung. We can estimate which would be the electron acceleration efficiency, $\eta_{\rm el}$, needed to have a significant bremsstrahlung emission in the GeV energy band. The luminosity due to bremsstrahlung for photons above energy $E_0$ can be written as
\begin{equation}
%    L_{\rm brem} = \eta_{\rm el}' L_w \frac{\tau_{\rm brem}}{\tau_{\rm loss}}
     L_{\rm brem} = \eta_{\rm el}' L_w \frac{\tau_{\rm loss}}{\tau_{\rm brem}}
\end{equation}
where $\tau_{\rm loss}$ is the time-scale for electron energy losses, $\tau_{\rm brem}$ is the time-scale for bremsstrahlung radiation and $\eta_{el}'$ is the fraction of wind luminosity converted into electrons with energy $>E_0$ and it is connected to the full efficiency through $\eta_{\rm el}' = \eta_{\rm el} (m_e c^2/E_0)^{s-4}$. 
Using $\tau_{\rm brem} = 5\times 10^{6} (n/8 \,{\rm cm^{-3}})^{-1}$\,yr \citep{Aharonian_book:2004}, $s=4.3$ and assuming that bremsstrahlung is responsible for the whole $\gamma$-ray emission in the Fermi-LAT band, $L_{1-100\,\rm GeV} = 9\times 10^{34}\, \rm erg \, s^{-1}$ \citep{ackermannCocoonFreshlyAccelerated2011a}, we estimate $\eta_{\rm el} \simeq 3\%$ for the reference value of the wind luminosity of $2 \times 10^{38}\, \rm erg\,s^{-1}$. Such an efficiency is about two orders of magnitude larger than what usually estimated for SNR shocks. Hence, it seems rather unlikely that bremsstrahlung can dominate the $\gamma$-ray emission, even though it cannot be completely ruled out. 
Anyway, a detailed analysis of leptonic emission is ongoing and will be presented in a forthcoming article.

%{Aside from spectral considerations, also the spatial distribution of the emission is difficult to account for with leptonic models assuming particle acceleration at the WTS: due to energy losses, the emission would be concentrated close to the TS.The propagation length due to both advection and diffusion can be approximated using the solution for plane parallel shocks  with constant downstream velocity \citep{Berezhko+2003}: 
%\begin{equation} \label{eq:Lprop}
%    L_{\rm prop} \simeq \frac{D}{u_2} \left[ \sqrt{1+ \frac{4 D}{u_2^2 \, t_{\rm loss}}} - 1 \right]^{-1} \,.
%\end{equation}
%Using this expression together with Eq.~\eqref{eq:t_loss_e} for $E_{\max}$ estimated above, we find $L_{\rm prop}\simeq 0.25$\,pc and 3.7~pc for Bohm and Kraichnan diffusion, respectively. Hence, it is difficult to explain the emission observed from a region much larger than the size of the termination shock.

}

%%% SUBSECTION %%%
\subsection{Possible future tests} \label{sec:future}
Considering the results obtained so far, our model can account for the $\gamma$-ray emission from Cygnus OB2 both with the Kraichnan's and Bohm's turbulence description and associated transport, albeit with the caveats mentioned in the previous sections. From the point of view of the spectral analysis, the difference in the $\chi^2$ value is not statistically significant to make a case preferred with respect to the other. In terms of radial morphology, both models lead to a flat profile. 

In principle, other pieces of information can be used to break the degeneracy between the cases. First, good knowledge of the structure and size of the bubble generated by the cluster would provide additional constraints on the wind power and mass loss rate, potentially breaking the degeneracy between the two cases. This kind of information could be provided by a deep study of the thermal diffuse X-ray emission produced by the hot plasma in the bubble: X-ray data from {\it eRosita} would be ideal for such an analysis.

\begin{figure}
\centering
\includegraphics[width=\columnwidth]{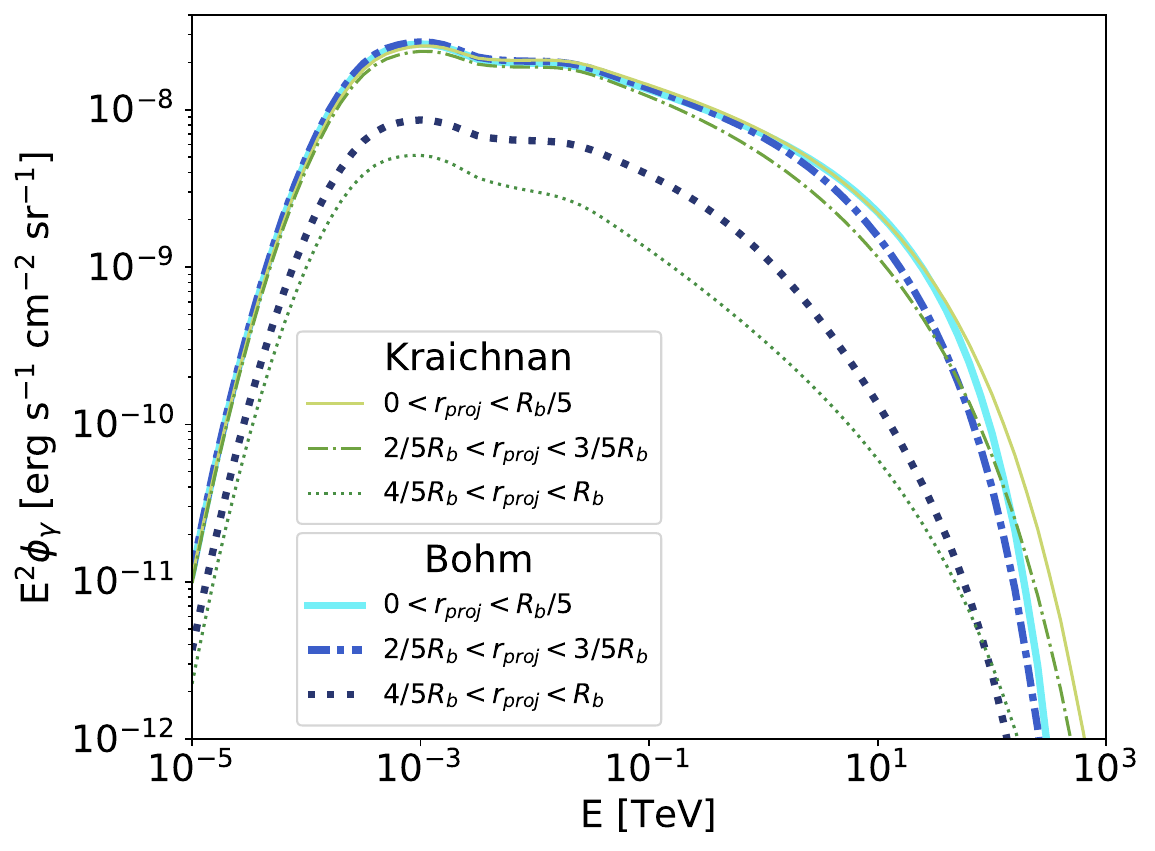}
\caption{Expected hadronic $\gamma$-ray spectra extracted from rings at projected distances of 0<r<$\frac{1}{5}R_{\rm b}$ (continuos lines), $\frac{2}{5}R_{\rm b}$<r<$\frac{3}{5}R_{\rm b}$ (dash-dotted lines), and $\frac{4}{5}R_{\rm b}$<r<$R_{\rm b}$ (dotted lines), for the Kraichnan (thin lines) and Bohm (thick lines) cases.}
\label{fig:SpectraPerRegion}
\end{figure}

Once the bubble is properly identified, one could analyze the $\gamma$-ray emission in regions close to its border. Figure~\ref{fig:SpectraPerRegion} shows the $\gamma$-ray spectrum extracted at different projected distances from the center of the bubble. The spectral shape from a region close to the bubble border could be used to discriminate between different diffusion coefficients. In this regard, data gathered by wide-field telescopes like CTA and LHAASO would play a pivotal role.

%Once the bubble is properly identified, a second possibility is to use the $\gamma$-ray emission in regions close to the bubble border. Figure~\ref{fig:SpectraPerRegion} shows the $\gamma$-ray spectrum extracted at different projected distances from the center of the bubble. The spectral shape from a region close to the bubble border could be used to discriminate between different diffusion coefficients. In this regard, data gathered by wide-field telescopes like CTA and LHAASO  would play a pivotal role.

\begin{figure}
\centering
\includegraphics[width=\columnwidth]{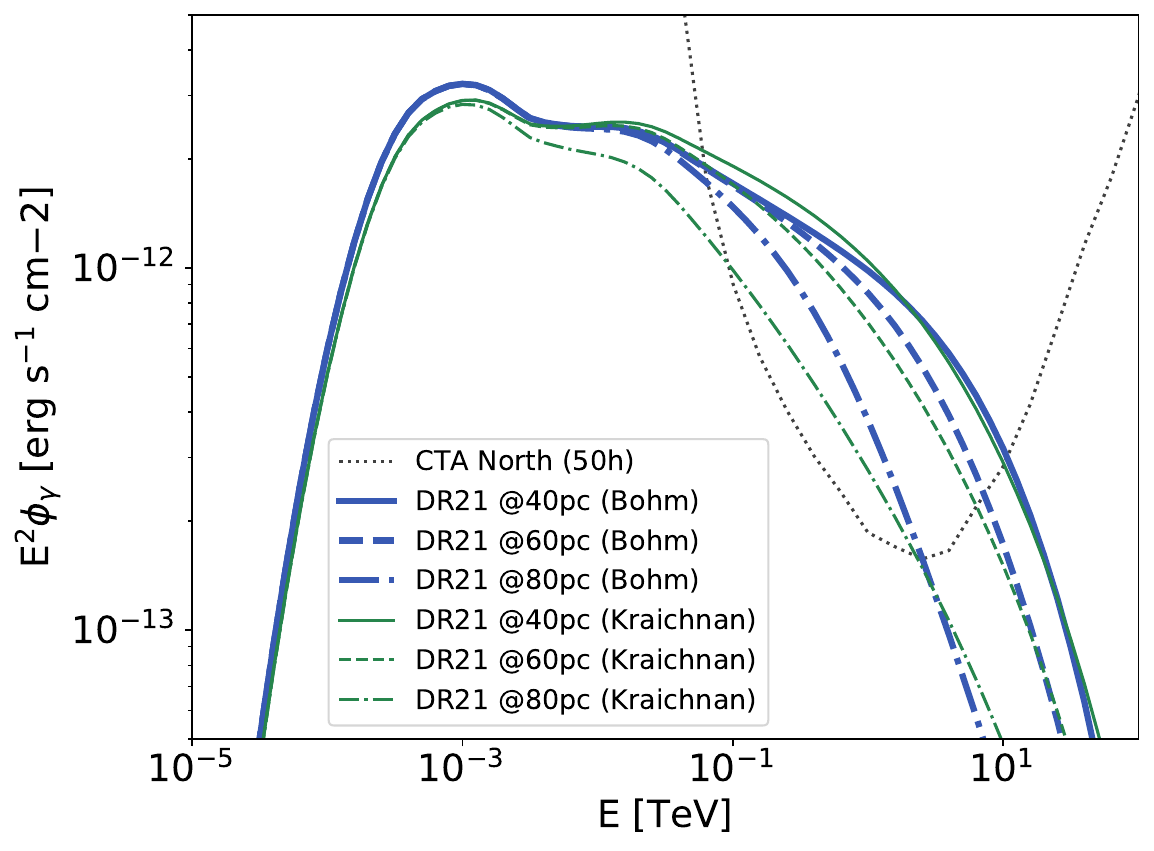}
\caption{Expected hadronic $\gamma$-ray spectra from the molecular cloud DR21 considering the two possible CR distributions given by the Kraichnan (thin lines) and Bohm (thick lines) cases. The spectra are calculated considering different relative distances from Cyg~OB2: 40 pc (continuous line), 60 pc (dashed line), and 80 pc (dot-dashed line). The dotted line shows the CTA North point source sensitivity for 50h of exposure considering the Alpha Configuration.}
\label{fig:DR21Emission}
\end{figure}

Another possible way to discriminate between the two cases is to consider the $\gamma$-ray emission originating from dense molecular clouds that are possibly found inside the bubble. As mentioned in \S~\ref{sec:CygOB2}, Cyg~OB2 is surrounded by a significant number of molecular clouds, some of which are observed to be in direct interaction with the stellar winds and radiation of the cluster. Among the clouds close to Cyg~OB2, DR21, thanks to its mass of $3.4 \times 10^4 \ \rm M_{\odot}$ \citep{SchneiderDR21StarFormation2010}, represents probably the most promising target to observe. Unfortunately, the exact distance between DR21 and Cyg~OB2 is not well constrained due to the uncertainty in the position of the latter. Figure~\ref{fig:DR21Emission} shows the expected $\gamma$-ray spectrum from this cloud, considering different possible distances from Cyg~OB2, and assuming that Cyg~OB2 is located at 1.4 kpc. Below $\sim 100$ GeV the spectra are similar. However, for $E \gtrsim 1$  TeV, the spectral shape in the two cases is significantly different (especially for large distances), so that it should be possible to discriminate between the two. More precisely, Bohm's turbulence is expected to produce a softer spectrum than Kraichnan's. Future observations with CTA will play a crucial role for this type of investigation thanks to the excellent angular resolution.

%%% SECTION $$$$
\section{Conclusions}
\label{sec:conc}

YMSCs probably represent a population of galactic CR sources that can accelerate particles up to very-high energies, generating sizeable extended $\gamma$-ray emission. In order to assess their properties as accelerators, an accurate study of this emission that passes through robust modelling of the CR propagation in their neighbourhood is needed. In this paper, we have considered the specific case of Cygnus OB2, which is associated to extended $\gamma$-ray emission, observed by several experiments, from a few GeVs up to hundreds of TeVs. We studied this emission under the assumption that it is purely hadronic, produced by CRs accelerated at the TS of the cluster wind \citep{Morlino+2021}. 
In our model, the spectral and morphological properties of the emission depend on several properties of the system: the wind luminosity, $L_{\rm w}$, the target density $n_{\rm gas}$, the diffusion coefficient, $D$, and the particle acceleration efficiency, $\epsilon_{\rm cr}$. The latter has no other diagnostics than the comparison between the computed and observed emission and can only be estimated at the end of our calculation, once meaningful estimates are used for the other parameters.  

To constrain the wind luminosity we have studied the stellar population of Cyg~OB2 and, based on empirical relations for the properties of stellar wind, we have inferred a range $L_{\rm w}\in [1.5\--2.9]\times 10^{38}\, \rm erg\, s^{-1}$. To estimate the gas density we have used the column density of neutral atomic and molecular gas, being the ionized component negligible. We have obtained $n_{\rm gas} \simeq 8.5\, \rm cm^{-3}$. 

Finally, the type of diffusion is a crucial ingredient, as it regulates the maximum energy, the shape of the cut-off as well as the CR spatial profile. 
The diffusion properties depend on the level of magnetic turbulence in the wind-blown bubble. We assume that turbulence is produced by MHD instabilities that convert a fraction $\eta_{\rm B}\sim 0.1$ of the wind luminosity into magnetic fields. The turbulence power spectrum is unknown, hence, we decided to explore three different models, commonly adopted in the astrophysics literature: Kolmogorov, Kraichnan and Bohm. 

From the spectral analysis of the observed $\gamma$-ray emission, we rule out the Kolmogorov case, as the wind luminosity needed to reproduce the spectrum is $L_{\rm w} \gtrsim 5 \times 10^{39}$ erg\,s$^{-1}$, which is more than one order of magnitude higher than the one inferred from the stellar population study. Likewise, also the Kraichnan model requires a high wind luminosity to reproduce the observed flux. However, if the value of $\eta_{\rm B}$ is increased up to $\sim 0.3$ (a rather extreme but not impossible value), the $\gamma$-ray emission can be still reproduced. Differently from the Kolmogorov and Kraichnan cases, Bohm-like turbulence, being more efficient at scattering particles, guarantees better confinement and requires a much smaller CR luminosity of the wind. 
Hence, the $\gamma$-ray spectrum can be explained assuming a standard value of $L_{\rm w}=2 \times 10^{38}$ erg\,s$^{-1}$ and a lower efficiency in generating magnetic field (i.e. $\epsilon_{\rm cr}=2.2\%$ and $\eta_{\rm B}=2.4 \times 10^{-3}$ instead of $\epsilon_{\rm cr}=13\%$ and $\eta_{\rm B}=0.1$). The spectral analysis alone is insufficient to discriminate between the Kraichnan and Bohm cases, as both models can provide a reasonable fit of the observed data. Obviously, there is also the possibility that the actual diffusion is something in between the Bohm and the Kraichnan type. Numerical MHD simulation could in principle shed light on this point. 

From the study of the $\gamma$-ray emission morphology, we found that the projected profile in all cases is almost constant in radius, which is consistent with HAWC observations. A flat profile is also expected in the Fermi-LAT domain. Yet, in this case, the results are in contrast with the analysis of Fermi-LAT data by \cite{aharonianMassiveStarsMajor2019a}, which shows a centrally peaked morphology. Such a morphology is indeed expected in a diffusion dominated regime but we have shown that it is unlikely to be realized in the region close to YMSC where strong winds are present, resulting in an advection dominated regime, especially for low energy particles. A possible explanation of the observed center peaked morphology in the Fermi-LAT energy range could be related to particle acceleration taking place in the cluster core, maybe in the wind-wind collision regions.

%\gionote{vedere dove mettwere questo pezzo}
%\gio{Before proceeding further, a note of caution is mandatory. 
%The analysis procedure implemented here is not completely rigorous. A better approach would require a robust multi-instrument joint analysis accounting for the systematics between different experiments which, however, are not available.}

The $\gamma$-ray morphology and spectrum do not allow at present to discriminate between Kraichnan and Bohm transport. This could become possible with upcoming new data and experiments, as we discussed in \S~\ref{sec:disc_profile}. While X-ray data by {\it e-Rosita} could provide useful constraints, by shedding light on the bubble structure, the most promising direct diagnostics for CR acceleration comes from high spatial resolution $\gamma$-ray observations, that could show emission from the molecular clouds in the cluster vicinity or spectral variation of the emission with distance from the cluster core. These will be possible with CTA and ASTRI-MiniArray.

 %In light of the results obtained from both the $\gamma$-ray morphological and spectral study, it is impossible to have clear discrimination between the \nb{these two} cases. In view of possible future new observations or analyses, especially with the arrival of the new Cherenkov Telescope Array (CTA), we have proposed some observables that could be employed to discriminate different types of turbulence in plasma. These include observations of molecular clouds in the vicinity of Cyg~OB2, which is a suitable task for current and future Imaging Air Cherenkov Telescopes, thanks to their small angular resolution. Alternatively, the study of the $\gamma$-ray spectrum at different distances, a technique that can potentially be used in instruments with large field of view and low angular resolution, such as HAWC and LHAASO. Finally, a good knowledge of the bubble structure and its dimension could provide an additional constrain on the cluster wind parameter, aiding in the discrimination between different diffusion regimes. In this regard, the upcoming observations of eRosita could be used to better trace the dimension of the wind bubble. 

\gio{
In this work we have neglected a possible contribution due to leptonic emission. The main reason is that a dominant leptonic contribution to VHE $\gamma$-ray emission is disfavoured by different pieces of evidence, analyzed in recent works \citep{Guevel+2023, Neronov-neutrini:2023}. 
In addition to this, we have shown in Sec.\ref{sec:IC} that electrons accelerated at the wind TS cannot reach energies much larger than $\sim 100$ TeV, implying that they cannot produce the highest energy photons detected by LHAASO. A possible leptonic contribution at lower energy remains possible, but requires an unusually high electron acceleration efficiency (see \S~\ref{sec:IC}). Better quantification of how much, and where in the spectrum, leptons con contribute to the $\gamma$-ray emission of Cyg-OB2 will be the subject of a forthcoming article.
}

The results obtained in this work rely on the assumption that the age of Cyg~OB2 is 3 Myr. As reported in \S~\ref{sec:CygOB2}, \cite{WrightMassiveStarPopOB22015} estimate an age ranging between $3 \-- 5$ Myr. If the age of Cyg~OB2 is 5 Myr, the energy injected by supernova explosions should be of the same order of that of stellar winds (see appendix~\ref{App:Alternative_LwMdot}). In such a scenario, our model for particle acceleration should be revised to account for the CR acceleration at SNR shocks \citep{VieuCRinSuperbubble2022}, which will likely dominate the CR production. However, this requires a detailed description of SNR-wind bubble interaction able to account for all the relevant physical ingredients, which is not a straightforward task. Further work in this direction is necessary.
In any case, we notice that at the moment there are no clear signatures of SN explosions in Cyg~OB2.

%Here we discuss three points:
%\begin{itemize}
%    \item Contribution from the central annuli (close to the TS). Estimate the shell size where electrons survive. Conclusions: the shell emitting in the FermiLAT energy band is too large ($\sim 50 pc$ depending on the diffusion coefficient)
%    \item Show which is the most relevant photon background: CMB, IR (from observed map) or starlight ($\sim r^{-2}$). Conclusion: If the IR or CMB are the most relevent, the emission cannot be peaked at the SC center
%    \item The maximum energy of electrons is $\lesssim 10$ TeV. They cannot contribute to the emission observed above $\sim 100$ GeV.
%    \item The IC contribution could be relevant below $\sim 100$ GeV only if the electron-to-proton ratio, $K_{\rm ep}$, is of the order unity, which is far from theoretical expectation and from what is inferred in SNR shock from multi-wavelength analysis (in fact in the case of SNR $K_{\rm ep} \lesssim 10^{-3}$.
%\end{itemize}
%Conclusion: the IC emission is probably negligible at all wavelengths.

\begin{acknowledgements}
This work has been partially funded by the
European Union - Next Generation EU, through
PRIN-MUR 2022TJW4EJ. We also acknowledge the support by INAF through grant PRIN INAF 2019 {\it ‘‘From massive stars to supernovae and supernova remnants: driving mass, energy and cosmic rays in our Galaxy''}.
SM and GM are partially supported by INAF through Theory Grant 2022 {\it ‘‘Star Clusters As Cosmic Ray Factories''} and Mini Grant 2023 {\it ‘‘Probing Young Massive Stellar Cluster as Cosmic Ray Factories''}.
\end{acknowledgements}

\bibliographystyle{aa} % style aa.bst
\bibliography{Bibliography} % your references Yourfile.bib

\begin{thebibliography}{97}
\expandafter\ifx\csname natexlab\endcsname\relax\def\natexlab#1{#1}\fi

\bibitem[{{Abdollahi} {et~al.}(2020){Abdollahi}, {Acero}, {Ackermann},
  {Ajello}, {Atwood}, {Axelsson}, {Baldini}, {Ballet}, {Barbiellini},
  {Bastieri}, {Becerra Gonzalez}, {Bellazzini}, {Berretta}, {Bissaldi},
  {Blandford}, {Bloom}, {Bonino}, {Bottacini}, {Brandt}, {Bregeon}, {Bruel},
  {Buehler}, {Burnett}, {Buson}, {Cameron}, {Caputo}, {Caraveo}, {Casandjian},
  {Castro}, {Cavazzuti}, {Charles}, {Chaty}, {Chen}, {Cheung}, {Chiaro},
  {Ciprini}, {Cohen-Tanugi}, {Cominsky}, {Coronado-Bl{\'a}zquez}, {Costantin},
  {Cuoco}, {Cutini}, {D'Ammando}, {DeKlotz}, {de la Torre Luque}, {de Palma},
  {Desai}, {Digel}, {Di Lalla}, {Di Mauro}, {Di Venere}, {Dom{\'\i}nguez},
  {Dumora}, {Fana Dirirsa}, {Fegan}, {Ferrara}, {Franckowiak}, {Fukazawa},
  {Funk}, {Fusco}, {Gargano}, {Gasparrini}, {Giglietto}, {Giommi}, {Giordano},
  {Giroletti}, {Glanzman}, {Green}, {Grenier}, {Griffin}, {Grondin}, {Grove},
  {Guiriec}, {Harding}, {Hayashi}, {Hays}, {Hewitt}, {Horan},
  {J{\'o}hannesson}, {Johnson}, {Kamae}, {Kerr}, {Kocevski}, {Kovac'evic'},
  {Kuss}, {Landriu}, {Larsson}, {Latronico}, {Lemoine-Goumard}, {Li},
  {Liodakis}, {Longo}, {Loparco}, {Lott}, {Lovellette}, {Lubrano}, {Madejski},
  {Maldera}, {Malyshev}, {Manfreda}, {Marchesini}, {Marcotulli},
  {Mart{\'\i}-Devesa}, {Martin}, {Massaro}, {Mazziotta}, {McEnery}, {Mereu},
  {Meyer}, {Michelson}, {Mirabal}, {Mizuno}, {Monzani}, {Morselli},
  {Moskalenko}, {Negro}, {Nuss}, {Ojha}, {Omodei}, {Orienti}, {Orlando},
  {Ormes}, {Palatiello}, {Paliya}, {Paneque}, {Pei}, {Pe{\~n}a-Herazo},
  {Perkins}, {Persic}, {Pesce-Rollins}, {Petrosian}, {Petrov}, {Piron}, {Poon},
  {Porter}, {Principe}, {Rain{\`o}}, {Rando}, {Razzano}, {Razzaque}, {Reimer},
  {Reimer}, {Remy}, {Reposeur}, {Romani}, {Saz Parkinson}, {Schinzel},
  {Serini}, {Sgr{\`o}}, {Siskind}, {Smith}, {Spandre}, {Spinelli}, {Strong},
  {Suson}, {Tajima}, {Takahashi}, {Tak}, {Thayer}, {Thompson}, {Tibaldo},
  {Torres}, {Torresi}, {Valverde}, {Van Klaveren}, {van Zyl}, {Wood},
  {Yassine}, \& {Zaharijas}}]{AbdollahiFermi4FGL2020}
{Abdollahi}, S., {Acero}, F., {Ackermann}, M., {et~al.} 2020, \apjs, 247, 33

\bibitem[{Abeysekara {et~al.}(2021)Abeysekara, Albert, Alfaro, Alvarez,
  Camacho, Arteaga-Velázquez, Arunbabu, Rojas, Solares, Baghmanyan,
  Belmont-Moreno, BenZvi, Blandford, Brisbois, Caballero-Mora, Capistrán,
  Carramiñana, Casanova, Cotti, León, De~la Fuente, Hernandez, Dingus,
  DuVernois, Durocher, Díaz-Vélez, Ellsworth, Engel, Espinoza, Fan, Fang,
  Fleischhack, Fraija, Galván-Gámez, Garcia, García-González, Garfias,
  Giacinti, González, Goodman, Harding, Hernandez, Hinton, Hona, Huang,
  Hueyotl-Zahuantitla, Hüntemeyer, Iriarte, Jardin-Blicq, Joshi, Kieda, Lara,
  Lee, Vargas, Linnemann, Longinotti, Luis-Raya, Lundeen, Malone, Martinez,
  Martinez-Castellanos, Martínez-Castro, Matthews, Miranda-Romagnoli,
  Morales-Soto, Moreno, Mostafá, Nayerhoda, Nellen, Newbold, Nisa,
  Noriega-Papaqui, Olivera-Nieto, Omodei, Peisker, Pérez~Araujo,
  Pérez-Pérez, Ren, Rho, Rosa-González, Ruiz-Velasco, Salazar, Greus,
  Sandoval, Schneider, Schoorlemmer, Serna, Smith, Springer, Surajbali,
  Tollefson, Torres, Torres-Escobedo, Ureña-Mena, Weisgarber, Werner, Willox,
  Zepeda, Zhou, De~León, \& Álvarez}]{HAWCcoll:2021}
Abeysekara, A.~U., Albert, A., Alfaro, R., {et~al.} 2021, Nature Astronomy

\bibitem[{Abramowski {et~al.}(2012)Abramowski, Acero, Aharonian, Akhperjanian,
  Anton, Balzer, Barnacka, Barres~de Almeida, Becherini, Becker, Behera,
  Bernlöhr, Birsin, Biteau, Bochow, Boisson, Bolmont, Bordas, Brucker, Brun,
  Brun, Bulik, Büsching, Carrigan, Casanova, Cerruti, Chadwick, Charbonnier,
  Chaves, Cheesebrough, Chounet, Clapson, Coignet, Cologna, Conrad, Dalton,
  Daniel, Davids, Degrange, Deil, Dickinson, Djannati-Ataï, Domainko, Drury,
  Dubois, Dubus, Dutson, Dyks, Dyrda, Egberts, Eger, Espigat, Fallon, Farnier,
  Fegan, Feinstein, Fernandes, Fiasson, Fontaine, Förster, Füßling, Gallant,
  Gast, Gérard, Gerbig, Giebels, Glicenstein, Glück, Goret, Göring,
  Häffner, Hague, Hampf, Hauser, Heinz, Heinzelmann, Henri, Hermann, Hinton,
  Hoffmann, Hofmann, Hofverberg, Holler, Horns, Jacholkowska, de~Jager, Jahn,
  Jamrozy, Jung, Kastendieck, Katarzyński, Katz, Kaufmann, Keogh, Khangulyan,
  Khélifi, Klochkov, Kluźniak, Kneiske, Komin, Kosack, Kossakowski, Laffon,
  Lamanna, Lennarz, Lohse, Lopatin, Lu, Marandon, Marcowith, Masbou, Maurin,
  Maxted, Mayer, McComb, Medina, Méhault, Moderski, Moulin, Naumann,
  Naumann-Godo, de~Naurois, Nedbal, Nekrassov, Nguyen, Nicholas, Niemiec,
  Nolan, Ohm, de~Oña~Wilhelmi, Opitz, Ostrowski, Oya, Panter, Paz~Arribas,
  Pedaletti, Pelletier, Petrucci, Pita, Pühlhofer, Punch, Quirrenbach, Raue,
  Rayner, Reimer, Reimer, Renaud, de~los Reyes, Rieger, Ripken, Rob,
  Rosier-Lees, Rowell, Rudak, Rulten, Ruppel, Sahakian, Sanchez, Santangelo,
  Schlickeiser, Schöck, Schulz, Schwanke, Schwarzburg, Schwemmer, Sheidaei,
  Sikora, Skilton, Sol, Spengler, Stawarz, Steenkamp, Stegmann, Stinzing,
  Stycz, Sushch, Szostek, Tavernet, Terrier, Tluczykont, Valerius, van Eldik,
  Vasileiadis, Venter, Vialle, Viana, Vincent, Völk, Volpe, Vorobiov, Vorster,
  Wagner, Ward, White, Wierzcholska, Zacharias, Zajczyk, Zdziarski, Zech, \&
  Zechlin}]{abramowskiDiscoveryExtendedVHE2012}
Abramowski, A., Acero, F., Aharonian, F., {et~al.} 2012, Astronomy \&
  Astrophysics, 537, A114

\bibitem[{Ackermann {et~al.}(2011)Ackermann, Ajello, Allafort, Baldini, Ballet,
  Barbiellini, Bastieri, Belfiore, Bellazzini, Berenji, Blandford, Bloom,
  Bonamente, Borgland, Bottacini, Brigida, Bruel, Buehler, Buson, Caliandro,
  Cameron, Caraveo, Casandjian, Cecchi, Chekhtman, Cheung, Chiang, Ciprini,
  Claus, Cohen-Tanugi, de~Angelis, de~Palma, Dermer, do~Couto~e Silva, Drell,
  Dumora, Favuzzi, Fegan, Focke, Fortin, Fukazawa, Fusco, Gargano, Germani,
  Giglietto, Giordano, Giroletti, Glanzman, Godfrey, Grenier, Guillemot,
  Guiriec, Hadasch, Hanabata, Harding, Hayashida, Hayashi, Hays, Johannesson,
  Johnson, Kamae, Katagiri, Kataoka, Kerr, Knodlseder, Kuss, Lande, Latronico,
  Lee, Longo, Loparco, Lott, Lovellette, Lubrano, Martin, Mazziotta, McEnery,
  Mehault, Michelson, Mitthumsiri, Mizuno, Monte, Monzani, Morselli,
  Moskalenko, Murgia, Naumann-Godo, Nolan, Norris, Nuss, Ohsugi, Okumura,
  Orlando, Ormes, Ozaki, Paneque, Parent, Pesce-Rollins, Pierbattista, Piron,
  Pohl, Prokhorov, Raino, Rando, Razzano, Reposeur, Ritz, Parkinson, Sgro,
  Siskind, Smith, Spinelli, Strong, Takahashi, Tanaka, Thayer, Thayer,
  Thompson, Tibaldo, Torres, Tosti, Tramacere, Troja, Uchiyama, Vandenbroucke,
  Vasileiou, Vianello, Vitale, Waite, Wang, Winer, Wood, Yang, Zimmer, \&
  Bontemps}]{ackermannCocoonFreshlyAccelerated2011a}
Ackermann, M., Ajello, M., Allafort, A., {et~al.} 2011, Science, 334, 1103

\bibitem[{{Aguilar} {et~al.}(2015){Aguilar}, {Aisa}, {Alpat}, {Alvino},
  {Ambrosi}, {Andeen}, {Arruda}, {Attig}, {Azzarello}, {Bachlechner}, {Barao},
  {Barrau}, {Barrin}, {Bartoloni}, {Basara}, {Battarbee}, {Battiston}, {Bazo},
  {Becker}, {Behlmann}, {Beischer}, {Berdugo}, {Bertucci}, {Bigongiari},
  {Bindi}, {Bizzaglia}, {Bizzarri}, {Boella}, {de Boer}, {Bollweg},
  {Bonnivard}, {Borgia}, {Borsini}, {Boschini}, {Bourquin}, {Burger}, {Cadoux},
  {Cai}, {Capell}, {Caroff}, {Casaus}, {Cascioli}, {Castellini}, {Cernuda},
  {Cerreta}, {Cervelli}, {Chae}, {Chang}, {Chen}, {Chen}, {Cheng}, {Chen},
  {Cheng}, {Chou}, {Choumilov}, {Choutko}, {Chung}, {Clark}, {Clavero},
  {Coignet}, {Consolandi}, {Contin}, {Corti}, {Gil}, {Coste}, {Creus},
  {Crispoltoni}, {Cui}, {Dai}, {Delgado}, {Della Torre}, {Demirk{\"o}z},
  {Derome}, {Di Falco}, {Di Masso}, {Dimiccoli}, {D{\'\i}az}, {von Doetinchem},
  {Donnini}, {Du}, {Duranti}, {D'Urso}, {Eline}, {Eppling}, {Eronen}, {Fan},
  {Farnesini}, {Feng}, {Fiandrini}, {Fiasson}, {Finch}, {Fisher},
  {Galaktionov}, {Gallucci}, {Garc{\'\i}a}, {Garc{\'\i}a-L{\'o}pez},
  {Gargiulo}, {Gast}, {Gebauer}, {Gervasi}, {Ghelfi}, {Gillard}, {Giovacchini},
  {Goglov}, {Gong}, {Goy}, {Grabski}, {Grandi}, {Graziani}, {Guandalini},
  {Guerri}, {Guo}, {Haas}, {Habiby}, {Haino}, {Han}, {He}, {Heil}, {Hoffman},
  {Hsieh}, {Huang}, {Huh}, {Incagli}, {Ionica}, {Jang}, {Jinchi}, {Kanishev},
  {Kim}, {Kim}, {Kirn}, {Kossakowski}, {Kounina}, {Kounine}, {Koutsenko},
  {Krafczyk}, {La Vacca}, {Laudi}, {Laurenti}, {Lazzizzera}, {Lebedev}, {Lee},
  {Lee}, {Leluc}, {Levi}, {Li}, {Li}, {Li}, {Li}, {Li}, {Li}, {Li}, {Li}, {Li},
  {Lim}, {Lin}, {Lipari}, {Lippert}, {Liu}, {Liu}, {Lolli}, {Lomtadze}, {Lu},
  {Lu}, {Lu}, {Luebelsmeyer}, {Luo}, {Lv}, {Majka}, {Ma{\~n}{\'a}},
  {Mar{\'\i}n}, {Martin}, {Mart{\'\i}nez}, {Masi}, {Maurin}, {Menchaca-Rocha},
  {Meng}, {Mo}, {Morescalchi}, {Mott}, {M{\"u}ller}, {Ni}, {Nikonov},
  {Nozzoli}, {Nunes}, {Obermeier}, {Oliva}, {Orcinha}, {Palmonari},
  {Palomares}, {Paniccia}, {Papi}, {Pauluzzi}, {Pedreschi}, {Pensotti},
  {Pereira}, {Picot-Clemente}, {Pilo}, {Piluso}, {Pizzolotto}, {Plyaskin},
  {Pohl}, {Poireau}, {Postaci}, {Putze}, {Quadrani}, {Qi}, {Qin}, {Qu},
  {R{\"a}ih{\"a}}, {Rancoita}, {Rapin}, {Ricol}, {Rodr{\'\i}guez},
  {Rosier-Lees}, {Rozhkov}, {Rozza}, {Sagdeev}, {Sandweiss}, {Saouter},
  {Sbarra}, {Schael}, {Schmidt}, {von Dratzig}, {Schwering}, {Scolieri}, {Seo},
  {Shan}, {Shan}, {Shi}, {Shi}, {Shi}, {Siedenburg}, {Son}, {Spada},
  {Spinella}, {Sun}, {Sun}, {Tacconi}, {Tang}, {Tang}, {Tang}, {Tao},
  {Tescaro}, {Ting}, {Ting}, {Tomassetti}, {Torsti}, {T{\"u}rko{\v{g}}lu},
  {Urban}, {Vagelli}, {Valente}, {Vannini}, {Valtonen}, {Vaurynovich},
  {Vecchi}, {Velasco}, {Vialle}, {Vitale}, {Vitillo}, {Wang}, {Wang}, {Wang},
  {Wang}, {Wang}, {Wang}, {Weng}, {Whitman}, {Wienkenh{\"o}ver}, {Wu}, {Wu},
  {Xia}, {Xie}, {Xie}, {Xiong}, {Xin}, {Xu}, {Xu}, {Yan}, {Yang}, {Yang}, {Ye},
  {Yi}, {Yu}, {Yu}, {Zeissler}, {Zhang}, {Zhang}, {Zhang}, {Zhang}, {Zheng},
  {Zhuang}, {Zhukov}, {Zichichi}, {Zimmermann}, {Zuccon}, {Zurbach}, \& {AMS
  Collaboration}}]{AMS02-protons:2015}
{Aguilar}, M., {Aisa}, D., {Alpat}, B., {et~al.} 2015, PRL, 114, 171103

\bibitem[{{Aharonian} {et~al.}(2022){Aharonian}, {Ashkar}, {Backes}, {Barbosa
  Martins}, {Becherini}, {Berge}, {Bi}, {B{\"o}ttcher}, {de Bony de Lavergne},
  {Bradascio}, {Brose}, {Brun}, {Bulik}, {Burger-Scheidlin}, {Cangemi},
  {Caroff}, {Casanova}, {Cerruti}, {Chand}, {Chandra}, {Chen}, {Chibueze},
  {Cristofari}, {Damascene Mbarubucyeye}, {Djannati-Ata{\"\i}}, {Ernenwein},
  {Feijen}, {Fichet de Clairfontaine}, {Fontaine}, {Funk}, {Gabici}, {Gallant},
  {Ghafourizadeh}, {Giavitto}, {Giunti}, {Glawion}, {Glicenstein}, {Goswami},
  {Grondin}, {H{\"a}rer}, {Haupt}, {Hinton}, {H{\"o}rbe}, {Hofmann}, {Holch},
  {Holler}, {Horns}, {Jamrozy}, {Joshi}, {Jung-Richardt}, {Kasai},
  {Katarzy{\'n}ski}, {Katz}, {Kh{\'e}lifi}, {Klu{\'z}niak}, {Komin}, {Kosack},
  {Kostunin}, {Kukec Mezek}, {Lang}, {Le Stum}, {Lemi{\`e}re},
  {Lemoine-Goumard}, {Lenain}, {Leuschner}, {Lohse}, {Luashvili}, {Lypova},
  {Mackey}, {Majumdar}, {Malyshev}, {Marandon}, {Marchegiani}, {Marcowith},
  {Mart{\'\i}-Devesa}, {Marx}, {Maurin}, {Meyer}, {Mitchell}, {Moderski},
  {Mohrmann}, {Montanari}, {Moulin}, {Muller}, {Murach}, {Nakashima}, {de
  Naurois}, {Nayerhoda}, {Niemiec}, {Ohm}, {Olivera-Nieto}, {de Ona Wilhelmi},
  {Ostrowski}, {Panny}, {Panter}, {Parsons}, {Peron}, {Prokhorov},
  {P{\"u}hlhofer}, {Punch}, {Quirrenbach}, {Rauth}, {Reichherzer}, {Reimer},
  {Reimer}, {Renaud}, {Reville}, {Rieger}, {Rowell}, {Rudak}, {Ruiz-Velasco},
  {Sahakian}, {Salzmann}, {Sanchez}, {Santangelo}, {Sasaki}, {Sch{\"u}ssler},
  {Schutte}, {Schwanke}, {Shapopi}, {Specovius}, {Spencer}, {Stawarz},
  {Steenkamp}, {Steinmassl}, {Steppa}, {Sushch}, {Suzuki}, {Takahashi},
  {Tanaka}, {Terrier}, {Thorpe-Morgan}, {Tsirou}, {Tsuji}, {Tuffs}, {Unbehaun},
  {van Eldik}, {van Soelen}, {Vecchi}, {Veh}, {Venter}, {Vink}, {Wagner},
  {White}, {Wierzcholska}, {Wun Wong}, {Zacharias}, {Zargaryan}, {Zdziarski},
  {Zhu}, {Zouari}, {{\.Z}ywucka}, {Blackwell}, {Braiding}, {Burton}, {Cubuk},
  {Filipovi{\'c}}, {Tothill}, \& {Wong}}]{HESSWesterlund12022}
{Aharonian}, F., {Ashkar}, H., {Backes}, M., {et~al.} 2022, arXiv e-prints,
  arXiv:2207.10921

\bibitem[{Aharonian {et~al.}(2019)Aharonian, Yang, \&
  de~Oña~Wilhelmi}]{aharonianMassiveStarsMajor2019a}
Aharonian, F., Yang, R., \& de~Oña~Wilhelmi, E. 2019, Nature Astronomy, 3, 561

\bibitem[{{Aharonian}(2004)}]{Aharonian_book:2004}
{Aharonian}, F.~A. 2004, {Very high energy cosmic gamma radiation : a crucial
  window on the extreme Universe}

\bibitem[{{Albacete-Colombo} {et~al.}(2023){Albacete-Colombo}, {Drake},
  {Flaccomio}, {Wright}, {Kashyap}, {Guarcello}, {Briggs}, {Drew}, {Fenech},
  {Micela}, {McCollough}, {Prinja}, {Schneider}, {Sciortino}, \&
  {Vink}}]{Albacete-Colombo+2023}
{Albacete-Colombo}, J.~F., {Drake}, J.~J., {Flaccomio}, E., {et~al.} 2023,
  \apjs, 269, 14

\bibitem[{{Astiasarain} {et~al.}(2023){Astiasarain}, {Tibaldo}, {Martin},
  {Kn{\"o}dlseder}, \& {Remy}}]{Astiasarain+2023}
{Astiasarain}, X., {Tibaldo}, L., {Martin}, P., {Kn{\"o}dlseder}, J., \&
  {Remy}, Q. 2023, \aap, 671, A47

\bibitem[{{Baade} \& {Zwicky}(1934)}]{BaadeCRsSNRsOrigin1934}
{Baade}, W. \& {Zwicky}, F. 1934, Physical Review, 46, 76

\bibitem[{{Bartoli} {et~al.}(2014){Bartoli}, {Bernardini}, {Bi}, {Branchini},
  {Budano}, {Camarri}, {Cao}, {Cardarelli}, {Catalanotti}, {Chen}, {Chen},
  {Creti}, {Cui}, {Dai}, {D'Amone}, {Danzengluobu}, {De Mitri}, {D'Ettorre
  Piazzoli}, {Di Girolamo}, {Di Sciascio}, {Feng}, {Feng}, {Feng}, {Gou},
  {Guo}, {He}, {Hu}, {Hu}, {Iacovacci}, {Iuppa}, {Jia}, {Labaciren}, {Li},
  {Liguori}, {Liu}, {Liu}, {Liu}, {Lu}, {Ma}, {Ma}, {Mancarella}, {Mari},
  {Marsella}, {Martello}, {Mastroianni}, {Montini}, {Ning}, {Panareo},
  {Perrone}, {Pistilli}, {Ruggieri}, {Salvini}, {Santonico}, {Shen}, {Sheng},
  {Shi}, {Surdo}, {Tan}, {Vallania}, {Vernetto}, {Vigorito}, {Wang}, {Wu},
  {Wu}, {Xue}, {Yang}, {Yang}, {Yao}, {Yuan}, {Zha}, {Zhang}, {Zhang}, {Zhang},
  {Zhang}, {Zhao}, {Zhaxiciren}, {Zhaxisangzhu}, {Zhou}, {Zhu}, {Zhu}, {Zizzi},
  \& {ARGO-YBJ Collaboration}}]{BartoliIdentificationTeVGammaray2014}
{Bartoli}, B., {Bernardini}, P., {Bi}, X.~J., {et~al.} 2014, \apj, 790, 152

\bibitem[{{Bell} {et~al.}(2013){Bell}, {Schure}, {Reville}, \&
  {Giacinti}}]{Bell:2013}
{Bell}, A.~R., {Schure}, K.~M., {Reville}, B., \& {Giacinti}, G. 2013, \mnras,
  431, 415

\bibitem[{{Berlanas} {et~al.}(2019){Berlanas}, {Wright}, {Herrero}, {Drew}, \&
  {Lennon}}]{BerlanasDisentanglingSpatialSubstructure2019}
{Berlanas}, S.~R., {Wright}, N.~J., {Herrero}, A., {Drew}, J.~E., \& {Lennon},
  D.~J. 2019, \mnras, 484, 1838

\bibitem[{Binns {et~al.}(2008)Binns, Wiedenbeck, Arnould, Cummings, {de Nolfo},
  Goriely, Israel, Leske, Mewaldt, Stone, \& {von Rosenvinge}}]{BinnsNe2008}
Binns, W., Wiedenbeck, M., Arnould, M., {et~al.} 2008, New Astronomy Reviews,
  52, 427, astronomy with Radioactivities. VI

\bibitem[{{Blasi} \& {Morlino}(2023{\natexlab{a}})}]{Blasi-Morlino:2023}
{Blasi}, P. \& {Morlino}, G. 2023{\natexlab{a}}, in preparation

\bibitem[{{Blasi} \& {Morlino}(2023{\natexlab{b}})}]{BlasiCygOB2Gamma2023}
{Blasi}, P. \& {Morlino}, G. 2023{\natexlab{b}}, \mnras, 523, 4015

\bibitem[{{Bolton}(1948)}]{BoltonDiscreteSourcesCyg1948}
{Bolton}, J.~G. 1948, \nat, 162, 141

\bibitem[{{Buzzoni}(2002)}]{BuzzoniTOtime2002}
{Buzzoni}, A. 2002, \aj, 123, 1188

\bibitem[{{Bykov} {et~al.}(2020){Bykov}, {Marcowith}, {Amato}, {Kalyashova},
  {Kruijssen}, \& {Waxman}}]{Bykov_review:2020}
{Bykov}, A.~M., {Marcowith}, A., {Amato}, E., {et~al.} 2020, \ssr, 216, 42

\bibitem[{{Cao} {et~al.}(2021){Cao}, {Aharonian}, {An}, {Axikegu}, {Bai},
  {Bao}, {Bastieri}, {Bi}, {Bi}, {Cai}, {Cai}, {Cao}, {Chang}, {Chang},
  {Chang}, {Chen}, {Chen}, {Chen}, {Chen}, {Chen}, {Chen}, {Chen}, {Chen},
  {Chen}, {Chen}, {Chen}, {Chen}, {Chen}, {Cheng}, {Cheng}, {Cui}, {Cui},
  {Cui}, {Dai}, {Dai}, {Dai}, {Danzengluobu}, {della Volpe}, {D'Ettorre
  Piazzoli}, {Dong}, {Fan}, {Fan}, {Fan}, {Fang}, {Fang}, {Feng}, {Feng},
  {Feng}, {Feng}, {Gao}, {Gao}, {Gao}, {Gao}, {Ge}, {Geng}, {Gong}, {Gou},
  {Gu}, {Guo}, {Guo}, {Guo}, {Guo}, {Han}, {He}, {He}, {He}, {He}, {He}, {He},
  {Heller}, {Hor}, {Hou}, {Hou}, {Hu}, {Hu}, {Hu}, {Hu}, {Huang}, {Huang},
  {Huang}, {Huang}, {Huang}, {Ji}, {Ji}, {Jia}, {Jiang}, {Jiang}, {Jin},
  {Kuleshov}, {Levochkin}, {Li}, {Li}, {Li}, {Li}, {Li}, {Li}, {Li}, {Li},
  {Li}, {Li}, {Li}, {Li}, {Li}, {Li}, {Li}, {Li}, {Li}, {Liang}, {Liang},
  {Lin}, {Liu}, {Liu}, {Liu}, {Liu}, {Liu}, {Liu}, {Liu}, {Liu}, {Liu}, {Liu},
  {Liu}, {Liu}, {Liu}, {Liu}, {Liu}, {Long}, {Lu}, {Lv}, {Ma}, {Ma}, {Ma},
  {Mao}, {Masood}, {Mitthumsiri}, {Montaruli}, {Nan}, {Pang},
  {Pattarakijwanich}, {Pei}, {Qi}, {Ruffolo}, {Rulev}, {S{\'a}iz}, {Shao},
  {Shchegolev}, {Sheng}, {Shi}, {Song}, {Stenkin}, {Stepanov}, {Sun}, {Sun},
  {Sun}, {Tam}, {Tang}, {Tian}, {Wang}, {Wang}, {Wang}, {Wang}, {Wang}, {Wang},
  {Wang}, {Wang}, {Wang}, {Wang}, {Wang}, {Wang}, {Wang}, {Wang}, {Wang},
  {Wang}, {Wang}, {Wang}, {Wang}, {Wang}, {Wang}, {Wei}, {Wei}, {Wei}, {Wen},
  {Wu}, {Wu}, {Wu}, {Wu}, {Wu}, {Xi}, {Xia}, {Xia}, {Xiang}, {Xiao}, {Xiao},
  {Xin}, {Xin}, {Xing}, {Xu}, {Xu}, {Xue}, {Yan}, {Yang}, {Yang}, {Yang},
  {Yang}, {Yang}, {Yang}, {Yang}, {Yao}, {Yao}, {Ye}, {Yin}, {Yin}, {You},
  {You}, {Yu}, {Yuan}, {Zeng}, {Zeng}, {Zeng}, {Zeng}, {Zha}, {Zhai}, {Zhang},
  {Zhang}, {Zhang}, {Zhang}, {Zhang}, {Zhang}, {Zhang}, {Zhang}, {Zhang},
  {Zhang}, {Zhang}, {Zhang}, {Zhang}, {Zhang}, {Zhang}, {Zhang}, {Zhang},
  {Zhang}, {Zhang}, {Zhao}, {Zhao}, {Zhao}, {Zhao}, {Zhao}, {Zheng}, {Zheng},
  {Zhou}, {Zhou}, {Zhou}, {Zhou}, {Zhou}, {Zhou}, {Zhu}, {Zhu}, {Zhu}, {Zhu},
  \& {Zuo}}]{CaoLHAASOPlaneSurvey2021}
{Cao}, Z., {Aharonian}, F.~A., {An}, Q., {et~al.} 2021, \nat, 594, 33

\bibitem[{{Cao} {et~al.}(2023){Cao}, {Li}, {Gau}, {Liu}, \&
  {Yang}}]{LHAASOcoll:2023}
{Cao}, Z., {Li}, C., {Gau}, C.~D., {Liu}, R.~Y., \& {Yang}, R.~Z. 2023, arXiv
  e-prints, arXiv:2310.10100

\bibitem[{{Cardillo} {et~al.}(2015){Cardillo}, {Amato}, \&
  {Blasi}}]{Cardillo:2015}
{Cardillo}, M., {Amato}, E., \& {Blasi}, P. 2015, Astroparticle Physics, 69, 1

\bibitem[{{Carroll} \& {Ostlie}(1996)}]{CarrollIntroModAstro1996}
{Carroll}, B.~W. \& {Ostlie}, D.~A. 1996, {An Introduction to Modern
  Astrophysics}

\bibitem[{{Casse} \& {Paul}(1980)}]{Casse-Paul:1980}
{Casse}, M. \& {Paul}, J.~A. 1980, \apj, 237, 236

\bibitem[{{Castor} {et~al.}(1975){Castor}, {McCray}, \&
  {Weaver}}]{CastorInterstellarBubbles1975}
{Castor}, J., {McCray}, R., \& {Weaver}, R. 1975, \apjl, 200, L107

\bibitem[{Cesarsky \& Montmerle(1983)}]{Cesarsky-Montmerle:1983}
Cesarsky, C.~J. \& Montmerle, T. 1983, Space Science Reviews, 36, 173

\bibitem[{{Comer{\'o}n} \& {Pasquali}(2012)}]{ComeronNewOB2Members2012}
{Comer{\'o}n}, F. \& {Pasquali}, A. 2012, \aap, 543, A101

\bibitem[{{Cristofari} {et~al.}(2022){Cristofari}, {Blasi}, \&
  {Caprioli}}]{Cristofari+2022}
{Cristofari}, P., {Blasi}, P., \& {Caprioli}, D. 2022, \apj, 930, 28

\bibitem[{Dame {et~al.}(2001)Dame, Hartmann, \&
  Thaddeus}]{dameMilkyWayMolecular2001}
Dame, T.~M., Hartmann, D., \& Thaddeus, P. 2001, The Astrophysical Journal,
  547, 792

\bibitem[{{Demircan} \& {Kahraman}(1991)}]{DemircanStarsMRR1991}
{Demircan}, O. \& {Kahraman}, G. 1991, \apss, 181, 313

\bibitem[{{Dickel} {et~al.}(1969){Dickel}, {Wendker}, \&
  {Bieritz}}]{DickelHII1969}
{Dickel}, H.~R., {Wendker}, H., \& {Bieritz}, J.~H. 1969, \aap, 1, 270

\bibitem[{{Downes} \& {Rinehart}(1966)}]{DownesMWObsCygX1966}
{Downes}, D. \& {Rinehart}, R. 1966, \apj, 144, 937

\bibitem[{{Draine}(2011)}]{DraineISMPhysics2011}
{Draine}, B.~T. 2011, {Physics of the Interstellar and Intergalactic Medium}

\bibitem[{{Drew} {et~al.}(2008){Drew}, {Greimel}, {Irwin}, \&
  {Sale}}]{DrewEarlyAStars2008}
{Drew}, J.~E., {Greimel}, R., {Irwin}, M.~J., \& {Sale}, S.~E. 2008, \mnras,
  386, 1761

\bibitem[{{Drury}(1983)}]{Drury:1983}
{Drury}, L.~O. 1983, Reports on Progress in Physics, 46, 973

\bibitem[{{Dwarkadas}(2023)}]{Dwarkadas:2023}
{Dwarkadas}, V.~V. 2023, Galaxies, 11, 78

\bibitem[{{Eker} {et~al.}(2018){Eker}, {Bak{\i}{\c{s}}}, {Bilir}, {Soydugan},
  {Steer}, {Soydugan}, {Bak{\i}{\c{s}}}, {Ali{\c{c}}avu{\c{s}}}, {Aslan}, \&
  {Alpsoy}}]{EkerMLR2018}
{Eker}, Z., {Bak{\i}{\c{s}}}, V., {Bilir}, S., {et~al.} 2018, \mnras, 479, 5491

\bibitem[{{El-Badry} {et~al.}(2019){El-Badry}, {Ostriker}, {Kim}, {Quataert},
  \& {Weisz}}]{El-Badry+2019}
{El-Badry}, K., {Ostriker}, E.~C., {Kim}, C.-G., {Quataert}, E., \& {Weisz},
  D.~R. 2019, \mnras, 490, 1961

\bibitem[{{Emig} {et~al.}(2022){Emig}, {White}, {Salas}, {Karim}, {van Weeren},
  {Teuben}, {Zavagno}, {Chiu}, {Haverkorn}, {Oonk}, {Orr{\'u}}, {Polderman},
  {Reich}, {R{\"o}ttgering}, \&
  {Tielens}}]{EmigFilamentaryStructuresIonizedGasCygX2022}
{Emig}, K.~L., {White}, G.~J., {Salas}, P., {et~al.} 2022, \aap, 664, A88

\bibitem[{{Funk}(2017)}]{Funk:2017}
{Funk}, S. 2017, in Handbook of Supernovae, ed. A.~W. {Alsabti} \& P.~{Murdin},
  1737

\bibitem[{{Green} {et~al.}(2019){Green}, {Schlafly}, {Zucker}, {Speagle}, \&
  {Finkbeiner}}]{GreenDust3dMaps2019}
{Green}, G.~M., {Schlafly}, E., {Zucker}, C., {Speagle}, J.~S., \&
  {Finkbeiner}, D. 2019, \apj, 887, 93

\bibitem[{{Guevel} {et~al.}(2023){Guevel}, {Beardmore}, {Page}, {Lien}, {Fang},
  {Tibaldo}, {Casanova}, \& {Huentemeyer}}]{Guevel+2023}
{Guevel}, D., {Beardmore}, A., {Page}, K.~L., {et~al.} 2023, \apj, 950, 116

\bibitem[{{Gupta} {et~al.}(2020){Gupta}, {Nath}, {Sharma}, \&
  {Eichler}}]{GuptaNeMSC2020}
{Gupta}, S., {Nath}, B.~B., {Sharma}, P., \& {Eichler}, D. 2020, \mnras, 493,
  3159

\bibitem[{{Haggerty} \& {Caprioli}(2020)}]{Haggerty-Caprioli:2020}
{Haggerty}, C.~C. \& {Caprioli}, D. 2020, \apj, 905, 1

\bibitem[{{Hanson}(2003)}]{HansonStudyCygOB22003}
{Hanson}, M.~M. 2003, \apj, 597, 957

\bibitem[{{Helder} {et~al.}(2012){Helder}, {Vink}, {Bykov}, {Ohira}, {Raymond},
  \& {Terrier}}]{HelderSNR2012}
{Helder}, E.~A., {Vink}, J., {Bykov}, A.~M., {et~al.} 2012, \ssr, 173, 369

\bibitem[{{Hey} {et~al.}(1946){Hey}, {Parsons}, \&
  {Phillips}}]{HeyFluctuationRF1946}
{Hey}, J.~S., {Parsons}, S.~J., \& {Phillips}, J.~W. 1946, \nat, 158, 234

\bibitem[{{Johnson} \& {Morgan}(1954)}]{JohnsonObscuredO-Association1954}
{Johnson}, H.~L. \& {Morgan}, W.~W. 1954, \apj, 119, 344

\bibitem[{Kafexhiu {et~al.}(2014)Kafexhiu, Aharonian, Taylor, \&
  Vila}]{Kafexhiu:2014}
Kafexhiu, E., Aharonian, F., Taylor, A.~M., \& Vila, G.~S. 2014, Physical
  Review D, 90, 123014, arXiv: 1406.7369

\bibitem[{{Kaur} {et~al.}(2020){Kaur}, {Sharma}, {Dewangan}, {Ojha},
  {Durgapal}, \& {Panwar}}]{Kaur+2020}
{Kaur}, H., {Sharma}, S., {Dewangan}, L.~K., {et~al.} 2020, \apj, 896, 29

\bibitem[{{Kelner} {et~al.}(2006){Kelner}, {Aharonian}, \&
  {Bugayov}}]{kelner2006}
{Kelner}, S.~R., {Aharonian}, F.~A., \& {Bugayov}, V.~V. 2006, \prd, 74, 034018

\bibitem[{{Khangulyan} {et~al.}(2014){Khangulyan}, {Aharonian}, \&
  {Kelner}}]{Khangulyan+:2014}
{Khangulyan}, D., {Aharonian}, F.~A., \& {Kelner}, S.~R. 2014, \apj, 783, 100

\bibitem[{{Kiminki} {et~al.}(2015){Kiminki}, {Kobulnicky}, {Vargas
  {\'A}lvarez}, {Alexander}, \&
  {Lundquist}}]{KiminkiPredictingGAIA'sParallax2015}
{Kiminki}, D.~C., {Kobulnicky}, H.~A., {Vargas {\'A}lvarez}, C.~A.,
  {Alexander}, M.~J., \& {Lundquist}, M.~J. 2015, \apj, 811, 85

\bibitem[{{Kn{\"o}dlseder}(2000)}]{KnodlsederOB22000}
{Kn{\"o}dlseder}, J. 2000, \aap, 360, 539

\bibitem[{{Koyama} {et~al.}(1995){Koyama}, {Petre}, {Gotthelf}, {Hwang},
  {Matsuura}, {Ozaki}, \& {Holt}}]{KoyamaXraySNR1995}
{Koyama}, K., {Petre}, R., {Gotthelf}, E.~V., {et~al.} 1995, \nat, 378, 255

\bibitem[{{Kroupa}(2001)}]{KroupaIMF2001}
{Kroupa}, P. 2001, \mnras, 322, 231

\bibitem[{{Kudritzki} \& {Puls}(2000)}]{KudritzkiWindMassiveStars2000}
{Kudritzki}, R.-P. \& {Puls}, J. 2000, \araa, 38, 613

\bibitem[{{Lancaster} {et~al.}(2021){Lancaster}, {Ostriker}, {Kim}, \&
  {Kim}}]{LancasterFragmentation2021}
{Lancaster}, L., {Ostriker}, E.~C., {Kim}, J.-G., \& {Kim}, C.-G. 2021, \apj,
  914, 89

\bibitem[{{Leitherer} {et~al.}(1992){Leitherer}, {Robert}, \&
  {Drissen}}]{LeithererDepMassMomEnergy1992}
{Leitherer}, C., {Robert}, C., \& {Drissen}, L. 1992, \apj, 401, 596

\bibitem[{{Li} {et~al.}(2023){Li}, {Qiu}, {Li}, {Wang}, {Cao}, {Liu}, {Ma}, \&
  {Yang}}]{LiCygXNFilament2023}
{Li}, C., {Qiu}, K., {Li}, D., {et~al.} 2023, \apjl, 948, L17

\bibitem[{{Lozinskaya} {et~al.}(2002){Lozinskaya}, {Pravdikova}, \&
  {Finoguenov}}]{LozinskayaShellSweptUpOB22002}
{Lozinskaya}, T.~A., {Pravdikova}, V.~V., \& {Finoguenov}, A.~V. 2002,
  Astronomy Letters, 28, 223

\bibitem[{{Massey} \& {Thompson}(1991)}]{MasseyMassiveStarsCYGOB21991}
{Massey}, P. \& {Thompson}, A.~B. 1991, \aj, 101, 1408

\bibitem[{{Menchiari}(2023)}]{MenchiariPhDThesis2023}
{Menchiari}, S. 2023, arXiv e-prints, arXiv:2307.03477

\bibitem[{Morlino {et~al.}(2021)Morlino, Blasi, Peretti, \&
  Cristofari}]{Morlino+2021}
Morlino, G., Blasi, P., Peretti, E., \& Cristofari, P. 2021, Monthly Notices of
  the Royal Astronomical Society, 504, 6096, arXiv: 2102.09217

\bibitem[{{Neronov} {et~al.}(2023){Neronov}, {Semikoz}, \&
  {Savchenko}}]{Neronov-neutrini:2023}
{Neronov}, A., {Semikoz}, D., \& {Savchenko}, D. 2023, arXiv e-prints,
  arXiv:2311.13711

\bibitem[{{Nieuwenhuijzen} \& {de Jager}(1990)}]{NieuwenhuijzenMdot1990}
{Nieuwenhuijzen}, H. \& {de Jager}, C. 1990, \aap, 231, 134

\bibitem[{{Nugis} \& {Lamers}(2000)}]{NugisMdotWR2000}
{Nugis}, T. \& {Lamers}, H.~J.~G.~L.~M. 2000, \aap, 360, 227

\bibitem[{{Orlando} {et~al.}(2008){Orlando}, {Bocchino}, {Reale}, {Peres}, \&
  {Pagano}}]{Orlando+2008}
{Orlando}, S., {Bocchino}, F., {Reale}, F., {Peres}, G., \& {Pagano}, P. 2008,
  \apj, 678, 274

\bibitem[{{Piddington} \& {Minnett}(1952)}]{PiddingtonRFRadCyg1952}
{Piddington}, J.~H. \& {Minnett}, H.~C. 1952, Australian Journal of Scientific
  Research A Physical Sciences, 5, 17

\bibitem[{{Prantzos}(2012)}]{Prantzos:2012}
{Prantzos}, N. 2012, \aap, 538, A80

\bibitem[{{Reddish} {et~al.}(1966){Reddish}, {Lawrence}, \&
  {Pratt}}]{ReddishOB21966}
{Reddish}, V.~C., {Lawrence}, L.~C., \& {Pratt}, N.~M. 1966, Publications of
  the Royal Observatory of Edinburgh, 5, 111

\bibitem[{Reimer {et~al.}(2006)Reimer, Pohl, \&
  Reimer}]{reimerNonthermalHighEnergy2006}
Reimer, A., Pohl, M., \& Reimer, O. 2006, The Astrophysical Journal, 644, 1118

\bibitem[{{Reipurth} \&
  {Schneider}(2008)}]{ReipurthStarFormationAndYCinCyg2008}
{Reipurth}, B. \& {Schneider}, N. 2008, in Handbook of Star Forming Regions,
  Volume I, ed. B.~{Reipurth}, Vol.~4, 36

\bibitem[{{Renzo} {et~al.}(2017){Renzo}, {Ott}, {Shore}, \& {de
  Mink}}]{RenzoSystematicSurveyMdot2017}
{Renzo}, M., {Ott}, C.~D., {Shore}, S.~N., \& {de Mink}, S.~E. 2017, \aap, 603,
  A118

\bibitem[{{Reynolds}(2008)}]{ReynoldsSNR2008}
{Reynolds}, S.~P. 2008, \araa, 46, 89

\bibitem[{Rygl {et~al.}(2012)Rygl, Brunthaler, Sanna, Menten, Reid, van
  Langevelde, Honma, Torstensson, \&
  Fujisawa}]{ryglParallaxesProperMotions2012}
Rygl, K. L.~J., Brunthaler, A., Sanna, A., {et~al.} 2012, Astronomy \&
  Astrophysics, 539, A79

\bibitem[{Saha {et~al.}(2020)Saha, Domínguez, Tibaldo, Marchesi, Ajello,
  Lemoine-Goumard, \& López}]{sahaMorphologicalSpectralStudy2020}
Saha, L., Domínguez, A., Tibaldo, L., {et~al.} 2020, The Astrophysical
  Journal, 897, 131, arXiv: 2006.00274

\bibitem[{Schneider {et~al.}(2006)Schneider, Bontemps, Simon, Jakob, Motte,
  Miller, Kramer, \& Stutzki}]{schneiderNewViewCygnus2006}
Schneider, N., Bontemps, S., Simon, R., {et~al.} 2006, Astronomy \&
  Astrophysics, 458, 855

\bibitem[{{Schneider} {et~al.}(2010){Schneider}, {Csengeri}, {Bontemps},
  {Motte}, {Simon}, {Hennebelle}, {Federrath}, \&
  {Klessen}}]{SchneiderDR21StarFormation2010}
{Schneider}, N., {Csengeri}, T., {Bontemps}, S., {et~al.} 2010, \aap, 520, A49

\bibitem[{{Subedi} {et~al.}(2017){Subedi}, {Sonsrettee}, {Blasi}, {Ruffolo},
  {Matthaeus}, {Montgomery}, {Chuychai}, {Dmitruk}, {Wan}, {Parashar}, \&
  {Chhiber}}]{Subedi+2017}
{Subedi}, P., {Sonsrettee}, W., {Blasi}, P., {et~al.} 2017, \apj, 837, 140

\bibitem[{Takekoshi {et~al.}(2019)Takekoshi, Fujita, Nishimura, Taniguchi,
  Yamagishi, Matsuo, Ohashi, Tokuda, \&
  Minamidani}]{takekoshiNobeyama45mCygnus2019}
Takekoshi, T., Fujita, S., Nishimura, A., {et~al.} 2019, The Astrophysical
  Journal, 883, 156, arXiv: 1907.12776

\bibitem[{{Tatischeff} {et~al.}(2021){Tatischeff}, {Raymond}, {Duprat},
  {Gabici}, \& {Recchia}}]{Tatischeff+2021}
{Tatischeff}, V., {Raymond}, J.~C., {Duprat}, J., {Gabici}, S., \& {Recchia},
  S. 2021, \mnras, 508, 1321

\bibitem[{Taylor {et~al.}(2003)Taylor, Gibson, Peracaula, Martin, Landecker,
  Brunt, Dewdney, Dougherty, Gray, Higgs, Kerton, Knee, Kothes, Purton,
  Uyaniker, Wallace, \& Willis}]{taylorCANADIANGALACTICPLANE2003}
Taylor, A.~R., Gibson, S.~J., Peracaula, M., {et~al.} 2003, 125, 20

\bibitem[{{Torres-Dodgen} {et~al.}(1991){Torres-Dodgen}, {Tapia}, \&
  {Carroll}}]{TorresDodgenPhotometryOB1991}
{Torres-Dodgen}, A.~V., {Tapia}, M., \& {Carroll}, M. 1991, \mnras, 249, 1

\bibitem[{{Uyan{\i}ker} {et~al.}(2001){Uyan{\i}ker}, {F{\"u}rst}, {Reich},
  {Aschenbach}, \& {Wielebinski}}]{UyanikerCygSB2001}
{Uyan{\i}ker}, B., {F{\"u}rst}, E., {Reich}, W., {Aschenbach}, B., \&
  {Wielebinski}, R. 2001, \aap, 371, 675

\bibitem[{{Vieu} {et~al.}(2022{\natexlab{a}}){Vieu}, {Gabici}, {Tatischeff}, \&
  {Ravikularaman}}]{VieuCRinSuperbubble2022}
{Vieu}, T., {Gabici}, S., {Tatischeff}, V., \& {Ravikularaman}, S.
  2022{\natexlab{a}}, \mnras, 512, 1275

\bibitem[{{Vieu} {et~al.}(2022{\natexlab{b}}){Vieu}, {Reville}, \&
  {Aharonian}}]{Vieu-Reville-Aharonian:2022}
{Vieu}, T., {Reville}, B., \& {Aharonian}, F. 2022{\natexlab{b}}, \mnras, 515,
  2256

\bibitem[{{Vink} {et~al.}(2001){Vink}, {de Koter}, \& {Lamers}}]{VinkMdot2001}
{Vink}, J.~S., {de Koter}, A., \& {Lamers}, H.~J.~G.~L.~M. 2001, \aap, 369, 574

\bibitem[{{Weaver} {et~al.}(1977){Weaver}, {McCray}, {Castor}, {Shapiro}, \&
  {Moore}}]{Weaver-McCray:1977}
{Weaver}, R., {McCray}, R., {Castor}, J., {Shapiro}, P., \& {Moore}, R. 1977,
  \apj, 218, 377

\bibitem[{{Weidner} \& {Kroupa}(2004)}]{WeidnerMaxStellarMassInYMSC2004}
{Weidner}, C. \& {Kroupa}, P. 2004, \mnras, 348, 187

\bibitem[{{Wiedenbeck} \& {Greiner}(1981)}]{WiedenbeckNe1981}
{Wiedenbeck}, M.~E. \& {Greiner}, D.~E. 1981, \prl, 46, 682

\bibitem[{{Wilson} {et~al.}(2009){Wilson}, {Rohlfs}, \&
  {H{\"u}ttemeister}}]{ToolsOfRadioAstronomy}
{Wilson}, T.~L., {Rohlfs}, K., \& {H{\"u}ttemeister}, S. 2009, {Tools of Radio
  Astronomy}

\bibitem[{{Wright} {et~al.}(2010){Wright}, {Drake}, {Drew}, \&
  {Vink}}]{WrightOB2SFH2010}
{Wright}, N.~J., {Drake}, J.~J., {Drew}, J.~E., \& {Vink}, J.~S. 2010, \apj,
  713, 871

\bibitem[{{Wright} {et~al.}(2015){Wright}, {Drew}, \&
  {Mohr-Smith}}]{WrightMassiveStarPopOB22015}
{Wright}, N.~J., {Drew}, J.~E., \& {Mohr-Smith}, M. 2015, \mnras, 449, 741

\bibitem[{Yang {et~al.}(2018)Yang, de~Oña~Wilhelmi, \&
  Aharonian}]{yangDiffuseRayEmissionin2018}
Yang, R.-z., de~Oña~Wilhelmi, E., \& Aharonian, F. 2018, Astronomy \&
  Astrophysics, 611, A77

\bibitem[{{Yungelson} {et~al.}(2008){Yungelson}, {van den Heuvel}, {Vink},
  {Portegies Zwart}, \& {de Koter}}]{YungelsonEvolutionFateMassiveStars2008}
{Yungelson}, L.~R., {van den Heuvel}, E.~P.~J., {Vink}, J.~S., {Portegies
  Zwart}, S.~F., \& {de Koter}, A. 2008, \aap, 477, 223

\end{thebibliography}
\begin{appendix} %First appendix
\section{Estimating the cluster wind parameters from the observed stellar population}
\label{Apndx:MassLossRatio}
In order to calculate the total mass loss rate and wind luminosity of Cyg~OB2, one needs to compute the mass loss rate $\dot{M}_i$ and the terminal wind speed $v_{\infty,i}$ for every i-th member of the association. Both the mass loss rate and the terminal wind speed are functions of global stellar properties: mass $M_i$, bolometric luminosity $L_i$, radius $R_{\star, i}$, and Temperature $T_{\rm eff,i}$. 
Mass, luminosity and effective temperature (together with the associated uncertainties) are taken from the sample of stars at the core of the association by \citet{WrightMassiveStarPopOB22015}, while to calculate the stellar radius we use the empirical relation provided by \cite{DemircanStarsMRR1991}:
\begin{equation}
\label{eq:MRR}
R_{\star}=0.85 \left(\frac{M_\star}{M_\odot} \right)^{0.67} R_\odot \, .
\end{equation} 
%$R_{\star, i}=0.125 (M_i/M_\odot)^{1.02}$ according to \citet{YungelsonEvolutionFateMassiveStars2008}.
The wind terminal velocity, following \citep{KudritzkiWindMassiveStars2000}, is given by:
\begin{equation}
\label{eq:StellarWindV}
 v_{\infty,i}=\mathcal{C} \, v_{\rm esc, \it i}
\end{equation}
where $\mathcal{C}=2.65$ for $T_{\rm eff,i}\geq 2.1\times 10^4$~K \citep{KudritzkiWindMassiveStars2000}, which is adequate for the considered stellar sample, $v_{\rm esc, \it i}=\sqrt{2GM_i(1-\Gamma_{\rm edd})/R_i}$ is the escape velocity, $G$ is the gravitational constant, and $\Gamma_{\rm edd}$ is the luminosity in units of the Eddington luminosity.

We use two different recipes to calculate $\dot{M}_i$. 
The first one is based on a theoretical formula given by \cite{YungelsonEvolutionFateMassiveStars2008}
\begin{equation} \label{eq:MdotYoungelson}
    \dot{M}_i=\frac{L_i}{v_{\infty, i} c} \frac{1}{(1-\Gamma)^{-0.25}}\,,
\end{equation}
while the second approach relies on an empirical relation for the mass loss rate provided by \cite{VinkMdot2001} and valid for stars with temperature 27500~K$<T_{\rm eff,i}<50000$~K (corresponding to stars with masses larger than $20\,M_{\odot}$) given by:
\begin{equation}
\label{eq:MdotRenzo}
    \begin{split}
        \log \left ( - \frac{\dot{M}_i}{M_\odot yr^{-1}} \right)=-6.668+2.210  \,\log \left (\frac{L_i}{10^5 \ L_\odot} \right ) \\
        -1.339 \,\log \left (\frac{M_i}{30 \ M_\odot} \right )-1.601 \,\log \left (\frac{v_{\infty, i}}{2 v_{\rm esc, \it i}} \right ) \\
        +1.07  \,\log \left (\frac{T_{\rm eff, i}}{40000  \ \textrm{K}} \right )+0.85 \,\log \left (\frac{Z_i}{Z_{\odot}} \right )
    \end{split}
\end{equation}
where $Z_i$ is the stellar metallicity (that we assume to be Solar). 
%When using the Equation \eqref{eq:MdotRenzo} we assume solar metallicity. %\begin{comment}For every star in its sample, \cite{WrightMassiveStarPopOB22015} provide an estimate for the stellar parameters $\tilde{L}_i$, $\tilde{T}_{\rm eff, i}$, and $\tilde{M}_i$ (we will refer to the quantities estimated by \cite{WrightMassiveStarPopOB22015} using the $\sim$ diacritic symbol) together with the associated parameters uncertainties ($\delta \tilde{L}_i$, $\delta \tilde{T}_{\rm eff, i}$, and $\delta \tilde{M}_i$). 
%\end{comment}
%The temperature validity range for Equation~\eqref{eq:MdotRenzo} is respected for all stars in the sample with masses larger than $20\,M_{\odot}$.

\begin{figure}
\centering
\includegraphics[width=\columnwidth]{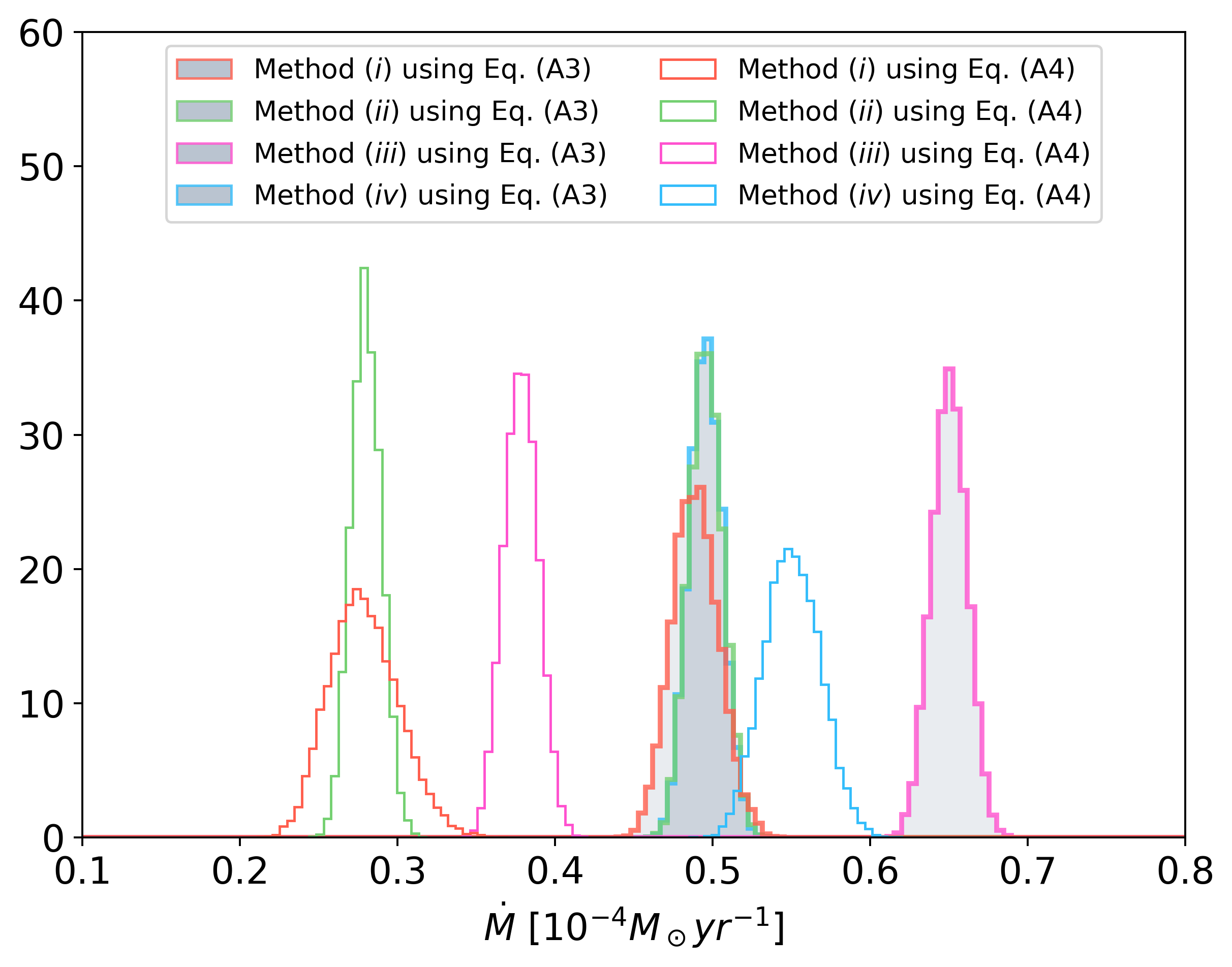}
\caption{Mass loss rate distribution for Cygnus OB2 considering the sample of stars investigated by \protect\cite{WrightMassiveStarPopOB22015}. The stellar parameters are estimated using four different recipes, from (i) to (iv), and for each one the mass loss rate is estimated using both Equation~\eqref{eq:MdotYoungelson} and \eqref{eq:MdotRenzo}. All estimates neglect the contribution due to WR stars.}
\label{fig:MdotNonBinaryStars}
\end{figure}

In order to evaluate the reliability of our estimate of the mass loss rate, we checked how the results depend on the approach adopted to evaluate the quantities that appear in Equation~\eqref{eq:MdotRenzo}. The different approaches are the following: 
%To evaluate the global consistency on the mass loss rate, we estimated $\dot{M}$, using four different approaches:
\begin{enumerate}
    \item We calculate the total mass loss rate using ${L}_i$, ${T}_{\rm eff, i}$, and ${M}_i$ provided by 
    \citet{WrightMassiveStarPopOB22015}.
    \item We use  ${L}_i$ and ${T}_{\rm eff, i}$ from \citet{WrightMassiveStarPopOB22015} while $M_i$ is recomputed as \cite{YungelsonEvolutionFateMassiveStars2008}:
    \begin{equation}
        \label{eq:StellarMFromL}
            \frac{M_i}{M_\odot}=\left [ 10^{-3.48} \left( \frac{L_i}{L_\odot} \right ) \right ]^{0.75} \, . 
        \end{equation}
    \item We use   ${M}_i$ and ${T}_{\rm eff, i}$ from \citet{WrightMassiveStarPopOB22015} while $L_i$ is recomputed inverting Equation~\eqref{eq:StellarMFromL}.
    \item We take only ${L}_i$ from \citet{WrightMassiveStarPopOB22015} while $T_{\rm eff, i}$ is calculated using the Boltzmann law:
        \begin{equation}
            T_{\rm eff, i}=\left (\frac{\tilde{L}_i}{4 \pi R_{\star,i}^2 \sigma_b} \right )^{1/4}
        \end{equation}
        and $M_i$ is recomputed from Equation~\eqref{eq:StellarMFromL} %\cite{YungelsonEvolutionFateMassiveStars2008}:
        %\begin{equation}
        %    T_{\rm eff, i}=\left (\frac{\tilde{L}_i}{4 \pi R_{\star,i}^2 \sigma_b} \right )^{1/4}
        %\end{equation}
        %begin{equation}
        %\label{eq:StellarMFromL}
        %    \frac{M_i}{M_\odot}=\left [ 10^{-3.48} \left( \frac{\tilde{L}_i}{L_\odot} \right ) \right ]^{0.75} \, . 
        %\end{equation}
   % \item As third trial, we use  $\tilde{L}_i$ and %$\tilde{T}_{\rm eff, i}$, while $M_i$ is obtained using %Equation~\eqref{eq:StellarMFromL}.
   % \item Finally, we use $\tilde{T}_{\rm eff, i}$ and %$\tilde{M}_{i}$ while $L_i$ is calculated inverting %equation \ref{eq:StellarMFromL} %\citep{YungelsonEvolutionFateMassiveStars2008}.
\end{enumerate}
To take into account the uncertainties we use a Montecarlo approach, generating $10^4$ different star samples where the parameters of every star have a Gaussian distribution around the mean with variance provided by \citet{WrightMassiveStarPopOB22015}.
%for which the used measured parameters ($\tilde{L}_i$, $\tilde{T}_{\rm eff, i}$, and $\tilde{M}_i$) are gaussian-fluctuated by a quantity corresponding to the associated error ($\delta \tilde{L}_i$, $\delta \tilde{T}_{\rm eff, i}$, and $\delta \tilde{M}_i$). 
Figure~\ref{fig:MdotNonBinaryStars} shows the result for the $\dot{M}$ calculation after considering only single stars (binary systems are not included) with $M_i>20 M_\odot$ and without the contribution of Wolf-Rayet (WR) stars. The value of $\dot{M}$ is found to lie in the range $[0.25 \times 10^{-4},\,0.65 \times 10^{-4}]$~M$_\odot$~yr$^{-1}$.

To further account for the the contribution of the three Wolf-Rayet  stars associated to Cyg~OB2 we adopt the  empirical relation by \cite{NugisMdotWR2000}: 
\begin{equation}
%\label{eq:MdotRenzo}
    \begin{split}
        \log \left ( - \frac{\dot{M}_{\textrm{WR},\ i}}{M_\odot yr^{-1}} \right)= -11.0+1.29  \,\log \left (\frac{L_{\textrm{WR}, \ i}}{10^5 \ L_\odot} \right ) \\
        +1.73  \,\log \left (\frac{Y_{\textrm{WR}, \ i}}{Y_\odot} \right )+0.47 \,\log \left (\frac{Z_{\textrm{WR}, \ i}}{Z_{\odot}} \right ) \,.
    \end{split}
\end{equation}
where both the Helium fraction ($Y_{\textrm{WR},i}$) and metallicity ($Z_{\textrm{WR},i}$) are set to the Solar value. We find a contribution to the mass loss rate from WR stars of $\sim 0.3 \times 10^{-4}$ M$_\odot$yr$^{-1}$
% \Sigma_i \dot{M}_{\textrm{WR}\ i}(\tilde{L}_{\textrm{WR}, \ i})\sim 0.3 \times 10^{-4}$ M$_\odot$yr$^{-1}$. 

The star sample of \cite{WrightMassiveStarPopOB22015} includes also 9 known binary systems, whose contribution to $\dot{M}$ is not straightforward to quantify. However, we can provide a rough estimate assuming that the total luminosity of the system results from two companions of equal luminosity. Then, for each of the stars, we adopt method iv described above for the calculation of $\dot{M}_i$ (using only ${L}_i$), we find $\dot{M}_{\textrm{Binary}}\in [0.18 \times 10^{-4}, \,0.6 \times 10^{-4}]$~M$_\odot$~yr$^{-1}$.
By summing up all the contributions, Cyg~OB2 mass loss rate should lie in a conservative range of $\dot{M}\in [0.7 \times 10^{-4},\, 1.5 \times 10^{-4}]$~M$_\odot$~yr$^{-1}$. 

\begin{figure}
\centering
\includegraphics[width=\columnwidth]{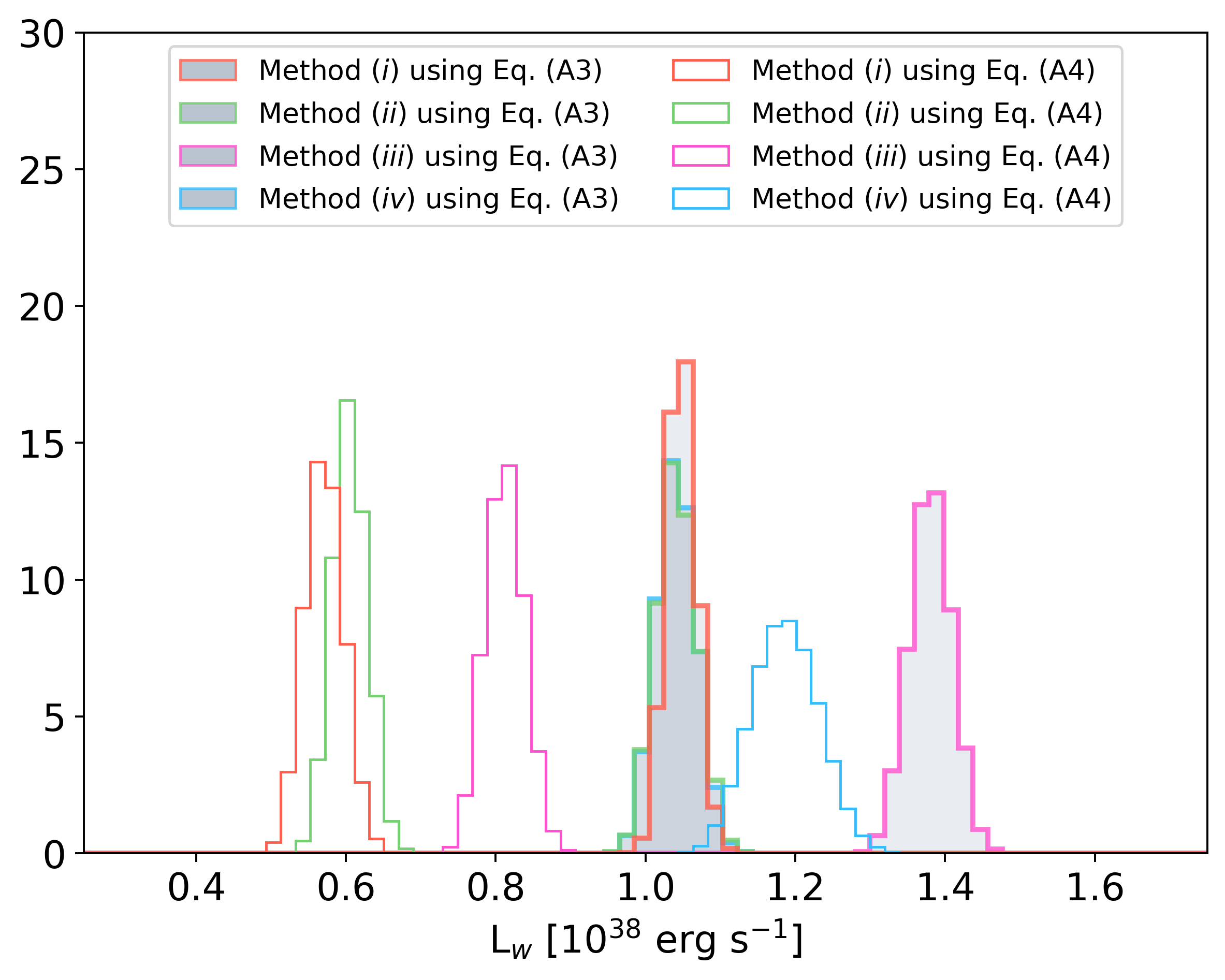}
\caption{Wind luminosity of Cygnus OB2 inferred using the estimated mass loss rates (see text).}
\label{fig:LwNonBinaryStars}
\end{figure}

Once the mass loss rate of every star in the cluster is known, the associated wind speed is calculated from momentum conservation:
\begin{equation}
\label{eq:LwindStar}
 v_{\rm w} = \frac{\sum_i \dot{M}_i \,v_{\rm w, \it i}}{\sum_i \dot{M}} \,.
\end{equation}
The cluster wind luminosity is then:
\begin{equation}
\label{eq:LwindCluster}
 L_{\rm w} = \frac{1}{2} v_{\rm w}^2 \, \sum_i \dot{M_i}  \,.
\end{equation}
The contribution of single, non WR stars, accounting for both the values of $\dot{M}_i$ inferred using the theoretical and empirical recipes of \cite{YungelsonEvolutionFateMassiveStars2008} and \cite{VinkMdot2001}, is in the range $L_{\rm w}^{\rm nbs} \in [0.4\times 10^{38}, 1.6 \times 10^{38}]$~erg~\,s$^{-1}$ (see Figure~\ref{fig:LwNonBinaryStars}). The wind luminosity contribution from the three Wolf-Rayet stars included in the sample of \cite{WrightMassiveStarPopOB22015} (assuming an average wind speed of 2000 km s$^{-1}$ is $L_{\rm w}^{\rm WR} \simeq 0.6 \times 10^{38}$ erg\,s$^{-1}$. Finally, the contribution of the 9 binary systems is $L_{\rm w}^{\rm bs} \in [0.55\times 10^{38}, \, 0.7  \times 10^{38}]$ erg\,s$^{-1}$. Accounting for all contributions, Cyg~OB2 wind luminosity should lie in the range $L_{\rm w} \in [1.55\times 10^{38},\, 2.9\times 10^{38}]$~erg~s$^{-1}$.

It is worth noticing that our estimate is compatible with the results of other works where different approaches have been used. For example, \cite{LozinskayaShellSweptUpOB22002} used the total mass of the association and the average wind luminosity per solar mass provided by \cite{LeithererDepMassMomEnergy1992}, obtaining $L_{\rm w} \simeq 1 \times 10^{39}$~erg~s$^{-1}$. However this value is likely overestimated as it is based on the stellar population study done by \cite{KnodlsederOB22000}, which likely overestimated the number of stars. Following the same approach and rescaling the mass to that measured by \cite{WrightMassiveStarPopOB22015}, the expected wind luminosity is $\sim 3.3 \times 10^{38}$~erg~s$^{-1}$. Similarly, \cite{ackermannCocoonFreshlyAccelerated2011a} evaluate a wind luminosity of $L_{\rm w} \in [2\times 10^{38},\, 3 \times 10^{38}]$~erg~s$^{-1}$ considering a different sample of stars that includes the presence of 17 stars with $M_\star > 35 M_\odot$ and 5 Wolf-Rayet. 
Finally, in a recent analysis of the diffuse X-ray emission, \cite{Albacete-Colombo+2023} provide an estimate of the total wind luminosity of $\approx 1.5 \times 10^{38}$~erg~s$^{-1}$.


%%% Section %%%
\section{Estimating the cluster wind parameters from the stellar IMF}
\label{App:Alternative_LwMdot}
An alternative approach to estimate the cluster wind parameters based only on the cluster stellar mass and age was presented by \cite{MenchiariPhDThesis2023}. This approach is especially useful if the stellar population is not known. The method can be summarized as follows: knowing the cluster mass, a mock stellar population is simulated by random sampling an assumed initial mass function. Afterward, given the age of the cluster, all stars massive enough to have exploded as supernova are removed from the population. The cluster wind properties are then calculated after estimating the wind parameters for all the mock stars.
This method can be used to cross check the results obtained in Appendix \ref{Apndx:MassLossRatio} for the main sequence stars, with the additional advantage that we can estimate the number of supernovae exploded within the cluster. The latter is useful to understand whether the energetics of the bubble is dominated by winds or rather by supernovae explosions. Finally, we apply this alternative method to NGC 6910.

We here assume the initial mass function by \cite{KroupaIMF2001}, which is suitable for Cyg~OB2 \citep{WrightMassiveStarPopOB22015}. 
Stars are sampled in the mass range between $M_{\star,\min}=0.08$ M$_\odot$ and $M_{\star,\max}=150$ M$_\odot$. The lower limit on the mass range is associated to the minimum mass required to start hydrogen nuclear burning \citep{CarrollIntroModAstro1996}, while the maximum mass is an empirical upper limit fixed by observations \citep{WeidnerMaxStellarMassInYMSC2004}. Being $\xi_\star (M_\star)$ the initial mass function, the total number of stars in the cluster and the cluster mass are:
\begin{align} 
  & N_\star(M_{\rm SC})= \int_{M_{\star,\min}}^{M_{\star,\max}} \xi_\star(M_\star)dM_\star  \,, \label{eq:N_stars} \\
  & M_{\rm SC} = \int_{M_{\star,\min}}^{M_{\star,\max}} M_{\star} \, \xi_\star(M_\star)dM_\star  \,. \label{eq:Msc}
\end{align}
Once the initial stellar population has been generated, all stars that have left the main sequence are removed from the sample (i.e. we neglect any evolutionary phase out of the main sequence), assuming the turnover time ($t_{\rm TO}$) for a star of mass $M_\star$, given by \citep{BuzzoniTOtime2002}:
\begin{equation}
\label{eq:tTO}
 \log \left(\frac{t_{\rm TO}}{1 \rm \ yr} \right) = 0.825 \log^2 \left(\frac{M_{\star}}{120 \rm \ M_\odot} \right) + 6.43 \ .
\end{equation}
For a given stellar mass we set the luminosity through the mass-luminosity relation presented in \cite{MenchiariPhDThesis2023}. This consists of a smoothed broken power law joining two different empirical mass-luminosity relations: the first one provided by \cite{EkerMLR2018} and valid between 0.179--31~M$_\odot$, and the second one provided by \cite{YungelsonEvolutionFateMassiveStars2008} and valid for very massive stars ($M_\star \gtrsim 35$ M$_\odot$): 
\begin{equation}
\label{eq:MergedMLR}
L_\star =  
\begin{cases}
  L_{\rm b1} \left(\frac{M_\star}{M_{\rm b1}} \right)^{\alpha_1} \left[ \frac{1}{2}+ \frac{1}{2} \left( \frac{M_\star}{M_{\rm b1}} \right )^{1/\Delta_1}\right ]^{(-\alpha_1+\alpha_2)\Delta_1} \\
  \hspace{4cm} \text{for } 2.4 \leq \frac{M_\star}{\rm M_\odot} < 12  \\ \\
  L_{\rm b2} \left(\frac{M_\star}{M_{\rm b2}} \right)^{\alpha_2} \left[ \frac{1}{2}+ \frac{1}{2} \left( \frac{M_\star}{M_{\rm b2}} \right )^{1/\Delta_2}\right ]^{(-\alpha_2+\alpha_3)\Delta_2}  \\
  \hspace{4cm} \text{for } M_\star \geq 12 \text{ M}_\odot 
\end{cases}
\end{equation}
where $L_{\rm b1}=3191$ L$_{\odot}$, $L_{\rm b2}=301370$ L$_{\odot}$, $M_{\rm b1}=7$ M$_\odot$ and $M_{\rm b2}=36.089$ M$_\odot$. The value of $M_{\rm b2}$ is the intersection point between the two mass luminosity relations of \cite{YungelsonEvolutionFateMassiveStars2008} and \cite{EkerMLR2018}. The power law indexes are $\alpha_1=3.97$, $\alpha_2=2.86$, and $\alpha_3=1.34$. The parameters $\Delta_1$ and $\Delta_2$ are used to smooth the transition between the different power law components. We fix the two parameters to 0.01 and 0.15, respectively. Note that in the interval 2.4--12 M$_\odot$ the mass luminosity relation is taken from \cite{EkerMLR2018} and adapted so as to ensure continuity at $M_{\rm b1}=7$ M$_\odot$.

%an adaptation of the one provided by \cite{EkerMLR2018} in the form of a smoothed power law. We decided to use this form because the original relation is not continuous at $M_{\rm b1}=7$ M$_\odot$.

To calculate the stellar radius, we use the same relation \eqref{eq:MRR} adopted in Appendix~\ref{Apndx:MassLossRatio}.
Given that the mock stellar population includes also stars with masses below 20 M$_\odot$, to calculate the mass loss rate, we use the empirical relation by \cite{NieuwenhuijzenMdot1990}:
\begin{align}
\label{eq:MdotNieu}
 \log \left( \frac{\dot{M}_{\star}}{\rm M_\odot yr^{-1}} \right) = -14.02+ 1.24 \log \left( \frac{L_\star}{\rm L_\odot} \right)  \nonumber +\\
    \hspace{2cm} + 0.16 \log \left( \frac{M_\star}{\rm M_\odot} \right) + 0.81 \left( \frac{R_\star}{\rm R_\odot} \right) \ ,
\end{align}
which is valid for stars with temperatures above 5000~K, ($\dot{M}_{\star}\gtrsim 3$ M$_\odot$). To this purpose, when calculating the wind properties we only account for stars with $M_\star > 3$ M$_\odot$, corresponding to O and B type stars. The contribution to the wind power and mass loss rate for stars with lower masses is totally negligible \citep{MenchiariPhDThesis2023}. 

%In Equation~\eqref{eq:MdotNieu}, $R_\star$ and $L_\star$ represent the stellar radius and bolometric luminosity, respectively. Both parameters are calculated using ad hoc empirical relations. For the stellar luminosity, we consider the mass-luminosity relation presented in \cite{MenchiariPhDThesis2023}. This consists of a smoothed broken power law mixing two different empirical mass-luminosity relations: the first one provided by \cite{EkerMLR2018} and valid between 0.179--31 M$_\odot$, and the second one provided by \cite{YungelsonEvolutionFateMassiveStars2008} valid for very massive stars ($M_\star \gtrsim 35$ M$_\odot$). The final expression is:
%\begin{equation}
%\label{eq:MergedMLR}
%L_\star =  
%\begin{cases}
%  L_{\rm b1} \left(\frac{M_\star}{M_{\rm b1}} \right)^{\alpha_1} \left[ \frac{1}{2}+ \frac{1}{2} \left( \frac{M_\star}{M_{\rm b1}} \right )^{1/\Delta_1}\right ]^{(-\alpha_1+\alpha_2)\Delta_1} \\
%  \hspace{4cm} \text{for } 2.4 \leq \frac{M_\star}{\rm M_\odot} < 12  \\ \\
%  \mathcal{K}\, L_{\rm b2} \left(\frac{M_\star}{M_{\rm b2}} \right)^{\alpha_2} \left[ \frac{1}{2}+ \frac{1}{2} \left( \frac{M_\star}{M_{\rm b2}} \right )^{1/\Delta_2}\right ]^{(-\alpha_2+\alpha_3)\Delta_2}  \\
%  \hspace{4cm} \text{for } M_\star \geq 12 \text{ M}_\odot 
%\end{cases}
%\end{equation}

Once the  parameters of each single star are known, the stellar wind power and velocity are calculated using Equations~\eqref{eq:StellarWindV} and \eqref{eq:LwindStar}. The collective cluster wind power and mass loss rate are then calculated by summing up all the contributions of the mock stars.

Cyg~OB2 has a total stellar mass of $16500^{+3800}_{-2800}$ M$_\odot$ and an age between 3\--5 Myr \citep{WrightMassiveStarPopOB22015}. Assuming an age of 3 Myr, after simulating $10^3$ different mock stellar populations, we found an average wind luminosity and mass loss rate of ${L_{\rm w}}=0.8 \times 10^{38}$ erg\,s$^{-1}$ and $\dot{M}=0.3 \times 10^{-4}$ M$_\odot$ yr$^{-1}$ (see Figure~\ref{fig:LwMdotSynth}), which are consistent with the estimates obtained in Appendix \ref{Apndx:MassLossRatio}, neglecting WR stars.
Moreover, as emphasized by \cite{BerlanasDisentanglingSpatialSubstructure2019}, 10\% of the stellar mass of Cyg~OB2 is likely associated to a foreground OB association. By rescaling the mass accordingly,  the wind luminosity and mass loss rate values change by only $\sim 10\%$, and are therefore consistent with the estimates provided in Appendix \ref{Apndx:MassLossRatio}.

One can wonder how the power injected by possible SNe compares to that of the wind. The above procedure can also be used to estimate the number of SNe exploded in the cluster by counting how many massive stars have a main sequence phase lasting less than the cluster age. For $\sim 3$\,Myr the expected number of supernova is $7 \pm 2.5$. Assuming that each SN releases an energy of $10^{51}$ erg, the total is $\sim 7\times 10^{51}$ erg,  a factor $\sim 2.5$ smaller than the total energy injected by the stellar winds. The latter can be found by integrating in time the cluster wind luminosity, and is $\sim 1.8 \times 10^{52}$ erg. We, hence, expect that the bubble dynamics and particle acceleration process is still dominated by the power of stellar winds. However, this is not true anymore if the age of Cyg~OB2 is $\sim 5$\,Myr. In such a case, the expected number of SNe rises to $26 \pm 5$, resulting into total power slightly larger than the wind power. For the sake of completeness, we also mention that from an observational point of view there are no clear indications of the presence of recent SN explosions.

Finally, as mentioned in the Introduction, close to Cyg~OB2 there is a second cluster, NGC~6910. By using the same method described above we can calculate its wind luminosity. The cluster has an estimated age of $\sim 4.5$~Myr and a lower limit to the mass of $\sim 775$~$M_{\odot}$ \citep{Kaur+2020}. NGC~6910 is also characterized by an initial mass function scaling as to $M_\star^{-1.7}$ \citep{Kaur+2020}, harder than the one by \cite{KroupaIMF2001}. Using these values we get an average wind luminosity and mass loss rate of ${L_{\rm w}}=1.8 \times 10^{36}$ erg\,s$^{-1}$ and ${\dot{M}}=7.6 \times 10^{-7}$ M$_\odot$ yr$^{-1}$. The expected number of supernovae is, instead, $\sim 4 \pm 2$. Even doubling the value of the estimated cluster mass, the final luminosity of NGC~6910 will remain negligible compared to Cyg~OB2. 

\begin{figure}
\centering
\includegraphics[width=\columnwidth]{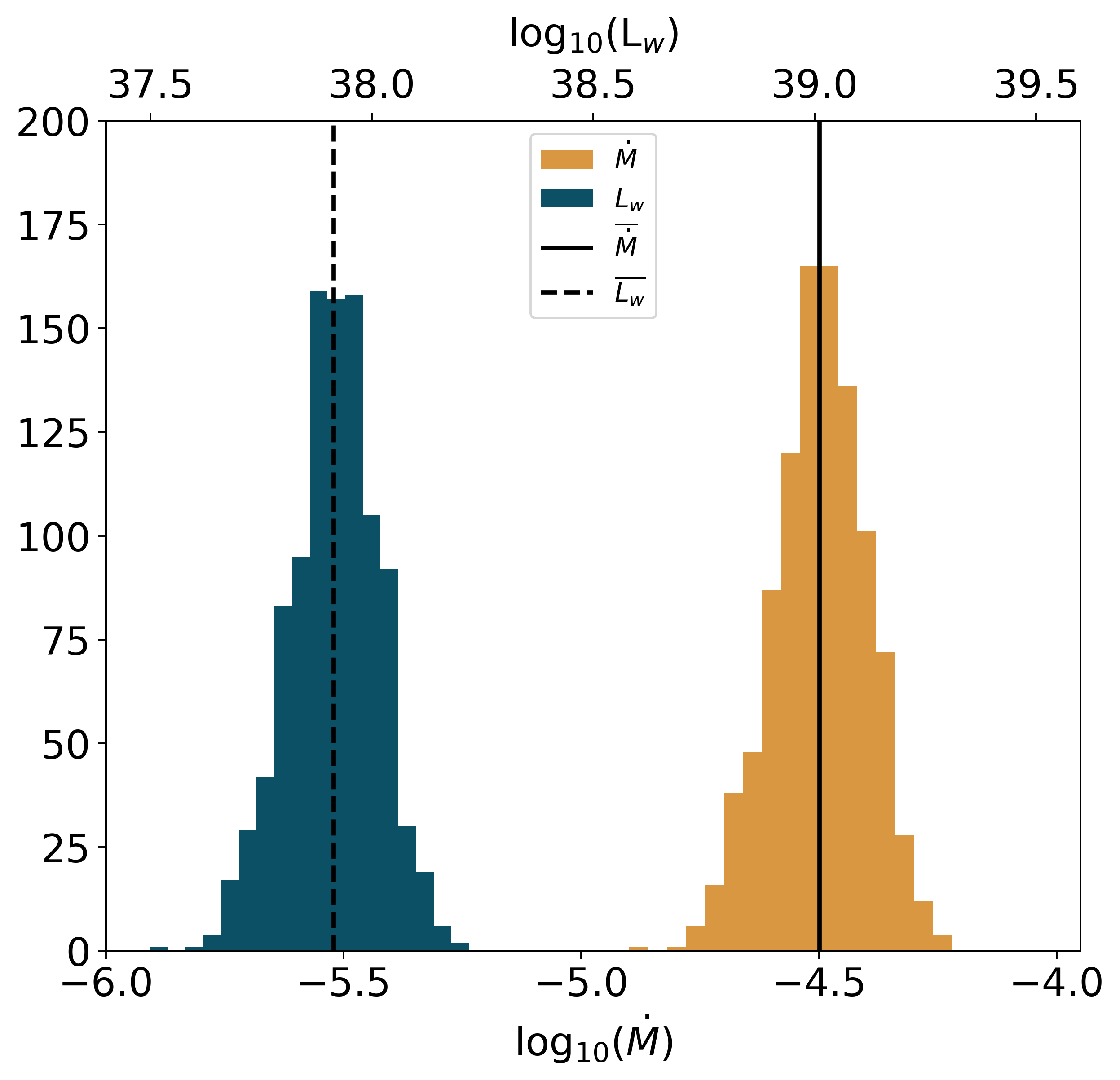}
\caption{Distribution of wind luminosity (left side histogram) and mass loss rate (right side histogram) of Cygnus OB2 inferred using a montecarlo method based on a mock stellar population. The values are obtained after generating 1000 different stellar populations. The dashed and solid lines represent the average wind luminosity and mass loss rate respectively.}
\label{fig:LwMdotSynth}
\end{figure}
\end{appendix}

\end{document}